\documentclass{article}
\usepackage[utf8]{inputenc}

\usepackage[margin=1in,marginparwidth=65pt,marginparsep=5pt]{geometry}
\usepackage{color}
\usepackage{graphicx}
\usepackage{epsfig}
\usepackage{epstopdf}
\usepackage{subfig}
\usepackage{multirow}
\usepackage[font=small,skip=0pt]{caption}

\usepackage[dvipsnames]{xcolor}

\usepackage{amsmath}
\usepackage{lipsum}
\usepackage{amssymb}
\usepackage{multirow}
\usepackage{amsthm}

\usepackage{rotating}
\usepackage{verbatim}
\usepackage{hyperref}
\usepackage{bm}
\usepackage[normalem]{ulem}
\usepackage[sort,authoryear]{natbib}
\usepackage{bbm}
\usepackage{array}
\usepackage{enumerate}
\usepackage{float}

\usepackage[flushleft]{threeparttable}
\usepackage{tablefootnote}
\usepackage[linesnumbered,ruled,vlined]{algorithm2e}

\usepackage{ifthen}
\usepackage{url}
\usepackage{mathtools}
\usepackage{authblk}
\title{Randomization Tests for Adaptively Collected Data}
\author[]{Yash Nair}
\author[]{Lucas Janson}
\affil[]{Department of Statistics, Harvard University}
\date{}
\newtheorem{theorem}{Theorem}[section]
\newtheorem{domain}{Problem Domain}
\newtheorem{proposition}{Proposition}[section]

\newtheorem{lemma}[theorem]{Lemma}
\newtheorem{remark}{Remark}

\newtheorem*{setup*}{Setup}
\newtheorem*{nullHypothesis*}{Null hypothesis}
\newtheorem*{constraints*}{Resampling constraints}

\newtheorem{env}{Environment}

\newcommand{\data}{D}
\newcommand{\sampledData}{\tilde{D}}

\newcommand{\decisionMaker}{\mathcal{A}}
\newcommand{\decisionProb}{\mathbb{P}_{\mathcal{A}}}

\newcommand{\numMCSamples}{m}
\newcommand{\sampledSet}{\mathfrak{D}}
\newcommand{\density}{f}
\newcommand{\proposal}{\tilde{q}}
\newcommand{\proposalNoTilde}{q}
\newcommand{\weight}{w^\proposalNoTilde}
\newcommand{\weightNoTilde}{w^{\proposal, \Sigma}}

\newcommand{\zeroM}{[0:\numMCSamples]}
\newcommand{\effnumMCSamples}{\numMCSamples_{\text{eff}}}
\newcommand{\horizon}{T}

\newcommand{\CondIndep}{\indep, g}

\newcommand{\id}{\textnormal{id}}

\newcommand{\pseudoorb}{\textnormal{pseudo-orb}}

\newcommand{\Stat}{\textnormal{S}}

\newcommand{\Loc}{\textnormal{Loc}}

\newcommand{\rt}{{\,:\,}} 
\newcommand{\orb}{\textnormal{orb}}

\newcommand{\indep}{\perp \!\!\! \perp}

\makeatletter
\renewcommand{\paragraph}{%
  \@startsection{paragraph}{4}%
  {\z@}{1.25ex \@plus 1ex \@minus .2ex}{-1em}%
  {\normalfont\normalsize\bfseries}%
}
\makeatother

\SetKwInput{KwInput}{Input}                % Set the Input
\SetKwInput{KwOutput}{Output}              % set the Output

\counterwithout*{footnote}{section}
\counterwithout*{footnote}{subsection}
\counterwithout*{footnote}{subsubsection}

\begin{document}

\maketitle

\begin{abstract}
Randomization tests (including permutation tests) are one of the most fundamental methods in statistics, enabling
a range of inferential tasks
such as testing for (conditional) independence of random variables, constructing confidence intervals in semiparametric location models, and constructing (by inverting a permutation test) model-free prediction intervals via conformal inference. 
Randomization tests are intuitive, easy to implement, and exactly valid for any sample size, but their use is generally confined to independent and/or exchangeable data.
Yet in many applications including clinical trials, online education, online advertising, and protein design, data is routinely collected \emph{adaptively}, meaning that the aspects of the data under the data collector's control (e.g., treatment assignments) are assigned at each time step via a (possibly randomized) algorithm that depends on all the data observed so far; such assignment algorithms include (contextual) bandit and reinforcement learning algorithms as well as adaptive experimental designs.
In this paper we present a general framework for randomization testing on adaptively collected data (despite its non-exchangeability), 
encompassing (and in some cases improving)
the few existing results on randomization testing and conformal inference for adaptively collected data,
as well as many other important settings. 
The key to our framework is the ability to compute likelihood-ratio-based weights involving known quantities based purely on the known adaptive assignment algorithm, as long as a certain proportionality condition is met. These weights can then be accounted for in our framework
to conduct an exact randomization test, but in order for the test to be powerful, resamples need to be diverse yet have weights as close to equal as possible. Thus, we additionally present novel computationally tractable resampling algorithms
for various popular adaptive assignment algorithms, data-generating environments, and types of inferential tasks. Finally, we demonstrate via a range of simulations our framework's 
power (in the case of hypothesis testing) and narrow widths (in the case of confidence or prediction intervals produced by inverting randomization tests).
\end{abstract}

\section{Introduction}
\label{introduction}
 
 \subsection{Motivation}\label{intro}
 Randomization tests form an important methodological framework that enjoys a wide array of uses across statistics, ranging from testing equality of distributions to testing (conditional) independence between covariates and response in supervised learning settings. Beyond these classical uses of randomization tests, the framework also encompasses permutation testing and (inversely) conformal inference \citep{alrw}. We briefly note here that, throughout this paper, we use the phrase ``randomization test'' not to refer to a test applied on data obtained via the physical act of randomization taken by an experimenter, but rather the randomization used by the test itself to compute a p-value. Specifically, we view a randomization test as a randomized procedure, taking the observed data set $\data$ as input, that samples $\numMCSamples$ copies $\sampledData^{(1)}, \ldots, \sampledData^{(\numMCSamples)}$ that are jointly exchangeable with $\data$ under the null, and then computes a (valid) p-value by taking a simple average of indicators comparing the resampled data to the observed data via a given test statistic. This interpretation of a randomization test has been referred to as a \emph{quasi-randomization} test by \cite{what-is-randomization-test}.

 While randomization tests can make very weak assumptions on the marginal distribution of observations, they generally make quite restrictive assumptions on the \emph{joint} distribution of the data.

 In particular, they usually require independent and/or exchangeable data \citep[e.g.,][]{candes2018panning,lehmann2005testing,edgington2007randomization,alrw,fisher1937design,pitman}, an assumption that is often violated in many real-world settings in which data are gathered in an adaptive fashion. One such example is drug discovery research, in which a scientist adaptively submits experiments serially, where the $t^{\textnormal{th}}$ experiment's design is based upon the results of the first $t-1$ \citep{popova2018deep}. This complex dependency will, in general, violate the exchangeability assumption when analyzing the data from all experiments at once.
 
Similar situations arise in experimental designs in the natural sciences more broadly in addition to mobile health, adaptive clinical trials, online education, and online advertising, to name a few.

Scenarios like those above naturally lead to a number of important inferential tasks that need to be performed on
adaptively collected data, all of which we show can be performed via a randomization testing-based framework: \begin{itemize}
    \item Testing if two or more treatments induce the same distribution over outcomes
    \item Detecting non-stationarity in the data (i.e., testing whether the outcome depends not only on the treatment, but also on time)
     \item Using past data to predict the outcomes of future samples with quantifiable uncertainty
     \item Constructing confidence intervals for the difference in locations of outcome distributions corresponding to different treatments in semiparametric models
 \end{itemize}
 We now briefly describe various settings in which the above tasks and generalizations thereof are both difficult and important problems and motivate them with real-world examples.

 \begin{domain}[Comparing response distributions of different covariates]\label{domain2}
 An interesting and challenging problem in adaptive data collection settings is to test whether two or more different treatments amongst a total of $K$ give rise to the same response distribution, i.e., for some pair $i,j \in \{1, \ldots, K\}$, is $(Y\mid X=i) \overset{d}{=} (Y\mid X=j)$? In the most extreme case, one might ask if there is even any dependency at all between $X$ and $Y$ (i.e., whether the treatment has any effect on the outcome). Beyond the mobile health examples delineated below in Problem Domain~\ref{domain1}, for which it is important to test if two different treatments actually have the same effect (or whether treatments have any effect at all), such testing is also useful in recommender systems \citep{li2010contextual,schafer1999recommender}, to determine if different content displays actually produce different outcomes. Further complicating this setting is that different users may react differently (e.g., as measured by clickthrough rate) to different content. Thus, one might ask: Given the `context' surrounding the user's preferences, do they react the same way to different advertisements?
 \end{domain}

  \begin{domain}[Testing non-stationarity]\label{domain1}
 An important problem that arises in domains in which actions are adaptively chosen and outcomes observed, is that of non-stationarity detection. That is, one may ask if the conditional distributions $Y_t\mid X_t$ in data gathered via some adaptive decision-making procedure change with $t$. One such scenario in which non-stationarity testing is important is in mobile health applications. For example, the Just-in-Time Adaptive Intervention (JITAI) \citep{jitai} is a mobile health framework designed to provide appropriate support to patients with dynamic, time-varying states (e.g., mood, location, etc.). JITAIs have been applied to, for instance, suicide prevention \citep{coppersmith_dempsey_kleiman_bentley_murphy_nock_2021}, smoking relapse prevention \citep{BATTALIO2021106534}, and addiction science \citep{addiction}. In situations like these, it is vital to be able to know whether or not a patient's response, $Y_t$, to the treatment, $X_t$, is varying over time, $t$, so as to facilitate administration of the most effective and appropriate care possible. Tests of non-stationarity provide a principled method of achieving this goal.
 \end{domain}

  \begin{domain}[Predicting with high confidence]\label{domain0}
In experimental design settings, providing prediction intervals for the results of future experiments can be instrumental in guiding researchers in decision-making. A motivating example comes from \cite{ALVARSSON202142}. In their work, the authors introduce conformal inference \citep{burnaev2014efficiency, shafer2008tutorial, vovk2013transductive, alrw, vovk} to researchers in the field of drug discovery and go on to give an example of how the methodology---which also assumes i.i.d.~or exchangeable data---can be applied to classify various types of ATP-Binding Cassette transporters at any user-specified uncertainty (i.e., significance) level. However, in light of the sequential and adaptive nature of many experimental designs in this field and in related fields like drug development \citep[e.g.,][]{adaptiveclinical,mahajan2010adaptive,GODFREY2013795}, there is need for provably valid test-time prediction intervals for these adaptively collected data. 
 \end{domain}

 \begin{domain}[Relating parameters in semiparametric models]\label{domain4}
 A final difficult problem in scenarios like those of the previous Problem Domains is to estimate how the parameters of different response distributions belonging to a semiparametric model relate to one another. For example, suppose that $Y\mid X=x \sim p_{\theta_{x}}$, where $\{p_\theta: \theta \in \mathcal{R}^d\}$ is a location family with parameter indexed by the treatment. How can we estimate $\theta_{x} - \theta_{x'}$ for treatments $x$ and $x'$? This problem setting is ubiquitous, for example, in the multi-armed bandit literature, where it is common to assume that the reward distributions of all arms are distributed according to a location family like the Normal \citep{LAI19854}. In such settings, being able to estimate parametric relationships between different treatments in finite samples is useful for a variety of real-world problems. As an example, as described in \cite{zhang2020inference}, performing \emph{post hoc} inference can be useful in many settings ranging from recommender systems \citep{mary2015bandits} to anomaly detection \citep{ban2020generic} to finance and portfolio selection \citep{huo2017risk}. In such settings, after-the-fact inference may be helpful to researchers and practitioners who wish to understand the differences between different treatments and potentially utilize such differences to guide future decision-making.

 \end{domain}

 A number of tasks that we consider in this paper---and, indeed, some of those presented in the Problem Domains above---involve the construction of prediction or confidence intervals. However, as we show, constructing such intervals reduces to hypothesis testing by inverting the corresponding test. In particular, although  not usually framed this way, conformal prediction is the inversion of a permutation test, a specific type of randomization test (see, e.g., \cite{chernozhukov2018exact}). More generally, although it is unusual to invert a randomization test, we show that by applying certain transformations to the data, our randomization tests can be inverted to construct confidence intervals in semiparametric models (see, e.g., \cite{rabideau2021randomization}). Section~\ref{invert-chap} delineates precisely how our tests can be inverted to produce conformal/confidence intervals. Due to this inverse relationship between prediction/estimation and testing, we generally (with the exception of Section~\ref{invert-chap}) restrict our attention \emph{solely} to hypothesis testing for clarity of presentation, and only state results in terms of the validity of such tests (and, in particular, the stochastic domination of the uniform distribution by their p-values).
 
 Finally, all of the problem domains which motivate this work are centered around adaptively collected data. Indeed, as we show in this paper, we will be able to answer all of these inferential questions on adaptively collected data via a general framework for randomization testing \emph{as long as} the adaptivity is known to the analyst. That is, we assume that the analyst has access to the (potentially non-deterministic) decision-making algorithm used to assign treatments in the data. We now introduce some notation to define this setup and formalize these notions.

\subsection{Notation}\label{notation}

The adaptive data collection settings we consider in this paper---of which popular settings like bandits, Markov Decision Processes (MDPs), reinforcement learning, and adaptive experimental designs are all special cases---comprise a data collection horizon of $\horizon$, the number of timesteps of data collected by the practitioner (we will treat $\horizon$ as deterministic in this paper). At each timestep $t$, a context (sometimes called a \emph{state}) $C_t \in \mathcal{C}$ is observed, an action (also referred to as a \emph{treatment}) $X_t \in \mathcal{X}$ is selected, and a response (called an \emph{outcome} or \emph{reward} in some situations) $Y_t\in\mathcal{Y}$ is subsequently observed. $\mathcal{Z} := \mathcal{C}\times\mathcal{X}\times\mathcal{Y}$ is the (assumed time-invariant) sample space of the triple $Z_t:=(C_t,X_t,Y_t)$---in this paper, we assume that $\mathcal{Z}$ is discrete; see Remark~\ref{discrete-remark} for an explanation of why as well as how to handle the case of continuous $\mathcal{Z}$. More formally, define the history at time $t$, $H_t$, to be the sequence  $((C_1, X_1, Y_1), \ldots, (C_t, X_t, Y_t))$; $H_0$ is simply defined to be empty $\varnothing$. Then the action $X_t$ is selected \emph{conditionally} on $H_{t-1}$ and $C_{t}$. These action-selection probabilities are encoded in the \emph{known} decision-making (i.e., adaptive assignment) algorithm $\decisionMaker$, where the probability of selecting action $x \in \mathcal{X}$ at the $t^{\text{th}}$ timestep given the $t^{\text{th}}$ context and prior history of context-action-response triples up until time $t$ is given by $\decisionProb(x| C_{t},H_{t-1})$. The joint density of the full data sequence $H_\horizon = ((C_1, X_1, Y_1), \ldots, (C_T, X_T, Y_T))$ is denoted by $\density$. Unless otherwise noted, we always consider Markovian systems, and thus assume that: \begin{equation}\label{markov-assum}Y_t \indep H_{t-1} \mid (C_t, X_t).\footnote{This assumption will only become relevant in Section~\ref{app-chap}. Additionally, while this Markovianity assumption is standard in the adaptive data collection literature \citep[e.g.,][]{bibaut2021risk,zhang2020inference,hahn2011adaptive}, we discuss in Remark~\ref{no-markov} how our procedure can be applied to test slightly different hypotheses when no Markovianity assumption is made.}\end{equation} In this paper, we also sometimes consider domains without contexts, in which case $C_t = \varnothing$ for all $t \in [\horizon]$. For a given data sequence $(C_1, X_1,\allowbreak Y_1),\allowbreak \ldots,\allowbreak (C_T, X_T,\allowbreak Y_T)$, set $\data := H_\horizon$ to be their concatenation; take $\mathcal{D} := \mathcal{Z}^\horizon$ to be the sample space of the random variable $\data{}$. Finally, as a note on general notation, we use $[n]$ to refer to the set $\{1, \ldots, n\}$ and write $[n:m]$ to denote $\{n,\ldots, m\}$ for integers $n \leq m$.

 \subsection{Contribution}\label{contribution}
 In this paper, we provide exactly valid randomization-based inference for adaptively collected data. We make the following contributions, outlined below:
  
  We derive a weighted Monte Carlo (MC) randomization testing framework. This test is able to resample data from essentially any arbitrary resampling distribution (as long as a certain proportionality hypothesis is satisfied) and, by weighting the samples according to certain likelihood ratios, produce a p-value. 
  %\change{(We note here that, upon completion of our work, the work of \cite{harrison} was brought to our attention, which develops and proves the validity of a test which is the same as the simplest form of our weighted MC randomization test in its simplest form (Theorem~\ref{weighted-mc-rand-valid}). The generalized version of our test, presented and proved in Appendix~\ref{non-iid-appendix}, subsumes their's, however.)} 
  Along the way, we show that we are also able to utilize an unweighted Markov Chain Monte Carlo (MCMC) randomization procedure developed in \cite{besag1989generalized,metropolized,cpt,RB-LJ:2020}, and applied in other settings, to perform inferential tasks on adaptively collected data. Both approaches offer exactly valid Type-I error control at essentially whatever computational cost the user desires (although, more computation generally results in higher power). Our novel weighted MC approach, however, is empirically no less powerful than the MCMC procedure and is, in fact, often more powerful in our simulations, especially when the number of resamples is small.
    
    We use both the weighted MC and unweighted MCMC frameworks to test against a number of general null hypotheses on adaptively collected data:
    
    \begin{enumerate}
        \item Letting $g$ be any function of $X$, we can test \begin{equation}\label{multi-dim-null}Y_t \indep X_t \mid (C_t,g(X_t))\end{equation} in particular settings by resampling copies of $X_{t}$ for each $t \in [\horizon]$ from \emph{any} distribution conditional on $g(X_t)$ and weighting these resamples by certain likelihood-based ratios. In the special case in which $g$ is a constant function, $X_t$ can sometimes be sampled in such a way such that no weighting is required. In particular, the prior works of \cite{pocock1975sequential,simon1979restricted,ham,bojinov2019time} are all able to handle this special setting by simply resampling the $X_t$'s by fixing the contexts and responses and re-running $\decisionMaker$. In general, however, when $g$ is non-constant, this unweighted procedure cannot be applied, yet our procedure always can be. One example of such a non-constant $g$ might map treatments $x$ in a multidimensional treatment space $\mathcal{X}$ to their first dimension $x_1$. The hypothesis being tested in equation \eqref{multi-dim-null} would then be of conditional independence of the response with all other dimensions of the treatment given the context and first treatment dimension. Additionally, our framework applies in challenging settings like MDPs, whereas that of prior work does not. Beyond this added generality of our testing procedure, it offers various improvements (in terms of power and potential computational efficiency) over the existing work in certain settings.
        
        \item We can test for non-stationarity in the conditional distributions $Y_t\mid X_t,C_t$ in various complex adaptive data collection settings. 
        \item Under the semiparametric assumption that $Y_t\mid X_t,C_t$ follows a location model with location parameter additive in $X_t$, our first test can be inverted to construct confidence regions for the location differences between actions as well. Similarly, our non-stationarity test can  be inverted to construct conformal prediction regions for $Y_\horizon$ under any adaptive assignment algorithm, considerably generalizing existing work which can only perform conformal inference under certain very specific adaptive assignment algorithms involving only a single round of adaptivity \citep{tibshirani2020conformal,feedbackcovshift} or can only do so approximately \citep{cpbeyond,gibbs2021adaptive,chernozhukov2018exact}.
    \end{enumerate}

     Within our framework, it is more desirable (in terms of power) to have the weight of each resample as close as possible to that of the observed data; developing resampling procedures which achieve this for the various inferential tasks and environments discussed above is challenging. As such, we devise a number of novel and computationally tractable resampling procedures to be used for our tests.
     
     Finally, to demonstrate the practicality of our randomization tests (which can be deployed at essentially any user-specified computational load) as well as the statistical efficiency of our resampling procedures, we perform a simulation study.  By considering the aforementioned inferential tasks in a variety of data-generating environments, and by using data collected by a number of decision-making algorithms---including both deterministic and randomized ones---we demonstrate that our resampling procedures produce a test which is both statistically efficient (in terms of power and average confidence/prediction interval length) and computationally tractable, while, of course, also guaranteeing exact validity.

 \subsection{Related work} 
% {\color{red}
% \subsection{Importance-weighted MC randomization testing}\label{harrison-related-work} The work of \cite{harrison} (which was brought to our attention following the preliminary online release of an earlier version of this paper). Our main theorem (Theorem~\ref{weighted-mc-rand-valid}) regarding the validity of the simplest version of our weighted MC randomization test is equivalent to their Theorem 2. While we make this connection precise in Appendix~\ref{harrison-connection}, we not here that, our generalized procedure, described in Appendix~\ref{non-iid-appendix}, and used for efficient construction of conformal prediction intervals in our simulations is, too, a generalization of theirs. Furthermore, to the best of our knowledge, ours is the first work which applies the weighted MC randomization testing procedure (both the version developed in \cite{harrison} as well as our more generalized version in Appendix~\ref{non-iid-appendix}) to the problem of inference on adaptively collected data.
% }
 \subsubsection{Randomization tests for adaptively collected data}\label{rand-tests-related-work}
As noted in \cite{rosenberger}, the works of \cite{pocock1975sequential} and \cite{simon1979restricted} consider randomization testing for adaptively collected data consisting of actions and outcomes in the Markovian setting of equation~\eqref{markov-assum} under a certain restricted set of assignment algorithms. Extending their approach, \cite{ham} are able to additionally handle context variables. \cite{bojinov2019time} also develop a randomization test that generalizes that of \cite{pocock1975sequential, simon1979restricted} primarily by relaxing the Markovianity assumption. The key to the approach put forth in all of the aforementioned papers is that they restrict the adaptivity of the data collection environment and consider null hypotheses such that simply fixing the contexts and responses and resampling the treatments via the known adaptive assignment algorithm produces exactly exchangeable copies of the data, thereby enabling them to conduct a standard \emph{unweighted} randomization test. On the other hand, our main result, the \emph{weighted} MC randomization test (Theorem~\ref{weighted-mc-rand-valid}) makes no such assumptions about the data-generating distribution nor the resampling distribution other than that they satisfy a certain proportionality hypothesis. Also, as we empirically demonstrate in Section~\ref{sim-chap}, our approach can be powerful even on data generated by deterministic assignment algorithms in the Markovian setting, while that of \cite{simon1979restricted,pocock1975sequential,ham,bojinov2019time} is provably powerless. For more details regarding the comparison of our work to prior work see Remarks~\ref{prior-work} and \ref{no-markov}.

%\paragraph{Inference in bandits and reinforcement learning} 
Other existing works for performing inference in reinforcement learning and adaptive data collection settings are either asymptotic \citep[e.g.,][]{zhang2020inference,zhang2021,hadad2021confidence,deshpande2018accurate,KZ-ea:2022,lai1982least,bugni2018inference,ma2015testing} or, if not, are conservative in their finite-sample bounds \citep[e.g.,][]{howard2021time,kaufmann2021mixture}. As opposed to these works, all our hypothesis tests are able to maintain exact and non-conservative validity in finite samples.

\subsubsection{MCMC, conditional and importance-weighted permutation/randomization tests}\label{mcmc-related}
The connection between sampling exchangeable samples and MCMC was first developed by \cite{besag1989generalized} and has more recently been utilized by \cite{metropolized, cpt,RB-LJ:2020}, for the purposes of efficiently running a conditional permutation test, generating knockoffs, and approximate co-sufficient sampling, respectively. The MCMC randomization test developed by \cite{besag1989generalized} which we outline in Section~\ref{wcrt-chap} takes the same approach:
it generates exchangeable samples using base draws from a Metropolis--Hastings Markov Chain. However, to the best of our knowledge, our work is the first that applies this MCMC approach to adaptively collected data. 

Another related line of work is that of \cite{harrison} (which was brought to our attention following the preliminary online release of an earlier version of this paper). An importance special case of our main theorem (Theorem~\ref{weighted-mc-rand-valid}) discussed in Remark~\ref{comp-iid} regarding the validity of our weighted MC randomization test can be derived from their (conditional) importance-weighted randomization test \citep[Theorem 2]{harrison}; this connection is made precise in Appendix~\ref{harrison-connection}.
%We note here that, while essentially all of our simulation results---apart from those that efficiently construct conformal prediction intervals, which requires an even more generalized version of Theorem~\ref{weighted-mc-rand-valid} discussed in Appendix~\ref{non-iid-appendix}---use this special case procedure that is equivalent to \cite{harrison}'s, ours is, to the best of our knowledge, the first work which applies it to the problem of inference on adaptively collected data.}

\subsubsection{Conformal inference} Our weighted MC randomization test, when applied to non-stationarity testing and inverted, has close connections with conformal inference. 
The works of \cite{tibshirani2020conformal} and \cite{feedbackcovshift} are most related to our weighted MC randomization test when applied to non-stationarity testing and conformal inference. \cite{tibshirani2020conformal} first extend conformal inference from i.i.d.~data to the setting of covariate shift, under the assumption that the likelihood ratio between test and train covariates is known by developing a more general \emph{weighted} conformal prediction (WCP) procedure. Relatedly, \cite{hu2020distribution} apply WCP to give an asymptotically powerful non-stationarity test in the same setting. The work of \cite{feedbackcovshift} extends
\cite{tibshirani2020conformal} and shows how to handle dependence between train and test sets in the setting of feedback covariate shift.
These prior methods handle only a single round of adaptivity (i.e., the first $\horizon-1$ rounds of ``train-time'' data is fully i.i.d., but the ``test-time'' covariate distribution at timestep $\horizon$ may either be different \citep{tibshirani2020conformal,hu2020distribution} or depend on the first $\horizon-1$ datapoints via feedback covariate shift \citep{feedbackcovshift}). In contrast, our test for non-stationarity (the inversion of which yields a conformal prediction interval) is able to handle \emph{any number} of rounds of arbitrary analyst adaptivity.

\cite{hmm} also develop an exact conformal inference procedure for non-exchangeable data, in the setting of HMMs, by using the notion of partial exchangeability \citep{diaconis} and restricting to a subgroup of permutations which preserve certain transition structure. In the application of our (weighted) conformal prediction procedure to MDPs (described in Section~\ref{non-stationarity}), the same subgroup arises. Our method is, however, more general, as their work does not allow for adaptivity in treatment assignments and hence does not require the use of
%allows for random sampling from a restricted \emph{set} $\Pi$ of permutations (as opposed to iterating across an entire \emph{subgroup}) and also 
the likelihood ratio-based weights described in Section~\ref{contribution}.

A final related line of work is in handling fully non-exchangeable data; that is, in contrast to \cite{tibshirani2020conformal} and \cite{feedbackcovshift}, no exchangeability is assumed within any particular train set. In particular, \cite{chernozhukov2018exact}, \cite{gibbs2021adaptive}, \cite{xu2021conformal}, \cite{cpbeyond} all consider various versions of non-exchangeable data and extend conformal inference to these regimes. The methods presented in these works, however, are anti-conservative with bounds degrading as the degree of non-exchangeability grows. In contrast, the methods presented in this paper (using the adaptivity known to the analyst) provide exact validity in the adaptive (and hence potentially highly non-exchangeable) settings in which they apply.

\subsection{Outline} In Section~\ref{wcrt-chap} we introduce the weighted MC randomization test. We also briefly describe how to perform an unweighted randomization test by using MCMC sampling. We then, in Section~\ref{app-chap}, show how our approach can be applied to solve a wide range of difficult inferential problems on adaptively collected data. Then in Section~\ref{proposal-chap} we introduce a number of novel algorithms for generating resamples that can be used for a variety of inferential tasks and adaptive data collection environments. Finally, in Section~\ref{sim-chap}, we perform a simulation study of the methods introduced in Sections~\ref{wcrt-chap} and \ref{app-chap}, using the resampling procedures from Section~\ref{proposal-chap}, that both empirically validates our methods by illustrating their validity and computational tractability, and demonstrates their power in a variety of challenging settings.

\section{The weighted MC randomization test}\label{wcrt-chap}
In this section we introduce our main approach to randomization testing: a novel weighted MC randomization test. Additionally, at the end of this section, we briefly review an unweighted MCMC randomization testing framework introduced by \cite{besag1989generalized}, but, to the best of our knowledge, never before applied to adaptively collected data (as we do in Section~\ref{app-chap}). The key to both approaches is the ability to define likelihood-ratio-based weights that satisfy a certain proportionality hypothesis, which, as we show in Section~\ref{app-chap}, is possible in a range of inferential tasks in various adaptive data collection settings. Empirically, our weighted MC approach never performs worse than the unweighted MCMC test, and indeed often dominates it, especially when the number of resamples is small.

Our weighted MC randomization test is a randomized procedure, taking the data set $\data{}$ as input, which samples $\numMCSamples$ conditionally i.i.d.~MC draws $\sampledData^{(1)}, \ldots, \sampledData^{(\numMCSamples)}$ given $\data$. Define $\sampledSet$ to be the list $(\sampledData^{(0)},\sampledData^{(1)},\ldots, \sampledData^{(m)})$ where $\sampledData^{(0)} := \data$ and let $\proposalNoTilde(\sampledData^{(i)}|\data)$ denote the conditional probability of sampling $\sampledData^{(i)}$ from this conditional distribution given that the dataset $\data$ was observed. Setting $\hat{f}(\sampledData^{(i)}) := \prod_{t=1}^\horizon \decisionProb(\tilde{X}^{(i)}_t|\tilde{C}^{(i)}_{t}, \tilde{H}^{(i)}_{t})$, the procedure then calculates likelihood-ratio-based weights for each resample $\sampledData^{(i)}$ in $\sampledSet$: \begin{equation}\label{weight-eqn}\weight_{\sampledSet}(\sampledData^{(i)}) = \frac{\hat{f}(\sampledData^{(i)})\prod_{k \in \zeroM\backslash\{i\}}\proposalNoTilde(\sampledData^{(k)}|\sampledData^{(i)})}{\sum_{j =0}^\numMCSamples\hat{f}(\sampledData^{(j)})\prod_{k \in \zeroM\backslash\{j\}}\proposalNoTilde(\sampledData^{(k)}|\sampledData^{(j)})},\end{equation} where $\tilde{C}^{(i)}_t, \tilde{X}^{(i)}_t$ and $\tilde{H}_t^{(i)}$ are, respectively, the context at, action at, and history up until timestep $t$ of the $i^{\textnormal{th}}$ resample $\sampledData^{(i)}$. Finally, the procedure outputs the p-value \begin{equation}\label{p-val-def}p := \sum_{i=0}^\numMCSamples\weight_\sampledSet(\sampledData^{(i)})\mathbf{1}[S(\sampledData^{(i)}) \geq S(\data)],\end{equation} where $S: \mathcal{D} \rightarrow \mathbb{R}$ is any test statistic.
We summarize this procedure in the pseudocode of Algorithm~\ref{weighted-mc-rand}

\begin{algorithm}[!ht]
  
  \KwInput{Data sequence $\data$, resampling distribution $\proposalNoTilde$, number of samples $\numMCSamples$, test statistic $S$}
  Sample $\sampledData^{(1)}, \ldots, \sampledData^{(\numMCSamples)} \mid \data \overset{\textnormal{i.i.d.}}{\sim} \proposalNoTilde(\cdot|\data)$\\
  $\sampledSet \gets (\sampledData^{(0)}, \sampledData^{(1)}, \ldots, \sampledData^{(\numMCSamples)})$ where $\sampledData^{(0)} := \data$\\
  Compute $\weight_{\sampledSet}(\sampledData^{(i)})$ for each $i \in \zeroM$ via equation \eqref{weight-eqn}\\
  \KwOutput{ $p$, calculated via equation \eqref{p-val-def}} 

\caption{Weighted Monte Carlo randomization test}
\label{weighted-mc-rand}
\end{algorithm}
The above weighting scheme may not always yield a valid p-value. We show, however, that as long as the hypothesis \begin{equation}\label{null-hypothesis}\mathcal{H}^\propto_0: \hat{f}(\sampledData^{(i)}) = Kf(\sampledData^{(i)}), \forall i \in \zeroM \textnormal{ for some constant } K \textnormal{ not depending on }i,\end{equation} holds, then $p$ is a valid p-value. Importantly, note that the left hand side of the proportionality~\eqref{null-hypothesis} is \emph{always} computable (since it depends only on $\decisionMaker$, which is known), whereas the right hand side is not in general since the density $\density$ is unknown. As we show in Section~\ref{app-chap}, $\mathcal{H}^\propto_0$ holds for a number of null hypotheses in various adaptive data collection settings, thereby allowing for the valid use of the above to test against such nulls. We now present the main result of this paper, which is that, under $\mathcal{H}^\propto_0$, $p$ is a valid p-value.

\begin{theorem}\label{weighted-mc-rand-valid}
The p-value defined in equation \eqref{p-val-def} stochastically dominates the uniform distribution under equation \eqref{null-hypothesis} for \emph{any} resampling distribution $\proposalNoTilde$ satisfying $\mathcal{H}^\propto_0$.
\end{theorem}

While we relegate the proof of Theorem~\ref{weighted-mc-rand-valid} to Appendix~\ref{AppendixA}, we note here that the proof is quite different
than those of randomization tests in the standard unweighted case. For example, the proof of validity of the standard (unweighted) randomization test for exchangeable data and resamples---which is essentially an immediate consequence of the joint exchangeability of the original data and all resamples (see, e.g., \cite{besag1989generalized,edgington2007randomization,candes2018panning,cpt,metropolized})---clearly will not apply in our case, since the data and resamples are not jointly exchangeable and the weighting scheme further makes it unclear how any such exchangeability could be used to prove validity of $p$. A second, more recent proof technique, given by \cite{hemerik} in the context of the MC permutation test on exchangeable data, is to condition on the set $\orb(\data)$ of all possible permutations of the observed data and use the fact that, conditionally on $\orb(\data)$, the original data is uniform over $\orb(\data)$ and the set of resampled permutations is also uniform to show that the test is valid. In our setting, however, due to the non-exchangeability, this conditional distribution need not be uniform.

Rather, the proof of Theorem~\ref{weighted-mc-rand-valid} is via the application of a simple yet new (to the best of our knowledge) application of Bayes’ theorem and proceeds by showing the conditional validity of $p$ given the list
$\mathfrak{D}$; see Appendix~\ref{AppendixA} for details. Furthermore, as discussed in Remark~\ref{non-iid}, the assumption that the resamples
$\sampledData^{(1)}, \ldots, \sampledData^{(\numMCSamples)}$ are drawn conditionally i.i.d.~given $\data$ can be relaxed and a more general test (with generalized
weights) can be used; see Appendix~\ref{non-iid-appendix} for details.

Finally, as discussed in Section~\ref{mcmc-related}, a special case of the above procedure discussed in Remark~\ref{comp-iid} can be derived from an importance-weighted MC randomization test presented in \cite{harrison} (although their proof technique is different and we were unable to extend it to prove our more general result). See Remark~\ref{comp-iid} and Appendix~\ref{harrison-connection} for details.
%We relegate both our proof of Theorem~\ref{weighted-mc-rand-valid} (which is slightly different than that presented in \cite{harrison} as we consider only the discrete setting) as well as its connection to \cite{harrison} to Appendix~\ref{AppendixA}. At a high-level, though, our proof of Theorem~\ref{weighted-mc-rand-valid} is via conditioning and an application of Bayes' theorem. On the other hand, the proof in \cite{harrison} is by viewing Algorithm~\ref{weighted-mc-rand} as a (conditional) importance sampling procedure (with target distribution $f$ and conditional proposal distribution $q(\cdot|A(D))$ for some statistic $A$ of the observed data) and utilizes change of measure arguments involving Radon-Nikodym derivatives induced by the target/proposal distribution likelihood ratio. For further details, see Appendix~\ref{harrison-connection}.

%\change{Finally, as discussed in Remark~\ref{non-iid}, the assumption that the resamples $\sampledData^{(1)}, \ldots, \sampledData^{(m)}$ are drawn conditionally i.i.d.~given $\data$ can be relaxed and a more general test (with generalized weights) can be used thereby extending both the procedure in Algorithm~\ref{weighted-mc-rand} as well as the corresponding importance-weighted MC randomization test of \cite{harrison}; see Appendix~\ref{non-iid-appendix} for details.}

\begin{remark}\label{discrete-remark}
The proof of Theorem~\ref{weighted-mc-rand-valid} given in Appendix~\ref{AppendixA} applies only to discrete data distributions $\density$ and resampling distributions $\proposalNoTilde$ (recall we assumed all data was discrete in Section~\ref{notation}). We focus only on such distributions as in any real-world application of our test, the computer on which our test is being run on has only finite precision, rendering both the original dataset as well as all resamples discrete. That is, any practitioner running our test on a computer is necessarily dealing with discrete data, \emph{not} due to any restrictions of the test itself, but rather the finite precision limitation of the computer. This is not to say, however, that continuous distributions should never be considered in any situation. For example, in an asymptotic theory, rates of convergence for discrete distributions may degrade as the resolution of discretization becomes finer. In such a case, a continuous approximation of a computer-rendered discrete distribution may be considerably more useful than directly handling the discrete distribution itself. The key difference in our case, however, is that our theory is \emph{exact in finite samples} for any discrete distribution. Hence, the guarantee of Theorem~\ref{weighted-mc-rand-valid} \emph{does not} worsen at finer discrete resolutions, but rather remains intact, thereby permitting our consideration of only discrete distributions. 

We note, however, that our procedure should apply much more broadly to continuous distributions, but perhaps not completely generally. See \cite[][Figure 5]{huang2020} for one such example in which our theory does not hold because Bayes' theorem is violated, by defining a dataset with $\horizon = 2$ as well as a resampling distribution with $\numMCSamples=1$ for which the marginal distribution of the resample $\mathbb{P}(\sampledData^{(1)})$ is not equal to $\proposal(\sampledData^{(1)}|\data)f(\data)$ by taking $X_1 \sim \text{Unif}(0,1)$ and defining $\proposalNoTilde$ to sample $(\tilde{X}^{(1)}_1, \tilde{X}^{(2)}_1) \sim L_{X_1}$ where $L_{X_1}$ is the segment of length $1$ orthogonal to the $x$-axis, intersecting it at the point $X_1$, with an angle of $(1-X_1)^{10}\pi/2$ with the $y$-axis.
\end{remark}

\begin{remark}[Randomizing for exact Type-I error control]\label{smoothed-p-vals}
Just as with a standard randomization test and, as a special case, conformal inference in which the procedure is called smoothing \citep[e.g.,][]{alrw}, one can define a ``lower'' p-value \[p^-:= \sum_{i=0}^\numMCSamples\weight_\sampledSet(\sampledData^{(i)})\mathbf{1}[S(\sampledData^{(i)}) > S(\data)]\] and randomize to obtain a test for Theorem~\ref{weighted-mc-rand-valid} which controls Type-I error at \emph{exactly} the nominal level. Specifically, the test which fails to reject when $p^- > \alpha$, rejects with probability $\frac{\alpha-p^-}{p - p^-}$ when $p^- \leq \alpha < p$, and otherwise always rejects, controls Type-I error at exactly the nominal level $\alpha$.
\end{remark}

While Theorem~\ref{weighted-mc-rand-valid} gives one great freedom and generality in computing valid randomization p-values by allowing for many choices of the conditional distribution $\proposalNoTilde$, two aspects in particular remain, at present, quite unclear: (a) in what situations and under what null hypotheses can we ensure that the proportionality hypothesis $\mathcal{H}^\propto_0$ holds, as well as (b) \emph{which} choices of $\proposalNoTilde$ one should sample from to obtain a powerful test. We address the first question in the context of adaptively collected data in Section~\ref{app-chap} and discuss the second in Sections~\ref{proposal-chap} and \ref{sim-chap}.

\paragraph{The unweighted MCMC randomization test} Above, we discussed a \emph{weighted} MC randomization test which weights resamples based on certain likelihood ratios. Here, we briefly outline an \emph{unweighted} MCMC randomization test developed by \cite{besag1989generalized}. While the test itself is not new (see, e.g., \cite{Berrett2019TheCP,RB-LJ:2020,metropolized} for more recent utilizations and extensions of the test), our application of it to adaptively collected data, to the best of our knowledge, is. The randomized procedure for the unweighted MCMC randomization test repeatedly takes Metropolis--Hastings steps, starting at $\data$ by, at each step $i$, sampling a proposal $\sampledData$ given $\sampledData^{(i-1)}$ from $\proposalNoTilde(\cdot|\sampledData^{(i-1)})$, and additionally computing an acceptance ratio under the stationary distribution $\density$ to decide if the Metropolis--Hastings step accepts the proposal or remains at $\sampledData^{(i-1)}$. Just as with Algorithm~\ref{weighted-mc-rand}, the acceptance ratio calculation actually only involves known quantities depending on $\hat{f}$ and $\proposalNoTilde$ and hence the validity of the MCMC test is again implied by the proportionality hypothesis $\mathcal{H}^\propto_0$. See \cite{besag1989generalized} for a delineation of the general test. We also note here that, in all of our simulations, our weighted MC test performs no worse than---and often dominates---the unweighted MCMC test. As such, exploring the utility of our weighted MC test in the settings of \cite{RB-LJ:2020,metropolized,cpt}, all of which use the unweighted MCMC procedure, may be of interest for future work.

\section{Randomization testing for adaptively collected data}\label{app-chap}
In this section, we apply the randomization tests from Section~\ref{wcrt-chap} to solve a host of challenging inferential tasks in adaptive data collection settings. In Section~\ref{wcrt-chap}, we saw that to perform a weighted MC randomization test we needed to be able to run Algorithm~\ref{weighted-mc-rand} which in turn required us to choose a $\proposalNoTilde$ that satisfies the hypothesis $\mathcal{H}^\propto_0$.
In this section, we show that, for a number of null hypotheses in various adaptive data collection settings, $\mathcal{H}^\propto_0$ can be satisfied with many non-trivial choices of $\proposalNoTilde$, thus allowing for the application of the weighted MC randomization test. We note here that, since the focus of this section is ensuring that $\mathcal{H}^\propto_0$ holds (under various null hypotheses and adaptive data collection settings), all results also immediately apply to the unweighted MCMC test.

We address two broad classes of tasks and describe each---as well as the adaptive data collection environments in which we consider them---in detail in the sections that follow:

\begin{enumerate}
    \item Testing conditional independence between treatment and response conditional on context and some function of the treatment (Section~\ref{distributional}). Applying certain transformations to the data and inverting such tests allows us to construct confidence intervals for response distribution parameters in certain semiparametric models (Section~\ref{confidence-inversion}).
    \item Non-stationarity testing (Section~\ref{non-stationarity}) and its application to conformal inference (Section~\ref{conformal-inversion})
\end{enumerate}

In Sections~\ref{distributional} and \ref{non-stationarity} below, we discuss a number of adaptive data collection environment setups. Each of these setups assumes some combination/subset of equations \eqref{markov-assum} (Markovianity), \eqref{stat-assum}, \eqref{non-react-assum}, \eqref{exch-assum}:

\begin{align}
Y_t\mid C_t, X_t \text{ does not depend on }t&  &\text{($Y$-Stationarity)} \label{stat-assum}
\end{align}

This first assumption asserts that the environment is \emph{$Y$-stationary} so that the conditional response distribution does not vary with time; this distinction will be important in whether or not additional randomization can be employed for a more powerful conditional independence test.

\begin{subequations}
\begin{align}
    %Y_t \indep H_{t-1} \mid (C_t, X_t)& &\text{(Markovianity)} \label{markov-assum}\\
    C_t \indep (X_1, \ldots, X_{t-1}) \mid (C_1, Y_1, \ldots, C_{t-1}, Y_{t-1})& &\text{(Weak non-reactivity)} \label{non-react-assum}\\
   \text{$C_t \mid H_{t-1}$ depends neither on $t$ nor $H_{t-1}$} & &\text{($C$-stationarity \& strong non-reactivity)} \label{exch-assum}
\end{align}
\end{subequations}

The assumptions of equations \eqref{non-react-assum} and \eqref{exch-assum} are related and describe the environment through the behavior of contexts. In particular, the weak non-reactivity assumption (equation \eqref{non-react-assum}) says that the actions prior to time $t$ do not affect $C_t$, conditionally on the previous contexts and responses; i.e., the future states of the environment are essentially (conditionally) \emph{non-reactive} to prior actions. This weak non-reactivity assumption is also made in \cite[Assumption~1]{ham}. The $C$-stationarity \& strong non-reactivity assumption (equation \eqref{exch-assum}) is a stricter version of this, and stipulates that each context is generated by the environment independently from the past and also that the distribution from which it is generated does not change with $t$.

Before proceeding, we pause here to emphasize that all results that ensue in this section hold for \emph{any} adaptive assignment algorithm $\decisionMaker$, as long as it is known. We present all results by first describing the null hypothesis being tested, then explaining the various environments in which it can be tested, and finally delineating any constraints on the resampling distribution $\proposalNoTilde$ in each of these environments so as to ensure that $\mathcal{H}^{\propto}_0$ holds under the null.

\subsection{Conditional independence testing}\label{distributional}

\paragraph{Null hypothesis} Let $g$ be any function of $x \in \mathcal{X}$, with unspecified codomain. We wish to perform a conditional independence test between treatment and response given both context and the $g$-evaluation of the treatment: \[\mathcal{H}^{\CondIndep}_0: Y_t \indep X_{t} \mid (C_t, g(X_t)), \forall t \in [\horizon].\] One example in which this hypothesis might be employed is in the problem of A/B testing in online advertising, where a company presents users with the same ad but with differing hues, saturations, and brightnesses (i.e., so the treatment space is $3$-dimensional). One concrete hypothesis that may be tested in this situation is if saturation and brightness are enough to predict user response alone (i.e., is user response independent of hue given both saturation and brightness?). Setting $g(X_t) = (X_{t, \text{saturation}}, X_{t, \text{brightness}})$ to map $X_t$ to its saturation and brightness components, $\mathcal{H}^{\CondIndep}_0$ is equivalent to this hypothesis. A second scenario is in testing if a particular subset of treatments induces the same response distribution. That is, letting $\mathcal{X} = \{0, 1, 2\}$ be the treatment space, the null hypothesis that $Y_t \indep X_t \mid (C_t, X_t \in \{0, 1\})$ is equivalent to $\mathcal{H}^{\CondIndep}_0$ with $g(X_t) = \mathbb{I}(X_t = 2)$. Finally, in the special case in which $g$ is a constant function, the hypothesis being tested is that of simple conditional independence between treatment and response given context, and, for this particular choice of $g$, the unweighted test of prior work \citep{pocock1975sequential,simon1979restricted,ham,bojinov2019time} can be employed in Environments~\ref{non-react-env} and \ref{stat-non-react-env} (but not in Environments~\ref{mdp-env} or \ref{stat-mdp-env}); as we describe below, however, our framework allows for a significant improvement in power over this prior work when employed in Environment~\ref{stat-non-react-env}.

\begin{env}[Non-reactive environment]\label{non-react-env}
The \emph{non-reactive environment} is one in which the Markovianity (equation \eqref{markov-assum}) and weak non-reactivity (equation \eqref{non-react-assum}) assumptions hold. Examples of a non-reactive environment include (contextual) bandits---a setting that is common in the mobile health literature \citep[e.g.,][]{tewari2017ads}---as well as many common adaptive experimental designs, such as adaptive crop yield experiments \citep[e.g.,][]{crop}.
\end{env}

\begin{env}[Markov Decision Process]\label{mdp-env}
The next environment we consider is an \emph{MDP}, which assumes only the Markovianity assumption (equation \eqref{markov-assum}) holds, and also stipulates that $Y_t := C_{t+1}$; note that this implies that Assumption \eqref{non-react-assum} also holds. One example of MDP data arises in electronic health records \citep{li2022electronic}.
\end{env}

Environments~\ref{stat-non-react-env} and \ref{stat-mdp-env} below are $Y$-stationary counterparts of Environments~\ref{non-react-env} and \ref{mdp-env}, respectively. 

\begin{env}[Stationary non-reactive environment]\label{stat-non-react-env}
The most restricted of the environments we consider, the \emph{stationary non-reactive environment}, assumes Markovianity (equation \eqref{markov-assum}), $Y$-stationarity (equation \eqref{stat-assum}), and $C$-stationarity \& strong non-reactivity (equation \eqref{exch-assum}). Examples of this environment arise in reinforcement learning as a stationary (contextual) bandit as well as in many common adaptive experimental designs---one example being in adaptive clinical trials \citep[e.g.,][]{giles2003adaptive}.\footnote{To distinguish this setting from the adaptive crop yield experiment example, note that in an adaptive clinical trial, patients are selected i.i.d.~from some population, thus obeying the $C$-stationarity \& strong non-reactivity (equation \eqref{exch-assum}). On the other hand, in an adaptive crop yield experiment, due to weather fluctuations (such as temperature and precipitation) over time, this assumption is violated. For the same reason, we expect adaptive clinical trial settings, but not adaptive crop yield experiments, to be $Y$-stationary.}
\end{env}

\begin{env}[Stationary MDP]\label{stat-mdp-env}
With the additional assumption of stationarity, the \emph{stationary MDP} is a special type of MDP in which $Y$-stationarity (equation \eqref{stat-assum}) also holds, thereby assuming that the MDP's transition distribution does not change across timesteps. An example of stationary MDP data is in patient admissions to the emergency room during surge demand \citep[e.g.,][]{lee2018markov}.\footnote{Again, this example differs from that of electronic health records in Environment~\ref{mdp-env} because here, we do not expect the transition dynamics to change over time, whereas in the previous example, administering certain medication to a patient may in fact change their internal state and thus alter the way in which they react to the same medication later on.} A second example arises in robotics, where the robot’s dynamics and interactions with its environment can be modeled as a stationary MDP \citep[e.g.,][]{suarez2019practical}.
\end{env}

\paragraph{Constraints on $\proposalNoTilde$} The resampling procedure $\proposalNoTilde$ may incorporate two sources of randomization: (a) randomizing the order of the data by some permutation in a certain environment-specific set $\Pi$, and (b) randomizing the actions conditional on their $g$-evaluations. More specifically, $\proposalNoTilde$ must be such that the sequences of contexts, responses, and $g$-evaluations of the treatment in each resample are equal to some (random) permutation---albeit, the same one---$\pi \in \Pi$ of their respective sequences in the original data: \begin{equation}\label{fixed}\left(\tilde{C}^{(i)}_t,g(\tilde{X}^{(i)}_{t}),\tilde{Y}^{(i)}_t\right) = \left(C_{\pi_i(t)},g(X_{\pi_i(t)}),Y_{\pi_i(t)}\right), \;\; \forall t \in [\horizon], \textnormal{ for some }\pi_i \in \Pi.\end{equation} Note that equation \eqref{fixed} does not force $\tilde{X}^{(i)}_t = X_{\pi_i(t)}$, and thus allows the treatments to be further randomized over $\mathcal{X}$, so long as the restriction $g(X_{\pi_i(t)}) = g(\tilde{X}^{(i)}_{t})$ is met. Finally, the restricted set $\Pi$ of allowable permutations is environment-specific; we first focus on Environments~\ref{non-react-env}--\ref{stat-non-react-env}, where the non-stationarity of Environments~\ref{non-react-env} and \ref{mdp-env} prevent us from permuting the data at all, while the stationarity of Environment~\ref{stat-non-react-env} allows for any permutation:

\begin{proposition}\label{cond-indep-prop}
Let $\Pi_i, i \in [3]$ denote the set of allowable permutations for Environments~\ref{non-react-env}-\ref{stat-non-react-env}, above. Then if $\Pi_1 = \Pi_2 = \{\id\}$, where $\id$ denotes the identity permutation, and $\Pi_3 = \Pi_{[\horizon]}$, the set of all permutations on $[\horizon]$, the proportionality hypothesis $\mathcal{H}^\propto_0$ is satisfied under $\mathcal{H}^{\CondIndep}_0$.
\end{proposition}

\begin{proof}
In Environments~\ref{non-react-env} and \ref{mdp-env}, the restriction that $\Pi_1 = \Pi_2 = \{\id\}$ ensures that

\begin{align*}
    \hat{f}(\sampledData^{(i)}) &= \prod_{t=1}^\horizon \decisionProb(\tilde{X}^{(i)}_t|\tilde{C}^{(i)}_{t}, \tilde{H}^{(i)}_{t-1})\\
    &= \frac{\prod_{t=1}^\horizon \mathbb{P}_t(Y_t|C_t, g(X_t), \tilde{X}^{(i)}_{t})\decisionProb(\tilde{X}^{(i)}_t|\tilde{C}^{(i)}_{t}, \tilde{H}^{(i)}_{t-1})\mathbb{P}(C_t|C_1, Y_1, \ldots, C_{t-1}, Y_{t-1})}{\prod_{t=1}^\horizon \mathbb{P}_t(Y_t|C_t, g(X_t), X_t)\mathbb{P}(C_t|C_1, Y_1, \ldots, C_{t-1}, Y_{t-1})} && \text{by $\mathcal{H}^{\CondIndep}_0$ and that $g(\tilde{X}^{(i)}_t) = g(X_t)$}\\
    &= \frac{\prod_{t=1}^\horizon \mathbb{P}_t(\tilde{Y}^{(i)}_t|\tilde{C}^{(i)}_t, \tilde{X}^{(i)}_{t})\decisionProb(\tilde{X}^{(i)}_t|\tilde{C}^{(i)}_{t}, \tilde{H}^{(i)}_{t-1})\mathbb{P}(\tilde{C}^{(i)}_t|\tilde{C}^{(i)}_1, \tilde{Y}^{(i)}_1, \ldots, \tilde{C}^{(i)}_{t-1}, \tilde{Y}^{(i)}_{t-1})}{\prod_{t=1}^\horizon \mathbb{P}_t(Y_t|C_t, g(X_t), X_t)\mathbb{P}(C_t|C_1, Y_1, \ldots, C_{t-1}, Y_{t-1})} && \text{as $\Pi_1=\Pi_2 = \{\id\}$ \& equation~\eqref{fixed}}\\
    &= \frac{1}{\prod_{t=1}^\horizon \mathbb{P}_t(Y_t|C_t, g(X_t), X_t)\mathbb{P}(C_t|C_1, Y_1, \ldots, C_{t-1}, Y_{t-1})}f(\sampledData^{(i)}), && \text{due to equation \eqref{non-react-assum}}
\end{align*}
where the subscript $t$ on $\mathbb{P}$ is used to emphasize that the conditional distribution $Y_t \mid C_t, X_t$ may depend on $t$. Hence, the proportionality hypothesis is satisfied with $K = \frac{1}{\prod_{t=1}^\horizon \mathbb{P}_t(Y_t|C_t, g(X_t), X_t)\mathbb{P}(C_t|C_1, Y_1, \ldots, C_{t-1}, Y_{t-1})}$.

On the other hand, in Environment~\ref{stat-non-react-env}, $\Pi_3 = \Pi_{[\horizon]}$ and any permutation is allowed since
\begin{align*}
    \hat{f}(\sampledData^{(i)}) &= \prod_{t=1}^\horizon \decisionProb(\tilde{X}^{(i)}_t|\tilde{C}^{(i)}_{t}, \tilde{H}^{(i)}_{t-1})\\
    &= \frac{\prod_{t=1}^\horizon \mathbb{P}(Y_{\pi_i(t)}|C_{\pi_i(t)}, X_{\pi_i(t)})\decisionProb(\tilde{X}^{(i)}_t|\tilde{C}^{(i)}_{t}, \tilde{H}^{(i)}_{t-1})\mathbb{P}(C_{\pi_i(t)})}{\prod_{t=1}^\horizon \mathbb{P}(Y_t|C_t, X_t)\mathbb{P}(C_t)} && \text{due to equations \eqref{stat-assum} and \eqref{exch-assum}}\\
    &= \frac{\prod_{t=1}^\horizon \mathbb{P}(Y_{\pi_i(t)}|C_{\pi_i(t)}, \tilde{X}^{(i)}_{t})\decisionProb(\tilde{X}^{(i)}_t|\tilde{C}^{(i)}_{t}, \tilde{H}^{(i)}_{t-1})\mathbb{P}(C_{\pi_i(t)})}{\prod_{t=1}^\horizon \mathbb{P}(Y_t|C_t, X_t)\mathbb{P}(C_t)} && \text{by $\mathcal{H}^{\CondIndep}_0$ and that $g(\tilde{X}^{(i)}_t) = g(X_{\pi_i(t)})$}\\
    &= \frac{\prod_{t=1}^\horizon \mathbb{P}(\tilde{Y}^{(i)}_t|\tilde{C}^{(i)}_t, \tilde{X}^{(i)}_{t})\decisionProb(\tilde{X}^{(i)}_{t}|\tilde{C}^{(i)}_t, \tilde{H}^{(i)}_{t-1})\mathbb{P}(\tilde{C}^{(i)}_t)}{\prod_{t=1}^\horizon \mathbb{P}(Y_t|C_t, X_t)\mathbb{P}(C_t)} && \text{by equation~\eqref{fixed}}\\
    &= \frac{1}{\prod_{t=1}^\horizon \mathbb{P}(Y_t|C_t, X_t)\mathbb{P}(C_t)} \cdot f(\sampledData^{(i)}), && \text{due to equation \eqref{exch-assum}}
\end{align*} where we have dropped the subscript $t$ on $\mathbb{P}$ used above (and also do not need to subscript $\mathbb{P}(C_{\pi_i(t)})$ due to the $Y$-stationarity assumption of equation \eqref{stat-assum} (resp.~the $C$-stationarity assumption of equation \eqref{exch-assum}). Hence, once again, the proportionality hypothesis holds, but now with $K = \frac{1}{\prod_{t=1}^\horizon \mathbb{P}(Y_t|C_t, X_t)\mathbb{P}(C_t)}$.
\end{proof}

The precise definition of the allowable set of permutations $\Pi_4$ (also introduced in \cite{hmm}) in Environment~\ref{stat-mdp-env} is somewhat more complicated due to the serial dependence between consecutive timesteps (i.e., $C_{t+1}$ is generated conditionally on $(C_t, X_t)$) that is not present in Environment~\ref{stat-non-react-env}. In particular, only permutations which preserve the property that the $t^{\text{th}}$ response is equal to the $(t+1)^{\text{th}}$ context are allowed. We formally state a Proposition here, but relegate its proof to Appendix~\ref{appendixC-1}:

\begin{proposition}\label{cond-indep-prop-mdp}
Setting \[\Pi_4 := \{\pi \in \Pi_{[\horizon]}: Y_{\pi(t)}=C_{\pi(t+1)}, \forall t \in [\horizon-1], \textnormal{ and } C_{\pi(1)} = C_1\},\] the proportionality hypothesis $\mathcal{H}^\propto_0$ is satisfied under $\mathcal{H}^{\CondIndep}_0$ in Environment~\ref{stat-mdp-env}.
\end{proposition}

We briefly note here that unless the MDP's state space $\mathcal{C}$ is relatively small, then the (data-dependent) permutation set $\Pi_4$ will typically be quite small and virtually no randomization in permutations can be performed. There are, however, settings in which the state space $\mathcal{C}$ is small, such as in the \emph{restless multi-armed bandit} considered by \cite{rmab1,mate2020collapsing} in the context of patient adherence monitoring and well-being, in which our test can be used effectively.

\begin{remark}[Comparison with existing work]\label{prior-work}
In Environment~\ref{non-react-env} (and Environment~\ref{stat-non-react-env}, which is a special case), when $g$ is a constant function, the unweighted randomization test of \cite{pocock1975sequential,simon1979restricted,ham,bojinov2019time} applies.
However, the testing procedure presented above has two advantages over those in prior work. First, by randomizing timesteps in addition to treatments, our procedure is more powerful than theirs in the stationary setting of Environment~\ref{stat-non-react-env}, especially for deterministic assignment algorithms for which their procedure is powerless; we empirically demonstrate this in Section~\ref{prior-work-comparison}. Second, their procedure requires that each resample rerun $\decisionMaker$ independently $\numMCSamples$ times with the same sequence of $C_t$ and $Y_t$ as in $\data$.\footnote{\cite{pocock1975sequential} and \cite{simon1979restricted} only describe how to do so for certain adaptive assignment algorithms, and both, as well as \cite{bojinov2019time}, only consider the non-contextual case.}
While this would seem to be the most natural and statistically powerful approach, it may not always be computationally feasible. In such a case, our method, by using a more easily computable resampling procedure, provides a computationally tractable workaround (albeit perhaps at the cost of degraded statistical efficiency per resample). As a concrete example, the adaptive assignment algorithm $\decisionMaker$ used to generate the original data could be based on Thompson sampling in a complex Bayesian model, thus rendering it too computationally burdensome to run for any but a very small number of MC samples. On the other hand, \emph{unnormalized} densities---which are all that our procedure requires, due to the proportionality assumption $\mathcal{H}^\propto_0$---are generally easy to compute, thus rendering computation of the weights in equation \eqref{weight-eqn} tractable under a less computationally intensive resampling procedure $\proposalNoTilde$.
\end{remark}

\begin{remark}[Relaxing structural assumptions]\label{no-markov}
In Appendix~\ref{appendixC-2}, we show the above testing procedure can be generalized to arbitrary adaptive data collection processes in which none of the assumptions of equations \eqref{stat-assum},\eqref{non-react-assum},\eqref{exch-assum} \emph{nor} the Markovianity assumption of equation \eqref{markov-assum} are made, and the adaptive data collection environment is assumed to be completely general. In such an environment, we show that one can, somewhat analogously, consider a sequence of functions $g_1, \ldots, g_\horizon$ and test simultaneous (over $t$) conditional independence between the $t^{\text{th}}$ context (as well as the $t^{\text{th}}$ response) and the prior sequence of actions given the prior \emph{sequences} of contexts and responses as well as the sequence comprising the $g_s$-evaluation of the $s^{\text{th}}$ action for $s \in [t]$. The special case of this hypothesis wherein all the $g_t$ are constant allows for unweighted randomization testing in the completely general environment described in this Remark as shown by the prior work of \cite{bojinov2019time}.
\end{remark}

\subsection{Testing for non-stationarity}\label{non-stationarity}

\begin{nullHypothesis*}
For our non-stationarity test, the null hypothesis $\mathcal{H}_0^{\Stat}$ is that the response distribution is stationary (but unknown). That is, the conditional distribution of response given the context and treatment is the same across all timesteps: \[\mathcal{H}_0^{\Stat}: Y_t \mid (C_t, X_t) \text{ does not depend on }t.\]
\end{nullHypothesis*}

In this section, we consider two environments. The first is a a special type of non-reactive environment (Environment~\ref{non-react-env}):

\begin{env}[$C$-stationary strongly non-reactive environment]\label{exch-non-react-env}
The \emph{$C$-stationary strongly non-reactive environment} is an environment in which equation \eqref{exch-assum} holds, in addition to the Markovianity assumption of equation \eqref{markov-assum}. Once again, (contextual) bandits and various adaptive experimental designs are special cases. One concrete example is in adaptive experimental designs studying the effects of job search assistance programs on helping job seekers find work \citep{caria2020adaptive} over short periods of time since, in these brief time intervals, we expect the ``context'' surrounding each individual (e.g., background, credentials, etc.) to be roughly i.i.d. One additional example of a $C$-stationary strongly non-reactive environment for which our theory also holds is an episodic MDP, in which each episode is viewed as a single time step, the reward sequence a single response, and the action sequence a single (high-dimensional) action.\footnote{This episodic MDP setting also falls under the category of a (stationary) non-reactive environment, and hence our tests of conditional independence presented in Section~\ref{distributional} also apply.}
\end{env}

The second environment we consider is simply the MDP of Environment~\ref{mdp-env}.

\paragraph{Constraints on $\proposalNoTilde$}By a proof similar to that of Proposition~\ref{cond-indep-prop} for Environment~\ref{stat-non-react-env}, it is straightforward to see that, as long as each draw from the resampling distribution $\proposalNoTilde$ is (any) permutation of the original data $\data$, the proportionality hypothesis $\mathcal{H}^\propto_0$ holds under $\mathcal{H}^{\Stat}_0$ in Environment~\ref{exch-non-react-env}. That is, as long as we set $\Pi = \Pi_{[\horizon]}$ \emph{and do not allow for any other randomization}, then $ \hat{f}(\sampledData^{(i)})\propto f(\sampledData^{(i)})$. Under the MDP setting of Environment~\ref{mdp-env}, we may use the same randomization set of permutations $\Pi_4$ stated in Proposition~\ref{cond-indep-prop-mdp} and again, do not include any additional randomization. We summarize this below:

\begin{proposition}\label{mdp-prop}
If $\Pi = \{\pi \in \Pi_{[\horizon]}: Y_{\pi(t)}=C_{\pi(t+1)}, \forall t \in [\horizon-1], \textnormal{ and } C_{\pi(1)} = C_1\}$ in Environment~\ref{mdp-env} and $\Pi = \Pi_{[\horizon]}$ in Environment~\ref{exch-non-react-env}, and the only randomization in $\proposalNoTilde$'s resampling consists solely of permutations drawn from $\Pi$, then the proportionality hypothesis $\mathcal{H}^\propto_0$ is satisfied under $\mathcal{H}^{\Stat}_0$.
\end{proposition}

\subsection{Inverting tests to construct confidence and prediction intervals}\label{invert-chap}
\subsubsection{Confidence regions in semiparametric models}\label{confidence-inversion}

Recall that the hypothesis tests in Section~\ref{distributional} were all nonparametric and focused on testing (conditional) independence and distributional equality. In this section, we turn to the problem of exact parametric inference in semiparametric models---focusing on semiparametric location models as a case study---and use a technique previously used in standard randomization testing to construct confidence intervals \citep[e.g.,][]{rabideau2021randomization}. Before proceeding, we emphasize that parametric inference for adaptively collected data is a challenging problem and, as far as we are aware, all examples in the literature are either asymptotic \citep[e.g.,][]{zhang2020inference,hadad2021confidence,deshpande2018accurate}, or conservative (see, e.g., \cite{howard2021time,kaufmann2021mixture}).

Now, consider the setting in which the response distribution is distributed according to a location family with locations determined by action. More precisely, letting $h_0$ denote a (unknown) base density, we assume that at each time-step $t \in [\horizon]$, $Y_t\mid (X_t=x)\sim h_{0}(y-\theta_x)$, where $x \mapsto \theta_x$ is some mapping of actions to location parameters. In such a setting it is quite natural to ask: how much better or worse is action $x$ compared to $x'$, in terms of their location parameters? More formally, how can we test against the null hypothesis $\mathcal{H}^{\Loc,\delta,x,x'}_0$ that $\theta_{x'} - \theta_{x} = \delta$ for actions $x, x' \in \mathcal{X}$ and $\delta \in \mathbb{R}$?

To test against $\mathcal{H}^{\Loc,\delta,x,x'}_0$, we modify the dataset $\data$ and then perform a conditional independence test. Specifically, by modifying the dataset by replacing $Y_t$ with $Y_t+\delta\cdot \mathbf{1}[X_t=x]$, we get that $\mathcal{H}^{\Loc,\delta,x,x'}_0$ implies that $x$ and $x'$ induce the same distribution over rewards in the modified data generating distribution. Hence a test against $\mathcal{H}^{\CondIndep}$ with $g(X_t) = \begin{cases}
\{x,x'\} \textnormal{ if } X_t \in \{x,x'\}\\
X_t \textnormal{ otherwise}
\end{cases}$ on this modified dataset serves as a test against $\mathcal{H}^{\Loc,\delta,x,x'}_0$. These same ideas can also be applied to scale families; see Appendix~\ref{simultaneous-inference-appendix} for details.

Most importantly, these tests can all be inverted to construct confidence regions for the parameter in question, namely $\theta_{x'} - \theta_x$. For example, consider constructing a confidence region for the difference in locations of treatments $x$ and $x'$ at nominal miscoverage rate $\alpha$. One can construct the acceptance region by simply running the test against $\mathcal{H}^{\Loc,\delta,x,x'}_0$ at level $\alpha$ at each $\delta$ in the parameter space $\Delta$: those $\delta$ for which the test fails to reject make up the acceptance region. In cases where the parameter space $\Delta$ is either continuous or too large to iterate over completely, approximate confidence regions can be constructed as follows: (a) grid the space into a small finite set of discrete points $\Delta' \subseteq \Delta$, (b) run the above procedure at each $\delta \in \Delta'$ to obtain an accepted set of grid points $A'$, and (c) include all $\delta \in \Delta$ within a certain user-specified distance of any of the accepted grid points of $A'$ in the confidence region.

\subsubsection{Prediction regions for $Y_\horizon$}\label{conformal-inversion}

Similar to the previous section, the tests of Section~\ref{non-stationarity} can be applied to construct prediction intervals for $Y_\horizon$ before it is observed. In particular, again, one can construct the acceptance region for $Y_\horizon$ by running the non-stationarity tests described in Section~\ref{non-stationarity} on the almost-fully realized dataset $((C_1, X_1, Y_1), \ldots, (C_\horizon, X_\horizon, y)),$ with $y$ ranging over $\mathcal{Y}$; just as with confidence intervals, in the case of large $\mathcal{Y}$, the gridding/rounding procedure described at the end of the previous section can be used to construct approximate prediction intervals by simply selecting a small discrete set $\mathcal{Y}' \subseteq \mathcal{Y}$. We do note, however, that it is not uncommon for $\mathcal{Y}$ to be small or even categorical in many (challenging) settings and thus, for such problems, we can grid $\mathcal{Y}$ directly to obtain exactly valid prediction regions. We now make a few remarks about some broader connections of our procedure when used in this fashion to conformal inference, as well as challenges and speedups in constructing intervals.

\begin{remark}[Conditional conformal inference]
While conditional conformal inference is impossible in general, even in the case of i.i.d.~data \citep{lei2014distribution}, Theorem~\ref{weighted-mc-rand-valid} admits a conditional version which may be useful when the covariate space $\mathcal{X}$ is finite and small. In particular, the validity of the test still holds when we replace $\density$ with the conditional density $\density(\cdot|X_\horizon)$, which conditions on the ``test-time'' covariate $X_\horizon = x_\horizon$; thus, when inverting the hypothesis test and constructing a conformal prediction interval, one can guarantee valid coverage conditional on $X_\horizon$ (i.e., $\mathbb{P}(Y_{\horizon} \in C^\alpha_\horizon(X_{\horizon})|X_\horizon) \geq 1-\alpha$, where $C^\alpha_m(X_{\horizon})$ denotes conformal band at $X_{\horizon}$ with nominal miscoverage rate $\alpha$). In practice, this is possible as long as each resample $\sampledData^{(i)}$ has $\tilde{X}^{(i)}_\horizon= X_\horizon$ (in addition to $\sampledData^{(i)}$ being an allowable permutation of $\data$), so that \[\frac{\density(\sampledData^{(i)}|X_\horizon)\prod_{k \in \zeroM\backslash\{i\}} \proposalNoTilde(\sampledData^{(k)}|\sampledData^{(i)})}{\sum_{j=0}^\numMCSamples\density(\sampledData^{(j)}|X_\horizon)\prod_{k \in \zeroM\backslash\{i\}} \proposalNoTilde(\sampledData^{(k)}|\sampledData^{(j)})} = \frac{\density(\sampledData^{(i)})\prod_{k \in \zeroM\backslash\{i\}} \proposalNoTilde(\sampledData^{(k)}|\sampledData^{(i)})}{\sum_{j=0}^\numMCSamples\density(\sampledData^{(j)})\prod_{k \in \zeroM\backslash\{i\}} \proposalNoTilde(\sampledData^{(k)}|\sampledData^{(j)})}.\]
 In turn, $p$'s validity once again ensues so long as the proportionality hypothesis $\mathcal{H}^\propto_0$ holds.
\end{remark}

\begin{remark}[Challenges with explicit construction of conformal bands]\label{approx-rounding}
Split conformal inference \citep{split} is a variant of conformal inference which involves data splitting in order to construct a conformal prediction region which can be explicitly and efficiently computed. 
Unfortunately, such an explicit construction of a conformal band using data splitting does not carry over using our procedure to test non-stationarity, since the weights, through $\hat{\density}(\sampledData^{(i)})$, may depend on the test-time response grid values $y$. Similarly, the discretization procedure of \cite{discretized} used to construct explicit bands in the i.i.d.~setting by rounding the response to a small discrete set is not valid in our framework since, in general, probabilities under $\decisionProb$ involving the \emph{rounded} data are either unknown or not efficiently computable.
\end{remark}

\begin{remark}[Sharing samples]\label{sharing}
To address the challenges of Remark~\ref{approx-rounding}, one can grid the space into a small finite set $\mathcal{Y}'$ and run the test at each $y \in \mathcal{Y}'$, as discussed above in the case of continuous or large $\mathcal{Y}$. Naively, however, this involves drawing $\numMCSamples$ resamples for \emph{each} $y \in \mathcal{Y}'$. To reduce the total number of resamples needed, we can however share resamples between different values of $y$. That is, for $y_1, y_2 \in \mathcal{Y}'$, resamples drawn in association with $y_1$ can be used to determine whether or not to accept $y_2$ and vice versa, thereby more effectively using each resample drawn. See Appendix~\ref{sharing-appendix} for details.
\end{remark}

\section{Resampling procedures}\label{proposal-chap}

In Section~\ref{app-chap}, we focused on an information-theoretic question: in what data regimes can we define a sampling procedure $\proposalNoTilde$ for which our testing procedure is valid? Here, we turn to the second key question posed at the end of Section~\ref{wcrt-chap}: which choices of $\proposalNoTilde$ are \emph{statistically} best? Thus, while the last section specified how to choose the the support of $\proposalNoTilde$ (i.e., which variables should be randomized over and which should be held fixed) depending on the null hypothesis being tested against and the adaptive data collection environment in which it is being tested, this section explores what \emph{distributions} to choose over these supports. In the remainder of this section, we provide a partial answer to this challenging question by proposing a number of resampling procedures that are compatible with the constraints outlined in Section~\ref{app-chap} and will be shown to be powerful and produce short confidence/prediction regions in Section~\ref{sim-chap}.

Recall that we are focusing solely on conditionally i.i.d.~resampling from procedures $\proposalNoTilde$ and thus all resampling/proposal distributions considered in this section can be applied to both the weighted MC and unweighted MCMC tests (with $\proposalNoTilde$ as the proposal distribution for the MCMC test). However, as mentioned in Section~\ref{wcrt-chap}, our weighted MC procedure does not in general require conditionally i.i.d.~resampling and hence the following remark describing how our test can be generalized in this case is in order:

\begin{remark}[Non-i.i.d.~resampling]\label{non-iid}

When the resamples $\sampledData^{(1)}, \ldots, \sampledData^{(m)}$ are \emph{not} generated in a conditionally i.i.d.~manner, the test in Algorithm~\ref{weighted-mc-rand} can be slightly generalized. In particular, letting $\Sigma$ be any subset of $\Pi_{\zeroM}$, the set of permutations on $\zeroM$, and $\proposal((\sampledData^{(1)}, \ldots, \sampledData^{(m)})|\data)$ denote the conditional probability of sampling $(\sampledData^{(1)}, \ldots, \sampledData^{(m)})$ given that $\data$ was observed, the procedure can be generalized by redefining the weights to be \[\weightNoTilde_{\sampledSet}(\sampledData^{(i)}) = \frac{\hat{f}(\sampledData^{(i)})\sum_{\pi \in \Sigma: \pi(0) = i}\proposal((\sampledData^{(\pi(1))}, \ldots, \sampledData^{(\pi(m))})|\sampledData^{(i)})}{\sum_{j =0}^\numMCSamples\hat{f}(\sampledData^{(j)})\sum_{\pi' \in \Sigma:\pi'(0)=j}\proposal((\sampledData^{(\pi'(1))}, \ldots, \sampledData^{(\pi'(m))})|\sampledData^{(j)})}.\] We relegate a description of this generalized procedure, as well as the proof of its validity, to Appendix~\ref{non-iid-appendix}.
\end{remark}

\begin{remark}[Computational issues with conditionally i.i.d.~resampling]\label{comp-iid}
We also note here that even with a conditionally i.i.d.~resampling scheme $\proposalNoTilde$, the computation of the p-value $p$ can sometimes take $\Omega(m^2)$ computations. 
This is because, even with a conditionally i.i.d.~resampling procedure, all pairs of conditional densities $\proposalNoTilde(\sampledData^{(i)}|\sampledData^{(j)})$ must be computed and incorporated into the p-value computation. 
On the other hand, if the conditional density draws samples $\sampledData^{(1)}, \ldots, \sampledData^{(\numMCSamples)}$ such that 
$\proposalNoTilde(\cdot|\sampledData^{(i)})$ is the same for all $i \in \zeroM$, then 
\begin{equation}\label{q-prop}\prod_{k \in \zeroM\backslash\{i\}}\proposalNoTilde(\sampledData^{(k)}|\sampledData^{(i)}) = \frac{\prod_{k=0}^{\numMCSamples}\proposalNoTilde(\sampledData^{(k)}|\sampledData^{(i)})}{\proposalNoTilde(\sampledData^{(i)}|\sampledData^{(i)})} \propto (\proposalNoTilde(\sampledData^{(i)}|\sampledData^{(i)}))^{-1}\end{equation} 
and so the p-value can be computed much more quickly in only $O(\numMCSamples)$ computations by replacing $\prod_{k \in \zeroM\backslash\{i\}}\proposalNoTilde(\sampledData^{(k)}|\sampledData^{(i)})$ with $(\proposalNoTilde(\sampledData^{(i)}|\sampledData^{(i)}))^{-1}$ in the weight equation \eqref{weight-eqn}; as a result, we focus on resampling procedures that have this property.

An equivalent way of characterizing this special case procedure of Algorithm~\ref{weighted-mc-rand} is that the conditional distribution $\proposalNoTilde(\cdot|\data)$ depends only on $\data$ through some statistic $A(\data)$ which is invariant under all resamples (i.e., $A(\sampledData^{(i)}) = A(\data),$ for all $i \in [\numMCSamples]$) and so the resulting p-value from Algorithm~\ref{weighted-mc-rand} is equal, under $\mathcal{H}^\propto_0$, to \[\sum_{i=0}^\numMCSamples\frac{f(\sampledData^{(i)}|A(\data))/\proposalNoTilde(\sampledData^{(i)}|A(\data))}{\sum_{j=0}^{\numMCSamples}f(\sampledData^{(j)}|A(\data))/\proposalNoTilde(\sampledData^{(j)}|A(\data))}\mathbf{1}[S(\sampledData^{(i)}) \geq S(\data)],\] which can be reduced, after conditioning on $A(\data)$, to the importance-weighted p-value of \cite[Theorem 2]{harrison}. For a more comprehensive explanation of the above connection as well as further details, see Appendix~\ref{harrison-connection}.
%At a high-level, though, our proof of Theorem~\ref{weighted-mc-rand-valid} is via conditioning and an application of Bayes' theorem. On the other hand, the proof in \cite{harrison} is by viewing Algorithm~\ref{weighted-mc-rand} as a (conditional) importance sampling procedure (with target distribution $f$ and conditional proposal distribution $q(\cdot|A(D))$ for some statistic $A$ of the observed data) and utilizes change of measure arguments involving Radon-Nikodym derivatives induced by the target/proposal distribution likelihood ratio. For further details, see Appendix~\ref{harrison-connection}

\end{remark}

The setup in this section is that we first consider resampling procedures used to test non-stationarity. We then go on to discuss resampling algorithms for testing the conditional independence null hypothesis $\mathcal{H}^{\CondIndep}_0$, some of which use some of the non-stationarity testing resampling algorithms as subprocedures in their sampling process. All resampling procedures presented in this section are evaluated in our simulations in Section~\ref{sim-chap}. Additionally, more detailed pseudocode outlines of the resampling procedures in this section can be found in Appendix~\ref{pseudocode}. Lastly, we make a brief note about the computation of probabilities under the resampling distribution $\proposalNoTilde$.

\begin{remark}
All resampling procedures we consider in this section---and in our simulations in Section~\ref{sim-chap}---either sample uniformly or involve some sort of sequential sampling procedure. In either case, computation of conditional probabilities under $\proposalNoTilde$ are tractable as they are constant in the former and, in the latter, can be calculated as the sample is generated by serially multiplying together the corresponding conditional probabilities of each sequential sample as it is generated. Of course, probabilities of the form $\proposalNoTilde(\data|\sampledData^{(i)})$ for $i \in [\numMCSamples]$ still must be computed (as the original dataset $\data$ is never resampled from $\proposalNoTilde$) from scratch, but this is simply done in the same manner as above: behaving as though $\data$ had indeed been sampled from $\proposalNoTilde$ conditionally on $\sampledData^{(i)}$ and sequentially multiplying conditional probabilities of each resampleed timestep.
\end{remark}

\subsection{Non-stationarity testing in a $C$-stationary strongly non-reactive environment}\label{non-stat-bandit-proposal}
We first describe four types of resampling procedures that can be used for non-stationarity testing in a $C$-stationary strongly non-reactive environment (i.e., Environment~\ref{exch-non-react-env}). As discussed in Section~\ref{non-stationarity}, all such distributions must only randomize timesteps by permuting the data sequence. Apart from uniform permutations, the other three procedures discussed in this section randomly permute the data in a way intended to mimic $\decisionMaker$ while also ensuring diverse random samples. We thus call these three resampling procedures $\text{imitation}_\pi$, $\text{re-imitation}_\pi$, and $\text{cond-imitation}_\pi$, the common word \emph{imitation} referencing the mimicking of $\decisionMaker$ (the prefixes \emph{re} and \emph{cond} will be explained later on in this section). All three procedures randomly permute the data by serially sampling not-yet-sampled timesteps from the original data sequence.

\paragraph{$\text{uniform}_\pi$ sampling} The $\text{uniform}_\pi$ sampling procedure simply selects a permutation of the data uniformly at random. Although simple, intuitively it may result in a diverse set of resamples.

\paragraph{$\text{imitation}_\pi$ sampling} This sampling procedure samples permutations by sequentially resampling timesteps from the original data, where the sampling distribution at time $t$ acts as though the first $t-1$ already-resampled timesteps were drawn according to the true data generating-process and selects the $t^{\text{th}}$ timestep proportionally to the probabilities dictated by $\decisionProb$. In other words, at each timestep $t$, letting $R$ denote the set of not-yet-sampled timesteps, the $\text{imitation}_\pi$ distribution draws a timestep $t'$ from $R$, where the probability of drawing any given $s \in R$ is proportional to $\decisionProb(X_s|C_s,\tilde{H}_{t-1})$; if $\decisionProb(X_s|C_s,\tilde{H}_{t-1}) = 0$ for all $s \in R$, the procedure is ended and an attempt at a new resample can be begun using the same process. Finally, $(C_{t'},X_{t'}, Y_{t'})$ is appended to $\tilde{H}_{t-1}$ and $t'$ is removed from $R$. Intuitively, the $\text{imitation}_\pi$ distribution behaves as $\decisionMaker$ would, feeding in the already-realized sequence of responses $(Y_1, \ldots, Y_\horizon)$, except that it may only sample amongst actions which correspond to not-yet-selected timesteps. See Algorithm~\ref{sim1-descr} for pseudocode.

The $\text{re-imitation}_\pi$ and $\text{cond-imitation}_\pi$ distributions which we describe below are intended \emph{only} for randomized decision-making algorithms $\decisionMaker$. When applied to a deterministic algorithm, they are both the same as $\text{imitation}_\pi$.

\paragraph{$\text{re-imitation}_\pi$ sampling} This distribution is similar to the $\text{imitation}_\pi$ distribution, except that, to incorporate more diversity, it independently regenerates the exogenous randomness that the randomized decision-making algorithm $\decisionMaker$ makes as it mimics it to draw resamples; it thus \emph{re}randomizes the randomness of $\decisionMaker$. More specifically, the $\text{re-imitation}_\pi$ distribution views $\mathcal{A}$ as a sequence of decision rules $\delta_t$ which take as input the tuple $(C_t, H_{t-1}, U_1, \ldots, U_{t})$, where $U_t$ is the exogenous random variable generated by $\mathcal{A}$ at time $t$, and output which action to take:
\begin{enumerate}
    \item Sample a permutation of $\data$ by sequentially resampling timesteps from the data, where the sampling distribution at time $t$ uses the $t-1$ already-resampled timesteps as well as the resampled exogenous randomness $\tilde{U}_1, \ldots, \tilde{U}_{t-1}$, and then generates the random variable $\tilde{U}_t$ from $U_t$'s distribution, but conditional on the remaining timesteps. Specifically, $\tilde{U}_t$ is sampled from the conditional distribution of $U_t$ given that \[X_s=\delta_t(C_s, \tilde{H}_{t-1}, \tilde{U}_1, \ldots, \tilde{U}_{t-1}, U_{t}) \text{ for at least one not-yet-selected timestep }s.\] If, however, this conditional distribution is degenerate (i.e., because there does not exist $\tilde{U}_t$ for which $X_s=\delta_t(C_s, \tilde{H}_{t-1}, \tilde{U}_1, \ldots, \tilde{U}_{t-1}, \tilde{U}_{t})$ for any remaining timesteps $s$), the process is terminated and sampling for a new resample is begun.
    \item Select the $t^{\text{th}}$ timestep uniformly over all those not-yet-sampled triples $(C_s, X_s, Y_s)$ with \begin{equation}\label{decision-rule}X_s=\delta_t(C_s, \tilde{H}_{t-1}, \tilde{U}_1, \ldots, \tilde{U}_{t}).\end{equation}
\end{enumerate}

The motivation behind the $\text{re-imitation}_\pi$ distribution, as opposed to $\text{imitation}_\pi$, is that, by incorporating more of the randomness used in the decision-making algorithm one may be able to obtain more diverse samples while also better imitating the decision-making mechanisms of $\decisionMaker$. See Algorithm~\ref{sim2-descr} for pseudocode.

\paragraph{$\text{cond-imitation}_\pi$ sampling} The $\text{cond-imitation}_\pi$ distribution is the same as $\text{re-imitation}_\pi$ except that instead of resampling the $\tilde{U}_t$'s, it \emph{cond}itions on them and thus uses the same sequence of exogenous randomness as $\text{re-imitation}_\pi$ does; this, of course, requires knowing the original sequence $U_1, \ldots, U_t$. Intuitively, this conditioning that $\text{cond-imitation}_\pi$ sampling performs should bias the weights closer to $(m+1)^{-1}$, resulting in more powerful resampling. See Algorithm~\ref{sim3-descr} for pseudocode.

\subsection{Non-stationarity testing in an MDP}\label{non-stat-mdp-proposal}
We finally briefly discuss the resampling procedures for non-stationarity testing in an MDP (Environment~\ref{mdp-env}). We consider essentially the same four types of resampling procedures used for non-stationarity testing in a $C$-stationary strongly non-reactive environment described in the last section. The only difference however, is that, as discussed in Section~\ref{non-stationarity}, only a subset of permutations are allowed in the MDP setting so as to ensure that the $\tilde{Y}^{(i)}_t = \tilde{C}^{(i)}_{t+1}$ condition that holds for $i = 0$ (i.e., in the observed data) also holds for each resample $\sampledData^{(i)}$. Thus, while the four procedures which we consider here---also called $\text{uniform}_\pi$, $\text{imitation}_\pi$, $\text{re-imitation}_\pi$, and $\text{cond-imitation}_\pi$---sample timesteps serially without replacement according to $\decisionMaker$ as their analogs did in the previous section, they do so over only a restricted subset of not-already-sampled timesteps at each round.\footnote{For $\text{uniform}_\pi$ permutations, we sequentially sample indices uniformly at random from these restricted subsets.} 

For this reason, the MDP data that we consider includes an additional action $X_{\horizon+1}$; hence the permutations which we consider can be viewed as permuting the $\horizon+1$ state-action pairs in this augmented dataset.\footnote{Note that, practically speaking, if the dataset $\data$ in consideration does \emph{not} have this additional action $X_{\horizon+1}$, then the analyst can add it with ease (because they know the adaptive assignment algorithm $\decisionMaker$) and can do so without affecting the null or alternative distribution from which the data was drawn (because the null/alternative distributions govern only the transition dynamics, and not action selection).} More specifically, suppose that at the end of round $t-1$, the state-action pair $(C_{s'}, X_{s'})$ had just been selected and appended to $\tilde{H}_{t-2}$. Then, the set of allowable timesteps which can be sampled at round $t$ are only those not-yet-sampled state-action pairs $(C_s, X_s)$ for which the state-action-next state triple $(C_{s'}, X_{s'}, C_s)$ is present in the original data sequence $\data$. The precise way in which the timestep is sampled is then in exact accordance with the weighting, randomization, and conditioning that correspond to the $\text{uniform}_\pi$, $\text{imitation}_\pi$, $\text{re-imitation}_\pi$, and $\text{cond-imitation}_\pi$ sampling procedures described in the previous section. See Algorithms~\ref{u-descr-mdp}, \ref{sim1-descr-mdp}, \ref{sim2-descr-mdp}, and \ref{sim3-descr-mdp} for pseudocode.

\subsection{Conditional independence testing}\label{distributional-proposal}
We now describe four types of resampling procedures that can be used in a stationary non-reactive environment (Environment \ref{stat-non-react-env}). Three of these resampling procedures randomize both timesteps and the action $X_{t}$ conditional on $g(X_t)$ allowed by the stationarity, as discussed in Section~\ref{distributional}. The procedure which does not both randomize timesteps and actions simply randomizes the actions alone and is called $\text{imitation}_X$; as such, it can also be applied to the non-reactive environment (Environment \ref{non-react-env}) as well as an MDP (Environment~\ref{mdp-env}). Of the three procedures which randomize both timesteps and actions, two randomize the timesteps and actions in two separate stages and therefore use some of the resampling procedures discussed in the previous section (as well as one more) in the first stage; we call these resampling procedures $\text{uniform}_{\pi}\text{+}\text{imitation}_X$ and $\text{restricted-uniform}_{\pi}\text{+}\text{imitation}_X$ (the latter of these two uses a permutation scheme that involves $g$ and was thus not discussed in the previous two sections). These resampling procedures can all be applied to a stationary MDP (Environment~\ref{stat-mdp-env}), by using the analogous MDP permutation distribution as described in Section~\ref{non-stat-mdp-proposal}. We note here that we do not consider the combinations $\text{imitation}_{\pi}\text{+}\text{imitation}_X$, $\text{re-imitation}_{\pi}\text{+}\text{imitation}_X$, and $\text{cond-imitation}_{\pi}\text{+}\text{imitation}_X$ because, while conditionally i.i.d., they violate the property discussed in Remark~\ref{comp-iid} that $\proposalNoTilde(\cdot|\sampledData^{(i)})$ is the same for all $i \in \zeroM$, and hence require $\Omega(\numMCSamples^2)$ computations rendering them somewhat computationally burdensome. The fourth resampling scheme combines the two stages \text{of permuting timesteps and randomizing $X_t$} into a single procedure and is thus referred to as $\text{combined}_{\pi,X}$; this procedure applies in the stationary non-reactive environment (Environment~\ref{stat-non-react-env}) and can be modified to work in a stationary MDP (Environment~\ref{stat-mdp-env}) by using the usual sequential permutation restriction described in Section~\ref{non-stat-mdp-proposal}. Finally, just as in the previous two sections, all of these distributions are based on the idea of trying to draw resamples in a way that mimics the behavior of $\decisionMaker$.

\paragraph{$\text{imitation}_X$ sampling} The $\text{imitation}_X$ distribution, at each timestep $t$, conditions on the $t-1$ already-resampled data points $((C_{1}, \tilde{X}_{1}, Y_{1}), \ldots, (C_{t-1}, \tilde{X}_{t-1}, Y_{t-1}))$ as well as $C_t$ and, treating them as though they were true data points sampled by $\decisionMaker$, samples $\tilde{X}_{t}$ amongst all $x \in \mathcal{X}$ with $g(x) = g(X_{t})$, weighting proportionally to the action-selection probabilities of $\decisionProb$---if all weights are $0$, then the process is exited and an attempt at a new resample can be begun. The intuition behind this sampling procedure is that it attempts to mimic $\decisionMaker$ by essentially feeding in the already-realized context, $g$-evaluation, and response sequences to generate the sequence of actions (each sampled conditionally on the $g$-evaluation at the corresponding timestep), thereby mimicking the true data-generating distribution induced by $\decisionMaker$ (resulting in weights closer to $(m+1)^{-1}$). We note here that the $\text{imitation}_{X}$ resampling algorithm when applied in Environments~\ref{non-react-env} and \ref{stat-non-react-env} yields precisely the same (unweighted) test as the prior work of \cite{bojinov2019time,ham,simon1979restricted,pocock1975sequential}. See Algorithm~\ref{s-descr} for pseudocode.

\paragraph{$\text{uniform}_{\pi}\text{+}\text{imitation}_X$ sampling} This resampling procedure proceeds in two stages: the first stage applies a uniform permutation using the $\text{uniform}_\pi$ sampling of Section~\ref{non-stat-bandit-proposal} and the second randomizes the treatment conditional on its $g$-evaluation in the permuted data sequence using the $\text{imitation}_X$ resampling procedure.
Similar to $\text{imitation}_X$, we intuitively expect such a sampling procedure to have weights near $(m+1)^{-1}$ while also incorporating significant diversity (resulting in more varied evaluated test statistics $S(\sampledData^{(i)})$) due to the initial uniform permutation. See Algorithm~\ref{us-descr} for pseudocode.

\paragraph{$\text{restricted-uniform}_{\pi}\text{+}\text{imitation}_X$ sampling} This procedure is identical to $\text{uniform}_{\pi}\text{+}\text{imitation}_X$ sampling, except that it modifies the uniform sampling in step 1. In particular, for this resampling procedure, the randomly sampled permutation in the first stage is a \emph{restricted uniform} permutation, which does \emph{not} permute the sequence of $g$-evaluations. In other words, only permutations $\pi$ for which $g(X_{t}) = g(X_{\pi(t)})$ for all $t \in [\horizon]$ are allowed, and the sampling is uniform over this restricted set. The intuition behind this sampling scheme is similar to that of the $\text{uniform}_{\pi}\text{+}\text{imitation}_X$ sampling procedure except that, by using restricted uniform permutations rather than fully uniform permutations, the sampled data appears more similar to the original data with the goal of making the weights closer to $(m+1)^{-1}$. See Algorithm~\ref{rus-descr} for pseudocode.

\paragraph{$\text{combined}_{\pi,X}$ sampling} As opposed to the previous two sampling schemes which permute timesteps and randomize treatments conditional on their $g$-evaluations in two separate stages, this sampling approach \emph{combines} both types of randomization into a single resampling stage. Specifically, at each timestep $t$, the $\text{combined}_{\pi,X}$ resampling procedure samples $t'$ according to the $\text{imitation}_\pi$ distribution from Section~\ref{non-stat-bandit-proposal} on $g$-evaluations over not-yet-selected timesteps, again, by conditioning on the already-resampled data, and then randomizes $X_{t'}$ conditional on $g(X_{t'})$. In other words, at timestep $t$, the $\text{combined}_{\pi,X}$ resampling proceeds by:
\begin{enumerate}
    \item Selecting a not-already-selected timestep $t'$ conditionally on $\tilde{H}_{t-1}$ via the $\text{imitation}_\pi$ distribution of Section~\ref{non-stat-bandit-proposal} applied to $g$-evaluations of actions (instead of simply the actions themselves). If $\decisionProb(g(X_s)|\tilde{H}_{t-1}, C_{s}) = 0$\footnote{By $\decisionProb(g(X_s)|\tilde{H}_{t-1}, C_{s})$ we mean the probability (induced by $\decisionProb$) that the $g$-evaluation at the $t^{\text{th}}$ timestep is equal to $g(X_s)$ given the history $\tilde{H}_{t-1}$ and context $C_s$.} for all not-yet-selected timesteps $s$, then no such sample $t'$ can be generated and so the sampling process is terminated and a new one may be begun.
    \item Once the timestep $(C_{t'}, X_{t'}, Y_{t'})$ is selected, $\tilde{X}_{t'}$ is sampled via $\decisionProb(\cdot|\tilde{H}_{t-1}, C_{t'}, g(X_{t'}))$.
\end{enumerate}
Similar to $\text{restricted-uniform}_{\pi}\text{+}\text{imitation}_X$ sampling, the intuition behind $\text{combined}_{\pi,X}$ sampling is that both randomization across timesteps and the treatments, conditional on their $g$-evaluations, are incorporated, except that here they are combined and both the timestep and first action component randomization mimic $\decisionMaker$. See Algorithm~\ref{c-descr} for pseudocode.

\section{Empirical Results}\label{sim-chap}
In Section~\ref{proposal-chap} we proposed a number of resampling procedures $\proposalNoTilde$. In this section, we implement these resampling procedures in a variety of adaptive data collection environments---which are all special cases of the more general environments discussed in Section~\ref{app-chap} for which our theory holds---and both demonstrate their validity and evaluate their statistical efficiency.

In our experiments, we first study the performance of our tests as the horizon $\horizon$ increases. Thus, we plot both average Type-I error (to empirically validate the Type-I error control guarantee of Theorem~\ref{weighted-mc-rand-valid}) as well as power under an alternative distribution, for values of $\horizon$ ranging from $10$ to $100$. This power plot, however, does not paint the whole picture. In each simulation, a combination of two factors influences the power of our method: 1.~the \emph{intrinsic} difficulty of the environment and 2.~the quality of our resampling algorithm. As an attempt to disentangle these two components, we additionally measure and plot the power of a standard randomization test on data collected via a baseline uniform i.i.d.~adaptive assignment algorithm which selects actions uniformly at random from $\mathcal{X}$ at each timestep $t$ independently from $H_{t-1}$. This baseline measurement captures the intrinsic difficulty of the environment: a less powerful test on data gathered by this baseline i.i.d.~treatment assignment indicates a more difficult environment. Now, while the power plots do illustrate the second factor regarding the quality of our resampling algorithms, this factor can be further decomposed into two more contributing components to paint a clearer picture: 1.~the effective sample size of the resampling procedure (i.e., ideally having weights close to $(m+1)^{-1}$) and 2.~the diversity of the resamples (i.e., if all resamples are equal---and equal to the original data---then all weights will be $(m+1)^{-1}$, but our test will be powerless). We disentangle these two components---and thereby give a more complete explanation of the accompanying power plots---by measuring effective sample size, \[\effnumMCSamples:= \frac{\left(\sum_{i=0}^\numMCSamples\weight_{\sampledSet}(\sampledData^{(i)})\right)^2}{\sum_{i=0}^\numMCSamples\weight_{\sampledSet}(\sampledData^{(i)})^2},\] and plotting $\frac{\effnumMCSamples}{\numMCSamples}$, the fractional effective sample size, under the alternative at the same increments as in the power plot. In Type-I error, power, and fractional effective sample size plots, we take $\numMCSamples$, the number of resamples drawn by our test, to be $100$. We also present plots showing how the power of our procedure grows with $\numMCSamples$ (taking values $10^2, 10^3,$ and $10^4$) for fixed $\horizon = 100$.

We note here that all randomization tests performed in our simulations use smoothed p-values (see Remark~\ref{smoothed-p-vals}) and thus provably control Type-I error (whose nominal rate is set at $\alpha = 0.05$ in all our experiments) at \emph{exactly} the nominal rate. In addition to testing, we also construct approximate confidence and conformal prediction intervals (at nominal miscoverage rate $0.05$), using the procedures described in Section~\ref{invert-chap}; as these inversion procedures involve non-smoothed $p$-values, we expect the coverage to be above $0.95$. All results are averaged over $1000$ independent trials and plotted with $\pm 2$ standard error bars. Finally, we only show plots involving the weighted MC version of our test (and its inversion for interval construction) in this section, because for each resampling procedure we consider, our weighted MC test \emph{is never outperformed by (in terms of power and length)}---and often in fact dominates---the unweighted MCMC test in all of our simulations. For analogous plots illustrating results for the unweighted MCMC test, see Appendix~\ref{mcmc-appendix}.

\paragraph{Computation} We also plot the computation times for each of our resampling algorithms in the various environments and adaptive data assignment algorithms considered here in Appendix \ref{comp-time-appendix}. In nearly all environments and assignment algorithms, we are able to run a powerful test on datasets of length $\horizon = 100$ within a matter of minutes (and often just seconds) in terms of serial computation time, and our test is of course embarrassingly parallelizable if desired. Generally, the computation times for each resampling procedure scale roughly linearly with $\horizon$. All code for our simulations is publicly available at \url{https://github.com/Yashnair123/RTs-for-AdaptiveData}.

\paragraph{Environments and adaptive assignment algorithms} We briefly describe the environments in and adaptive assignment algorithms for which we conduct our simulations. We consider a total of five different environments: two are examples of a stationary non-reactive environment (Environment~\ref{stat-non-react-env})---one with contexts and one without contexts---two are instantiations of a $C$-stationary strongly non-reactive environment (Environment~\ref{exch-non-react-env})---again, both contextual and contextless---and the last is an instance of an MDP (Environment~\ref{mdp-env}); recall that Environments~\ref{stat-non-react-env} and \ref{exch-non-react-env} are special cases of Environment~\ref{non-react-env}. The adaptive assignment algorithms we consider in these environments are summarized in Table \ref{fig:table}; note that, in each environment, we consider both a randomized and a deterministic adaptive assignment algorithm.

\begin{table}
\begin{tabular}{|p{9.5cm} ||p{6cm}| }
\hline
    Contextual stationary non-reactive environment & $\epsilon$-greedy, LinUCB\\
    \hline
    Contextless stationary non-reactive environment & $\epsilon$-greedy, UCB\\
    \hline
    Contextless $C$-stationary strongly non-reactive environment & $\epsilon$-greedy, UCB\\
    \hline
    Contextual $C$-stationary strongly non-reactive environment & $\epsilon$-greedy, LinUCB\\
    \hline
    Markov decision process & $\epsilon$-greedy $Q$-learning, greedy $Q$-learning\\
    \hline
\end{tabular}

\caption{Table of environments and adaptive assignment algorithms considered in each.}
\label{fig:table}
\end{table}

We now briefly describe each of these adaptive assignment algorithms. $Q$-learning \citep{watkins1989learning} maintains an estimate of the state-action value $Q$ function and selects actions at each timestep based on the current estimate; we consider one version in which this action selection is (deterministically) greedy and one in which it is $\epsilon$-greedy. The $\epsilon$-greedy algorithm in the contextless stationary non-reactive environment and contextless $C$-stationary strongly non-reactive environment both, at each timestep, determine the empirically best action and then select an action $\epsilon$-greedily. In the contextual stationary non-reactive environment, the $\epsilon$-greedy algorithm behaves similarly by maintaining a linear regressor $L_x$ for each action $x \in \mathcal{X}$, and progressively updating $L_x$ with the context-response (i.e., input-output) pair $(C_t, Y_t)$ when $x$ is selected at time $t$ (upon seeing $C_t$); the algorithm selects, at time $t$ after seeing $C_t$, the action $x$ with highest predicted response by the $L_x$'s. Finally, UCB \citep{auer2002finite} is a deterministic bandit algorithm which additively inflates each action's empirical value by a bound on its error with respect to the true value and selects the highest value action. The contextual analogue in which the response depends linearly on the context is LinUCB \citep{li2010contextual}.

We note here that essentially all of the adaptive assignment algorithms above are typically used in reinforcement learning as they all enjoy quite low regret. However, in comparison to much of the literature on asymptotic inference in reinforcement learning, which impose clipping constraints on these assignment algorithms that stipulate that action-selection probabilities cannot be too close to $0$ or $1$ \citep{zhang2020inference,deshpande2018accurate,hadad2021confidence}, our procedure makes no such assumptions and even allows for deterministic adaptive assignment algorithms.

\subsection{Conditional independence testing}\label{cond-indep-hypo-test-sim}
In this section we apply the conditional independence testing framework and corresponding resampling algorithms from Section~\ref{distributional-proposal} to two environments: a stationary strongly non-reactive environment (Environment~\ref{stat-non-react-env}) with contexts and one without contexts. Our simulations in the former environment demonstrate the power gain our framework has over prior work \citep{ham,bojinov2019time,pocock1975sequential,simon1979restricted} in a stationary environment by incorporating the additional randomization over permutations described in Section~\ref{distributional}. On the other hand, our simulations in the latter environment focus on the problem of testing if a subset of treatments induce the same response and thus demonstrates our test's power on an inferential test for which, to the best of our knowledge, no prior exact test exists.

\subsubsection{Conditional independence testing with constant $g$}\label{prior-work-comparison}
Our first series of simulations is in a contextual stationary strongly non-reactive environment, and involves testing against the null hypothesis $\mathcal{H}^{\CondIndep}$ with $g(X_t) := \emptyset$ (i.e., $Y_t \indep X_t \mid C_t$ for all $t \in [\horizon]$). While, as mentioned in Remark~\ref{prior-work}, this setting has been covered in prior work, our framework offers a more powerful test, as our simulation results illustrate. Additionally, we note that in this setting, the weighted MC and unweighted MCMC tests are identical. This is because all weights in the MC test are $(m+1)^{-1}$ and all acceptance ratios in the MCMC test are $1$ since our resampling procedures all involve the $\text{imitation}_{X}$ distribution which, under a constant $g$, is the same as sampling, conditionally on the context and response sequences, from $\mathcal{A}$. For this reason, we do not plot the fractional effective sample sizes, as they too are all equal to $1$.

The precise environment in which our simulations in this section are conducted has treatment space $\mathcal{X} = \{0,1\}$. Letting $I_k$ denote the $k\times k$ identity matrix and $\vec{1}_k$ the length-$k$ vector of all $1$'s, we consider null and alternative distributions involving $2$-dimensional contexts $C_t$ sampled i.i.d.~from $\mathcal{N}\left(\begin{pmatrix}
1\\
-1
\end{pmatrix}, I_2\right)$ and whose conditional response distributions are, respectively, given by:

\[Y_t \mid (C_t,X_t) \sim \mathcal{N}(C_t^\top \vec{1}_2, 1) \textnormal{ and } \; \; \; Y_t \mid (C_t,X_t) \sim \mathcal{N}(C_t^\top \vec{1}_2+X_t, 1).\] Finally, the test statistic used is simply the absolute value of the $t$-test statistic against $\beta_2 = 0$ in a Normal linear model with design matrix comprising rows of the form $(1, X_t, C_{t,1}, C_{t,2})$ and response vector $(Y_t)_{t=1}^\horizon$.

\begin{figure}
     \centering
     \includegraphics[width=5.1in, height=3.1in]{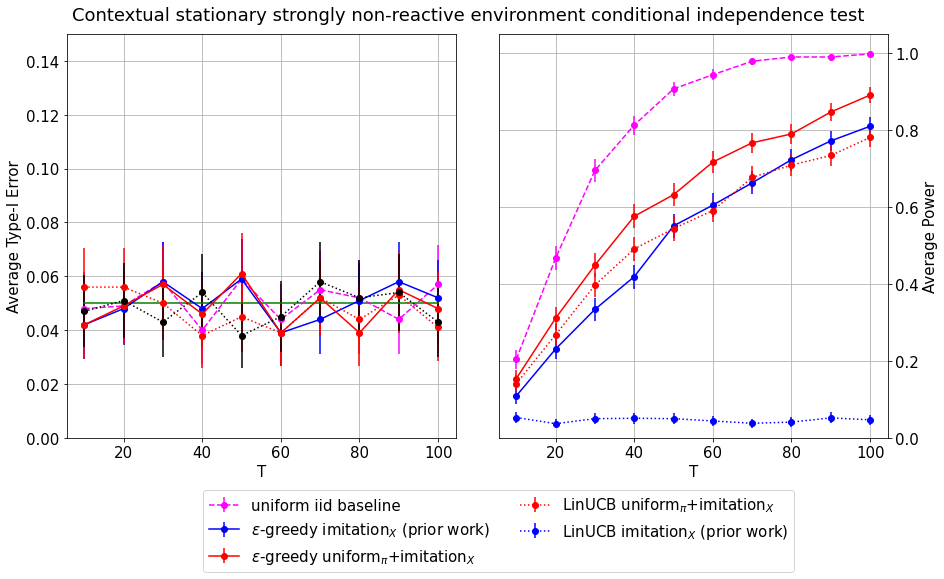}
     \caption{Type-I error (left) and power (right) of randomization tests at fixed $\numMCSamples = 100$ and varying $\horizon$ in a contextual stationary strongly non-reactive environment on data gathered via $\epsilon$-greedy and LinUCB.}
     %$\mathcal{X}= \{-1,1\}$ and $Y_i |X_i \sim \mathcal{N}(X_i,1), i \in [m-1]$. Data is collected using $\pi^{\egreedy}$ and $\pi^{\UCB}$, with $Y_m |X_m \sim \mathcal{N}(X_m,1)$ under the null $\mathcal{H}_0^{\Stat}$ and $Y_m |X_m \sim \mathcal{N}(4X_m,1)$ under the alternative $\mathcal{H}_1^{\Stat}$.}
     \label{fig:cont_dist_test}
 \end{figure}
 
Figure~\ref{fig:cont_dist_test} demonstrates that the added diversity of incorporating uniform permutations---as opposed to sampling only from the $\text{imitation}_X$ distribution---indeed results in an increase in power. Notably, when comparing our resampling scheme, $\text{uniform}_{\pi}\text{+}\text{imitation}_{X}$, against the $\text{imitation}_{X}$ resampling of prior work \citep{ham,bojinov2019time,pocock1975sequential,simon1979restricted} we observe that our approach is more powerful than theirs on data collected via an $\epsilon$-greedy treatment assignment. Perhaps even more strikingly, our approach when applied to LinUCB, a deterministic adaptive assignment algorithm, has high and increasing (with $\horizon$) power. This is in contrast to the usual $\text{imitation}_{X}$ resampling of the prior work, which is powerless. We relegate the power curves for increasing $\numMCSamples$ but finite $\horizon$ to Appendix~\ref{auxiliary-sims} as they are all approximately constant.

\subsubsection{Conditional independence testing with non-constant $g$}\label{dist-test-subset-sim}
We now discuss our simulations in the contextless stationary non-reactive setting of Environment~\ref{stat-non-react-env}, given as an example in the beginning of Section~\ref{distributional}. In particular, we consider a contextless instantiation of a stationary non-reactive environment with action space $\mathcal{X} = \{0, 1, 2\}$ and are testing against the null hypothesis $\mathcal{H}^{\CondIndep}_0$ with $g(X_t) := \mathbb{I}(X_t=2)$ (i.e., that $Y_t \indep X_t \mid X_t \in \{0, 1\}$ for all $t \in [\horizon]$).

In this environment, we specify the null and alternative distributions as \[Y_t \mid X_t \sim \begin{cases}
\mathcal{N}(0,1) \textnormal{ if } X_t \in \{0, 1\}\\
\mathcal{N}(2,1) \textnormal{ if } X_t = 2
\end{cases} \textnormal{ and } \; \; \; Y_t \mid X_t \sim \begin{cases}
\mathcal{N}(0,1) \textnormal{ if }X_t = 0\\
\mathcal{N}(3,1) \textnormal{ if }X_t = 1\\
\mathcal{N}(2,1) \textnormal{ if }X_t = 2
\end{cases},\] respectively.

\begin{figure}
     \centering
     \includegraphics[width=6.5in, height=2.5in]{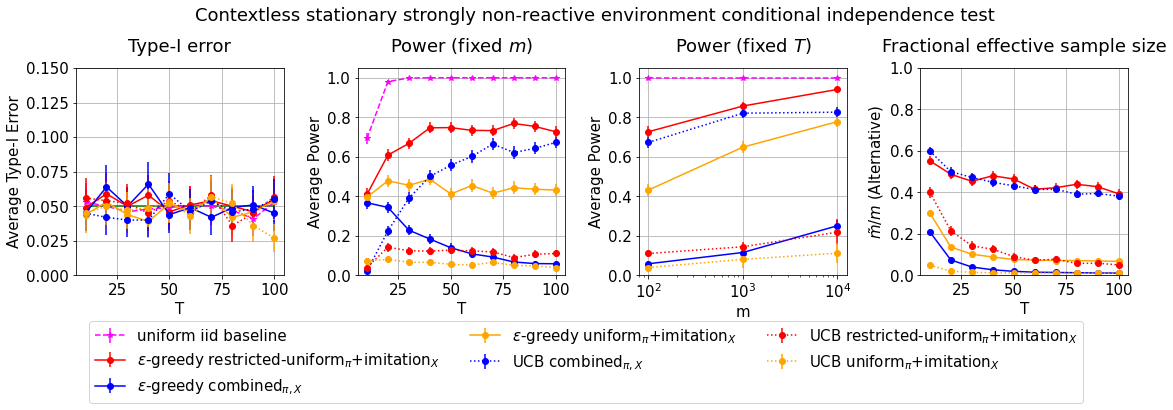}
     
     \caption{Type-I error rate (leftmost) and power (second from left) of the MC randomization test at fixed $\numMCSamples=100$ and varying $\horizon$ as well as power at fixed $\horizon=100$ and varying $\numMCSamples$ (third from left) and fractional effective sample size plots at fixed $\numMCSamples=100$ and varying $\horizon$ (rightmost) in a contextless stationary strongly non-reactive environment on data gathered via $\epsilon$-greedy, UCB, and the uniform i.i.d.~baseline.}
     \label{fig:dist_test}
 \end{figure}

In all our simulations, the test statistic $S(\data)$ we use is, similar to as in Section~\ref{prior-work-comparison}, simply the absolute value of the $t$-test statistic for the test against $\beta_2 = 0$ in a Normal linear model whose design matrix has $(1, \mathbb{I}(X_t=0), \mathbb{I}(X_t=2))$ as its $t^{\textnormal{th}}$ row and $(Y_t)_{t=1}^{\horizon}$ as the response vector. Figure~\ref{fig:dist_test} summarizes the results in this domain.

First, the Type-I error rates in Figure~\ref{fig:dist_test} are controlled at exactly the nominal level, validating the theoretical guarantee of Theorem~\ref{weighted-mc-rand-valid}. With regards to power, we notice that $\text{restricted-uniform}_{\pi}+\text{imitation}_X$ sampling performs better on data gathered via $\epsilon$-greedy while $\text{combined}_{\pi,X}$ sampling performs better under UCB. The fractional effective sample size plots in the same figure offer an explanation for why this is the case: the fractional effective sample sizes under $\epsilon$-greedy are larger with $\text{restricted-uniform}_{\pi}+\text{imitation}_X$ sampling than with $\text{combined}_{\pi,X}$ and, conversely, under UCB they are larger with $\text{combined}_{\pi,X}$ than $\text{restricted-uniform}_{\pi}+\text{imitation}_X$. We note here that this testing problem is much harder on data gathered via $\epsilon$-greedy and UCB than on the uniform i.i.d.~baseline (thus explaining the power gap between the latter and all resampling procedures applied on data gathered by the former two, especially for large $\horizon$). This is because $\epsilon$-greedy and UCB are low-regret adaptive assignment algorithms and thus, under the alternative, will select action $1$ very frequently, and all other actions much less frequently. Hence the problem of detecting if the response distributions induced by $X_t = 0$ and $X_t=1$ are the same becomes much more challenging, as action $0$ is sampled rarely under these two low-regret algorithms. On the other hand, it is sampled more frequently and at the same rate as action $1$ under the uniform i.i.d.~baseline. Despite this challenge, however, our method is still able to attain quite high power under both $\epsilon$-greedy and UCB adaptive assignment algorithms.

 \subsection{Non-stationarity testing}\label{sim-non-stat}
 In this section, we empirically evaluate our test of non-stationarity on three different environments: the first two are examples of the $C$-stationary strongly non-reactive setting of Environment~\ref{exch-non-react-env}---one with contexts and one without---and the third is an MDP (Environment~\ref{mdp-env}).

\subsubsection{Testing non-stationarity in a $C$-stationary strongly non-reactive environment}\label{stat-bandit}

Our simulations in this section are performed on a contextless $C$-stationary strongly non-reactive environment with action space $\mathcal{X}= \{-1,1\}$ and Gaussian rewards: the reward distribution for the first $\horizon-1$ steps is given by $Y_t \mid X_t \sim \mathcal{N}(X_t,1)$. Under the null hypothesis $\mathcal{H}^{\Stat}_0$, the reward distribution at the $\horizon^{\text{th}}$ timestep is unchanged and hence $Y_\horizon \mid X_\horizon \sim \mathcal{N}(X_\horizon,1)$. We analyze power under an alternative distribution that samples $Y_\horizon \mid X_\horizon \sim \mathcal{N}(4X_\horizon,1)$. Finally, as we are testing for non-stationarity, our test statistic $S$ is simply a non-conformity score, and we choose it to be the absolute residual: $S(\data) = |Y_\horizon - \hat{\mu}_{\data}(X_\horizon)|,$ where $\hat{\mu}_{\data}$ is the fitted ordinary least squares (OLS) model to $\data$ with an intercept term.

 \begin{figure}
     \centering
     \includegraphics[width=6.5in, height=2.5in]{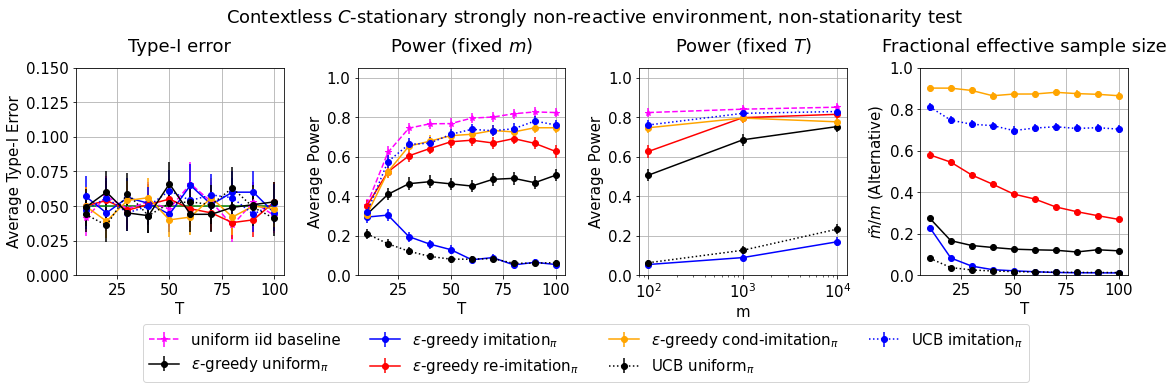}
     
     \caption{Type-I error rate (leftmost) and power (second from right) of the MC randomization test at fixed $\numMCSamples=100$ and varying $\horizon$ as well as power for fixed $\horizon=100$ and varying $\numMCSamples$ (third from right) and fractional effective sample size at fixed $\numMCSamples=100$ and varying $\horizon$ (rightmost) in a contextless $C$-stationary strongly non-reactive environment with data gathered via $\epsilon$-greedy, UCB, and the uniform i.i.d.~baseline.}
     \label{fig:nonstat}
 \end{figure}

Figure~\ref{fig:nonstat} shows the results for our simulations in this section. Indeed the Type-I error plots once again demonstrate the validity our randomization tests. In terms of power, Figure~\ref{fig:nonstat} shows that $\text{cond-imitation}_\pi$ sampling performs best under an $\epsilon$-greedy treatment assignment and $\text{imitation}_\pi$ performs best under UCB and both attain power close to that of the uniform i.i.d.~baseline for nearly all values of $\horizon$. Figure~\ref{fig:nonstat}, however, also shows that, with large enough $\numMCSamples$, $\text{re-imitation}_\pi$ sampling eventually performs better than $\text{cond-imitation}_\pi$ at $\horizon = 100$. Combining this information with the fractional effective sample size plots of Figure~\ref{fig:nonstat}, we thus see that while $\text{cond-imitation}_\pi$ has greater effective sample size than $\text{re-imitation}_\pi$, it has less diverse samples, leading to a more powerful procedure when $\numMCSamples$ is small (since, in this regime, the diversity plays a greater role), but a (slightly) less powerful one when $\numMCSamples$ is large.

\subsubsection{Testing non-stationarity in a contextual $C$-stationary strongly non-reactive environment}\label{context-stat-bandit}
We now describe our simulations in a contextual $C$-stationary strongly non-reactive environment. The specific environment we consider in this section involves a action space $\mathcal{X} = \{-1,1\}$ and $100$-dimensional contexts sampled i.i.d.~from $\mathcal{N}(\vec{1}_{100}, I_{100})$. The conditional response distribution during the first $\horizon-1$ timesteps is a sparse linear combination of the context vector and is given by \[Y_t \mid (C_t, X_t) \sim \mathcal{N}(-5X_t + \sum_{j=1}^{10}C_{t,j},1).\] The null conditional response distribution at time $\horizon$ is of course the same as the above whereas the alternative distribution we evaluate power under swaps the effects of the two treatments and is given by \[Y_\horizon \mid (C_\horizon, X_\horizon) \sim \mathcal{N}(5X_\horizon + \sum_{j=1}^{10}C_{\horizon,j},1).\] We regularize the regressor $L_x$ used in the $\epsilon$-greedy adaptive assignment by using Lasso regression with with penalty parameter $10$ as opposed to OLS. Finally, the test statistic used is the same non-conformity score as in the previous section, except that we again use Lasso---this time with penalty parameter chosen through $5$-fold cross-validation---instead of OLS, as $\hat{\mu}_{\data}$.

\begin{figure}
     \centering
     \includegraphics[width=6.5in, height=2.5in]{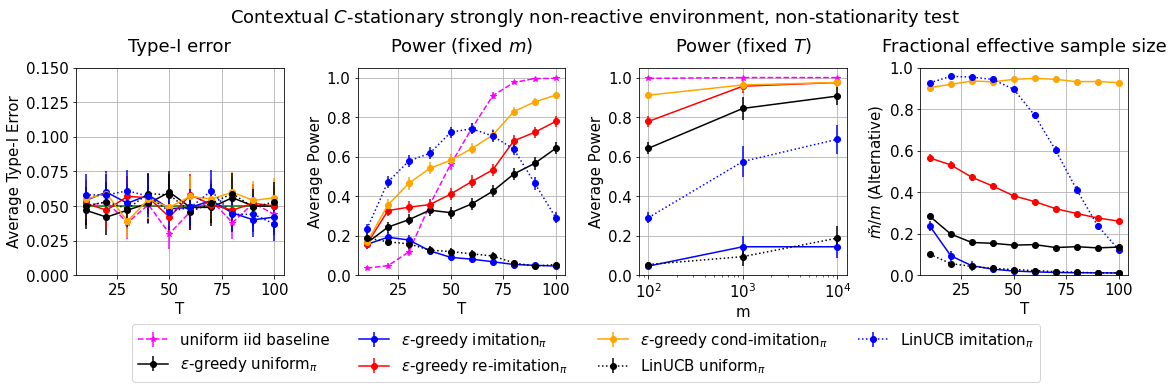}
     
     \caption{Type-I error rate (leftmost) and power (second from right) of the MC randomization test at fixed $\numMCSamples=100$ and varying $\horizon$ as well as power for fixed $\horizon=100$ and varying $\numMCSamples$ (third from right) and fractional effective sample size at fixed $\numMCSamples=100$ and varying $\horizon$ (rightmost) in a contextless $C$-stationary strongly non-reactive environment with data gathered via $\epsilon$-greedy, LinUCB, and the uniform i.i.d.~baseline.}
     \label{fig:nonstat2}
 \end{figure}

 Figure~\ref{fig:nonstat2} shows the Type-I error, power, and effective sample size curves for our simulations in this environment. The Type-I error is again controlled at the nominal level. In terms of power, similar conclusions to those drawn in the contextless $C$-stationary strongly non-reactive environment of the previous section apply here, too. In particular, while $\text{cond-imitation}_\pi$ sampling performs best under an $\epsilon$-greedy adaptive assignment at $\numMCSamples=100$ (and again exhibits, for large $\horizon$, power quite close to---and, in fact greater than, for small $\horizon$---that of the uniform i.i.d.~baseline) it has essentially the same power as $\text{re-imitation}_\pi$ at both $\numMCSamples = 10^3$ and $\numMCSamples=10^4$. Using the fractional effective sample size plots, we may therefore infer, once again, that while $\text{cond-imitation}_\pi$ sampling has a higher effective sample size than $\text{re-imitation}_\pi$ does under $\epsilon$-greedy, its samples exhibit less diversity than those of $\text{re-imitation}_\pi$. Under LinUCB, the contextual analog of the deterministic UCB assignment, however, we see that while $\text{imitation}_\pi$ has quite high power for small to moderate values of $\horizon$, the power drops for larger $\horizon$. The fractional effective sample size curve explains that this is caused by a corresponding drop in the effective sample size as $\horizon$ grows. We hypothesize that this is due to the extremely uneven action selection exhibited by LinUCB due to its very low regret (even as compared to $\epsilon$-greedy); we discuss why this low regret and the uneven action selection it causes---which we again emphasize is extreme in the case of LinUCB, even in comparison to $\epsilon$-greedy---renders the testing problem harder in the paragraph below. As such, we leave the problem of developing a powerful resampling scheme for deterministic algorithms like LinUCB in this type of contextual $C$-stationary strongly non-reactive environment to future work.
 
Lastly, we hypothesize that the shift in relative power of the uniform i.i.d.~baseline (from the least powerful adaptive assignment-resampling algorithm combination for small $\horizon$ to the most powerful for large $\horizon$) is again an artifact of the low-regret properties of $\epsilon$-greedy and LinUCB. This low regret results in the outlier context-action-response triple sampled at the $\horizon^{\text{th}}$ timestep to often have action agreeing with the action taken during most of the first $\horizon-1$ timesteps. Thus, for small $\horizon$, these first $\horizon-1$ context-treatment-response triples may outweigh the effect of the outlier at time $\horizon$ when training $\hat{\mu}_{\data}$ via Lasso with cross-validation, thereby resulting in better outlier detection than that under the uniform i.i.d.~base, wherein only around half of the first $\horizon-1$ timesteps (and thus very few, in total) will have action agreeing with that at the $\horizon^{\text{th}}$. The effect however, is diminished with large $\horizon$, most likely because in that regime even half of the data generated via $Y_t$'s true (null) conditional distribution, which agrees with the action taken at time $\horizon$, is enough to offset the effect of the outlier in training $\hat{\mu}_{\data}$. In particular, the remaining timesteps whose corresponding action differs from the one taken at time $\horizon$---of which there are more under the uniform i.i.d.~baseline than under $\epsilon$-greedy (and \emph{many} more under the baseline than under LinUCB) for large $\horizon$---may allow for a better fit $\hat{\mu}_{\data}$ through more effective learning of the dependence of $Y_t$ on $X_t$. As noted above, this may also explain the difficulty in attaining high power for large $\horizon$ under LinUCB in this environment. We empirically validate this hypothesis in Figure~\ref{fig:comparison} of Appendix~\ref{auxiliary-sims} by showing that the uniform i.i.d.~baseline exhibits precisley this same relative performance, with respect to $\horizon$, in comparison to a biased i.i.d.~assignment algorithm that selects action $-1$ with probability $0.9$ and otherwise selects action $1$.

\subsubsection{Testing non-stationarity in an MDP}
Our final set of hypothesis testing simulations is in an MDP, where recall that $Y_{t} = C_{t+1}$ and hence we drop the notation $Y_t$ and refer to $C_t$ as states instead of contexts.\footnote{As discussed in Section~\ref{non-stat-mdp-proposal}, the dataset we consider in these simulations is the usual dataset $\data$ along with one final action taken at time $T+1$: $((C_1, X_1, C_2), \ldots, (C_\horizon, X_\horizon, C_{\horizon+1}), X_{\horizon+1})$ as this allows us to simply permute the state-action pairs $(C_t, X_t)$.} The precise specifications of environment involve a state space of $\mathcal{C} = \{0,1,2\}$, action space $\mathcal{X} = \{-1,1\}$, and transition kernel during the first $\horizon-1$ transitions given by \[C_{t+1} \mid (C_t, X_t) \sim \begin{cases}
C_t + X_t\pmod{3} \textnormal{ with probability }0.95\\
C_t -X_t \pmod{3} \textnormal{ with probability }0.05
\end{cases}.\] Under the null hypothesis, the transition distribution at time $\horizon$ remains the same as above, but under the alternative, we swap the role of the two actions so that the alternative distribution is given by \[C_{\horizon+1} \mid (C_\horizon, X_\horizon) \sim \begin{cases}
C_\horizon - X_\horizon\pmod{3} \textnormal{ with probability }0.95\\
C_\horizon +X_\horizon \pmod{3} \textnormal{ with probability }0.05
\end{cases}.\] Finally, the test statistic that we use in these simulations trains a decision tree classifier on the data $\data$ and outputs the negative log likelihood loss of the trained model on the triple $(C_\horizon, X_\horizon, C_{\horizon+1})$ at time $\horizon$.
 
 \begin{figure}
     \centering
     \includegraphics[width=6.5in, height=2.5in]{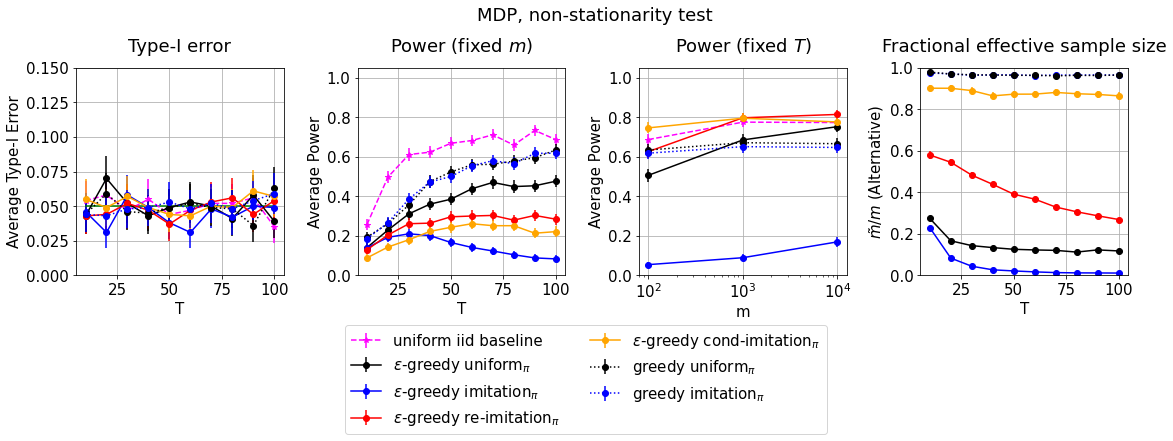}
     \caption{Type-I error rate (leftmost) and power (second from right) of the MC randomization test at fixed $\numMCSamples=100$ and varying $\horizon$ as well as power for fixed $\horizon=100$ and varying $\numMCSamples$ (third from right) and fractional effective sample size at fixed $\numMCSamples=100$ and varying $\horizon$ (rightmost) in a contextless $C$-stationary strongly non-reactive environment with data gathered via $\epsilon$-greedy and greedy $Q$-learning.}
     \label{fig:nonstat3}
 \end{figure}

Figure~\ref{fig:nonstat3} summarizes the results for our simulations in this setting. Once again, Type-I error is controlled at exactly the nominal level, as guaranteed by our theory. In a departure from the results of the previous two sections (most likely due to the sequential dependence in this environment not present in the last two as well as the modified sampling process that it requires), $\text{uniform}_{\pi}$ resampling has the greatest power for $Q$-learning with $\epsilon$-greedy action selection. Additionally, both $\text{uniform}_{\pi}$ and $\text{imitation}_\pi$ resampling perform similarly well for $Q$-learning with greedy action selection. Again, using both the fractional effective sample size plots and power plots with fixed $\horizon$ but varying $\numMCSamples$, we see that under $\epsilon$-greedy, $\text{re-imitation}_\pi$ resampling has lower effective sample size than $\text{uniform}_{\pi}$, but higher diversity, leading it to perform worse for smaller $\numMCSamples$ but better for large $\numMCSamples$. Finally, we note that, in comparison to the uniform i.i.d.~baseline\footnote{In contrast to the last section, the uniform i.i.d.~data used in this section is not actually gathered in the same environment as the other adaptive assignment algorithms, and in particular, the uniform i.i.d.~data is \emph{not} MDP data. Rather, the data comprises state-action-next state triples $(C,X,C')$ sampled i.i.d.~from a distribution in which both $C$ and $X$ are independent and uniform over $\mathcal{C}$ and $\mathcal{X}$ respectively, and $C'$ is sampled from the transition distribution described above, conditional on $(C,X)$.}, $\text{uniform}_\pi$ and $\text{imitation}_\pi$ are very competitive under the greedy $Q$-learning adaptive assignment, and $\text{imitation}_{\pi}$ also exhibits quite high power under $\epsilon$-greedy.

\subsection{Constructing confidence and prediction intervals}
In this section we apply our framework to constructing confidence and prediction intervals using the inversion procedures described in Section~\ref{invert-chap}. In particular, we plot both the coverage of the intervals as well as their average length, averaged over $1000$ trials for varying $\horizon$ at a fixed number of MC samples $\numMCSamples = 100$. For construction of conformal intervals, we also apply the sample sharing described in Remark~\ref{sharing} (in addition to the standard gridding procedure described in Section~\ref{invert-chap}) at both $\numMCSamples = 10$ and $\numMCSamples = 100$. Finally, we note that some of the resampling procedures in certain environments and adaptive assignment algorithms considered in this section exhibit overcoverage, in contrast to the last section, in which all tests attained Type-I error control at exactly the nominal level. We remind the reader that this is due to the fact that the gridding procedure described in Section~\ref{invert-chap} is based on the (approximate) inversion of a conservative test, rather than that of the smoothed test described in Remark~\ref{smoothed-p-vals}.

\subsubsection{Confidence intervals}
We now discuss our simulations constructing confidence intervals using the gridding procedure described at the end of Section~\ref{invert-chap}. We consider essentially the same environment as the contextless stationary strongly non-reactive environment discussed in Section~\ref{dist-test-subset-sim}, except that here we have \[Y_t \mid X_t \sim \begin{cases}
\mathcal{N}(0,1) \textnormal{ if }X_t = 0\\
\mathcal{N}(b_0,1) \textnormal{ if }X_t = 1\\
\mathcal{N}(2,1) \textnormal{ if }X_t = 2
\end{cases},\] with $b_0 = 4$; we construct a confidence interval for the location difference between $Y_t \mid (X_t=0)$, and $Y_t \mid (X_t=1)$, which simply corresponds to $b_0$, using a gridding set $\mathcal{Y}' = [-1:9]$.

\begin{figure}
     \centering
     \includegraphics[width=6in, height=3in]{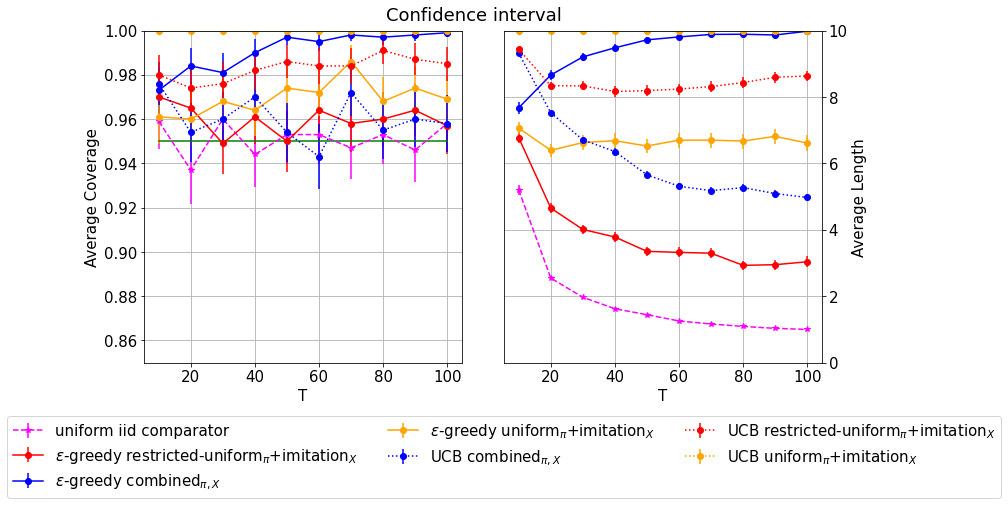}
     \caption{Coverage and average length of confidence intervals for location difference between $Y_t \mid (X_t=0)$, and $Y_t \mid (X_t=1)$ (i.e., $b_0=4$)} using the MC randomization test with data gathered via $\epsilon$-greedy, UCB, and the uniform i.i.d.~baseline.
     \label{fig:conf}
 \end{figure}

 Figure~\ref{fig:conf} summarizes the results for these simulations. We note there is no evidence in our simulations that the necessary (and in fact rather coarse) gridding results in any undercoverage.
 In terms of length, our results align with the power results presented in Section~\ref{dist-test-subset-sim}. In particular, under an $\epsilon$-greedy adaptive assignment, $\text{restricted-uniform}_{\pi}\text{+}\text{imitation}_X$ sampling produces the shortest average length intervals, whereas under UCB, $\text{combined}_{\pi,X}$ resampling does.

\subsubsection{Conformal prediction intervals}\label{conformal-sims}
Finally, we discuss our simulations constructing conformal prediction intervals. The environment considered in this section is the same as contextless $C$-stationary strongly non-reactive environment considered in Section~\ref{stat-bandit} except that we do not have access to $Y_\horizon$; it is the quantity for which we construct the conformal prediction region.

Our simulations in this section use the gridding procedure described at the end of Section~\ref{invert-chap} and assess the benefit of the sharing of samples described in Remark~\ref{sharing} and Appendix~\ref{sharing-appendix}. Without sample sharing, we fix $\numMCSamples = 100$ and set $\mathcal{Y}' = [-5:5]$ and plot coverage and length at varying $\horizon$. With sample sharing, we use the same gridding set $\mathcal{Y}'$ and study both $\numMCSamples = 10$ and $\numMCSamples = 100$ number of MC samples\footnote{For each $y \in \mathcal{Y'}$, $\numMCSamples$ samples are generated, but all $m|\mathcal{Y}'|$ samples are used to determine the membership of $y \in \mathcal{Y'}$ in the prediction interval} and varying $\horizon$. As discussed in Appendix~\ref{sharing-appendix} this sample sharing results in non-conditionally-i.i.d.~draws because the resamples generated corresponding to $y_1 \in \mathcal{Y}'$ may come from (and indeed do in our simulations) a different distribution than those sampled according to $y_2 \in \mathcal{Y}'$ for $y_1 \neq y_2$. As such, following Remark~\ref{non-iid} (and as discussed in further detail in Appendix~\ref{non-iid-appendix}), we choose a subset of permutations $\Sigma$ to be the set of $m+1$ permutations swapping $0$ and $i$ for each $i \in \zeroM$ and employ the test of Algorithm~\ref{weighted-mc-rand} with weights $\weightNoTilde_{\sampledSet}(\sampledData^{(i)})$.

\begin{figure}
     \centering
     \includegraphics[width=5.4in, height=2.7in]{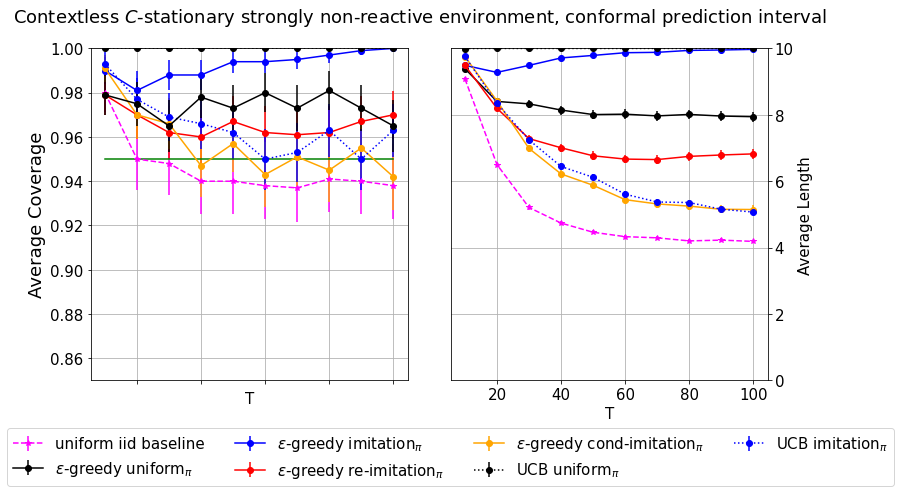}
     \caption{Coverage and average length of conformal prediction intervals for $Y_\horizon$ using the MC randomization test with data gathered via $\epsilon$-greedy, UCB, and the uniform i.i.d.~baseline.}
     \label{fig:conformal1}
 \end{figure}

 \begin{figure}
     \centering
     \includegraphics[width=6.5in, height=2.5in]{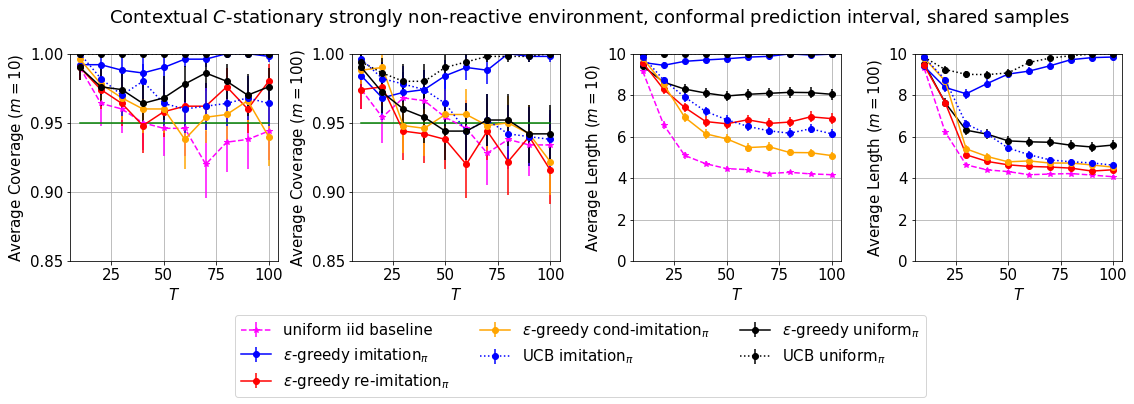}
     \caption{Coverage and average length of approximate conformal prediction intervals, with sample sharing, for $Y_\horizon$ using the MC randomization test with data gathered via $\epsilon$-greedy, UCB, and the uniform i.i.d.~baseline.}
     \label{fig:conformal2}
 \end{figure}

Figures~\ref{fig:conformal1} and \ref{fig:conformal2} summarize our simulation results constructing approximate conformal prediction intervals. Figure~\ref{fig:conformal1} illustrates both coverage and length as $\horizon$ grows by simply using the standard gridding procedure described at the end of Section~\ref{invert-chap}. With regards to average length, our results mirror those in Section~\ref{stat-bandit}, in which $\text{cond-imitation}_\pi$ resampling is best for an $\epsilon$-greedy adaptive assignment, and $\text{imitation}_\pi$ is best for UCB. In Figure~\ref{fig:conformal2}, we observe that any undercoverage, attributable to the gridding of $\mathcal{Y}$, is relatively minor, and should be fixable by using a finer-resolution grid; the reason that the uniform i.i.d.~baseline tends to exhibit the greatest undercoverage is most likely explained by that fact that we are not smoothing in this set of experiments, so that whereas all other procedures are conservative, the uniform i.i.d.~baseline is not. Additionally, we see essentially the same ranking of resampling procedures, although $\text{re-imitation}_\pi$ does perform slightly better than $\text{cond-imitation}_\pi$ when $100$ MC samples are used.
Finally, the shortest average length curves in Figure~\ref{fig:conformal2} using $\numMCSamples = 10$ are quite competitive in comparison to their counterparts in Figure~\ref{fig:conformal1}. In particular, $\text{imitation}_{\pi}$ under UCB produces only slightly wider intervals with $\numMCSamples=10$ and sample sharing than with $\numMCSamples=100$ and no sample sharing; for $\epsilon$-greedy, $\text{cond-imitation}_{\pi}$ performs essentially the same in both settings. This demonstrates that, by utilizing sample sharing, we are able to obtain conformal prediction intervals of nearly the same length at a fraction of the computational cost (because in the left hand column of Figure~\ref{fig:conformal2} we generate a \emph{total} of $110$ samples, whereas in Figure~\ref{fig:conformal1}, we generate $100$ samples \emph{for each} $y \in \mathcal{Y}'$). Indeed, as illustrated by Figures~\ref{fig:conformal1_time} and \ref{fig:conformal2_time} of Appendix~\ref{AppendixF}, the computation time required by the procedure using $\numMCSamples = 10$ and shared sampling is, for most resampling schemes and adaptive assignments, more than $5$ times faster than its $\numMCSamples = 100$ non-sample sharing counterpart.

\section{Discussion}
In this paper, we study the problem of performing various challenging inferential tasks on adaptively collected data. Through the development of a weighted MC randomization test (along with the novel application of the unweighted MCMC randomization test of \cite{besag1989generalized}), we show that such tasks can be performed with \emph{exact} Type-I error control with a great degree of flexibility in the choice of resampling procedure as long as the proportionality hypothesis $\mathcal{H}^\propto_0$ is satisfied.

The question, however, of how \emph{best} to perform these tests still remains, and is an interesting one for future research. In particular, while we have discussed some preliminary empirical results demonstrating powerful resampling algorithms in a number of different challenging environments, there are a number of interesting and potentially fruitful directions forward. In particular, as mentioned in Section~\ref{context-stat-bandit}, one important direction for future work is to develop resampling procedures that are powerful for deterministic algorithms like LinUCB in a contextual $C$-stationary strongly non-reactive setting. But more generally, it may be interesting to formally investigate any commonalities between powerful resampling algorithms in different environments in order to develop a more principled method for deciding which procedure to apply to a particular problem (i.e., combination of adaptive assignment algorithm, environment, and inferential task).

Along this same vein of how best to resample, it may be useful to also consider different resampling procedures which incorporate non-conditionally-i.i.d.~sampling. Utilization of these non-conditionally-i.i.d.~resampling schemes under the unweighted MCMC randomization testing framework has the potential to be especially advantageous as the full MCMC test will still require only $O(\numMCSamples)$ number of computations. On the other hand, as discussed in Remark~\ref{non-iid} and Appendix~\ref{non-iid-appendix}, when applying such sampling schemes to the weighted MC test, we may generalize and choose a subset $\Sigma$ of the full permutation set $\Pi_{\zeroM}$ different than $\Sigma_{\textnormal{swap}}$, and condition only on the data permutations induced by these sets rather than the full set of permutations.

As such, one final direction for future work that may also be of interest is to investigate how the power of the weighted MC test changes under different choices of $\Sigma$. Specifically, while we expect the test to be more powerful for choices of $\Sigma$ which are larger, we also expect such larger choices to require a greater amount of comptutation time. Hence, it may be worthwhile to study this tradeoff between the size of the sets $\Sigma$ and computation time. Just as in the potential investigation of more powerful resampling procedures mentioned above, exploring if there are more well-structured connections between not-too-large choices of $\Sigma$ that yield a powerful test and the specific problem at hand may allow for the development of a more coherent theory---or at least a more formal set of guidelines---for how best to select/develop a resampling algorithm for any given problem.

\paragraph{Acknowledgements} YN and LJ were partially supported by NSF CBET-21120 and LJ was partially supported by DMS-2045981. The authors would also like to thank Subhabrata Sen, Joe Blitzstein, and Henry Bosch for valuable discussions.

\bibliographystyle{plainnat}
\bibliography{references.bib}

\begin{thebibliography}{75}
\providecommand{\natexlab}[1]{#1}
\providecommand{\url}[1]{\texttt{#1}}
\expandafter\ifx\csname urlstyle\endcsname\relax
  \providecommand{\doi}[1]{doi: #1}\else
  \providecommand{\doi}{doi: \begingroup \urlstyle{rm}\Url}\fi

\bibitem[Alesso et~al.(2021)Alesso, Cipriotti, Bollero, and Martin]{crop}
Carlos~Agustín Alesso, Pablo~Ariel Cipriotti, Germán~Alberto Bollero, and
  Nicolas~Federico Martin.
\newblock Design of on-farm precision experiments to estimate site-specific
  crop responses.
\newblock \emph{Agronomy Journal}, 113\penalty0 (2):\penalty0 1366--1380, 2021.
\newblock \doi{https://doi.org/10.1002/agj2.20572}.
\newblock URL
  \url{https://acsess.onlinelibrary.wiley.com/doi/abs/10.1002/agj2.20572}.

\bibitem[Alvarsson et~al.(2021)Alvarsson, {Arvidsson McShane}, Norinder, and
  Spjuth]{ALVARSSON202142}
Jonathan Alvarsson, Staffan {Arvidsson McShane}, Ulf Norinder, and Ola Spjuth.
\newblock Predicting with confidence: Using conformal prediction in drug
  discovery.
\newblock \emph{Journal of Pharmaceutical Sciences}, 110\penalty0 (1):\penalty0
  42--49, 2021.
\newblock ISSN 0022-3549.
\newblock \doi{https://doi.org/10.1016/j.xphs.2020.09.055}.
\newblock URL
  \url{https://www.sciencedirect.com/science/article/pii/S002235492030589X}.

\bibitem[Auer et~al.(2002)Auer, Cesa-Bianchi, and Fischer]{auer2002finite}
Peter Auer, Nicolo Cesa-Bianchi, and Paul Fischer.
\newblock Finite-time analysis of the multiarmed bandit problem.
\newblock \emph{Machine learning}, 47\penalty0 (2):\penalty0 235--256, 2002.

\bibitem[Ban and He(2020)]{ban2020generic}
Yikun Ban and Jingrui He.
\newblock Generic outlier detection in multi-armed bandit.
\newblock In \emph{Proceedings of the 26th ACM SIGKDD International Conference
  on Knowledge Discovery \& Data Mining}, pages 913--923, 2020.

\bibitem[Barber et~al.(2022)Barber, Candes, Ramdas, and Tibshirani]{cpbeyond}
Rina Barber, Emmanuel Candes, Aaditya Ramdas, and Ryan Tibshirani.
\newblock Conformal prediction beyond exchangeability.
\newblock \emph{arXiv preprint arXiv:2202.13415}, 2022.

\bibitem[Barber and Janson(2022+)]{RB-LJ:2020}
Rina~Foygel Barber and Lucas Janson.
\newblock Testing goodness-of-fit and conditional independence with approximate
  co-sufficient sampling.
\newblock \emph{Annals of Statistics}, 2022+.
\newblock To Appear.

\bibitem[Bates et~al.(2021)Bates, Candès, Janson, and Wang]{metropolized}
Stephen Bates, Emmanuel Candès, Lucas Janson, and Wenshuo Wang.
\newblock Metropolized knockoff sampling.
\newblock \emph{Journal of the American Statistical Association}, 116\penalty0
  (535):\penalty0 1413--1427, 2021.
\newblock \doi{10.1080/01621459.2020.1729163}.
\newblock URL \url{https://doi.org/10.1080/01621459.2020.1729163}.

\bibitem[Battalio et~al.(2021)Battalio, Conroy, Dempsey, Liao, Menictas,
  Murphy, Nahum-Shani, Qian, Kumar, and Spring]{BATTALIO2021106534}
Samuel~L. Battalio, David~E. Conroy, Walter Dempsey, Peng Liao, Marianne
  Menictas, Susan Murphy, Inbal Nahum-Shani, Tianchen Qian, Santosh Kumar, and
  Bonnie Spring.
\newblock Sense2stop: A micro-randomized trial using wearable sensors to
  optimize a just-in-time-adaptive stress management intervention for smoking
  relapse prevention.
\newblock \emph{Contemporary Clinical Trials}, 109:\penalty0 106534, 2021.
\newblock ISSN 1551-7144.
\newblock \doi{https://doi.org/10.1016/j.cct.2021.106534}.
\newblock URL
  \url{https://www.sciencedirect.com/science/article/pii/S1551714421002706}.

\bibitem[Berrett et~al.(2019)Berrett, Wang, Barber, and
  Samworth]{Berrett2019TheCP}
Thomas~B. Berrett, Yi~Wang, Rina~Foygel Barber, and Richard~J. Samworth.
\newblock The conditional permutation test for independence while controlling
  for confounders.
\newblock \emph{Journal of the Royal Statistical Society: Series B (Statistical
  Methodology)}, 2019.

\bibitem[Berrett et~al.(2020)Berrett, Wang, Barber, and Samworth]{cpt}
Thomas~B. Berrett, Yi~Wang, Rina~Foygel Barber, and Richard~J. Samworth.
\newblock The conditional permutation test for independence while controlling
  for confounders.
\newblock \emph{Journal of the Royal Statistical Society: Series B (Statistical
  Methodology)}, 82\penalty0 (1):\penalty0 175--197, 2020.
\newblock \doi{https://doi.org/10.1111/rssb.12340}.
\newblock URL
  \url{https://rss.onlinelibrary.wiley.com/doi/abs/10.1111/rssb.12340}.

\bibitem[Besag and Clifford(1989)]{besag1989generalized}
Julian Besag and Peter Clifford.
\newblock Generalized monte carlo significance tests.
\newblock \emph{Biometrika}, 76\penalty0 (4):\penalty0 633--642, 1989.

\bibitem[Bibaut et~al.(2021)Bibaut, Kallus, Dimakopoulou, Chambaz, and van~der
  Laan]{bibaut2021risk}
Aur{\'e}lien Bibaut, Nathan Kallus, Maria Dimakopoulou, Antoine Chambaz, and
  Mark van~der Laan.
\newblock Risk minimization from adaptively collected data: Guarantees for
  supervised and policy learning.
\newblock \emph{Advances in Neural Information Processing Systems},
  34:\penalty0 19261--19273, 2021.

\bibitem[Bojinov and Shephard(2019)]{bojinov2019time}
Iavor Bojinov and Neil Shephard.
\newblock Time series experiments and causal estimands: exact randomization
  tests and trading.
\newblock \emph{Journal of the American Statistical Association}, 114\penalty0
  (528):\penalty0 1665--1682, 2019.

\bibitem[Bugni et~al.(2018)Bugni, Canay, and Shaikh]{bugni2018inference}
Federico~A Bugni, Ivan~A Canay, and Azeem~M Shaikh.
\newblock Inference under covariate-adaptive randomization.
\newblock \emph{Journal of the American Statistical Association}, 113\penalty0
  (524):\penalty0 1784--1796, 2018.

\bibitem[Burnaev and Vovk(2014)]{burnaev2014efficiency}
Evgeny Burnaev and Vladimir Vovk.
\newblock Efficiency of conformalized ridge regression.
\newblock In \emph{Conference on Learning Theory}, pages 605--622. PMLR, 2014.

\bibitem[Candes et~al.(2018)Candes, Fan, Janson, and Lv]{candes2018panning}
Emmanuel Candes, Yingying Fan, Lucas Janson, and Jinchi Lv.
\newblock Panning for gold:‘model-x’knockoffs for high dimensional
  controlled variable selection.
\newblock \emph{Journal of the Royal Statistical Society: Series B (Statistical
  Methodology)}, 80\penalty0 (3):\penalty0 551--577, 2018.

\bibitem[Caria et~al.(2020)Caria, Kasy, Quinn, Shami, Teytelboym,
  et~al.]{caria2020adaptive}
Stefano Caria, Maximilian Kasy, Simon Quinn, Soha Shami, Alex Teytelboym,
  et~al.
\newblock An adaptive targeted field experiment: Job search assistance for
  refugees in jordan.
\newblock 2020.

\bibitem[Carpenter et~al.(2020)Carpenter, Menictas, Nahum-Shani, Wetter, and
  Murphy]{addiction}
Stephanie Carpenter, Marianne Menictas, Inbal Nahum-Shani, David Wetter, and
  Susan Murphy.
\newblock Developments in mobile health just-in-time adaptive interventions for
  addiction science.
\newblock \emph{Current Addiction Reports}, 7, 09 2020.
\newblock \doi{10.1007/s40429-020-00322-y}.

\bibitem[Chen et~al.(2018)Chen, Chun, and Barber]{discretized}
Wenyu Chen, Kelli-Jean Chun, and Rina~Foygel Barber.
\newblock Discretized conformal prediction for efficient distribution-free
  inference.
\newblock \emph{Stat}, 7\penalty0 (1):\penalty0 e173, 2018.
\newblock \doi{https://doi.org/10.1002/sta4.173}.
\newblock URL \url{https://onlinelibrary.wiley.com/doi/abs/10.1002/sta4.173}.
\newblock e173 sta4.173.

\bibitem[Chernozhukov et~al.(2018)Chernozhukov, W{\"u}thrich, and
  Yinchu]{chernozhukov2018exact}
Victor Chernozhukov, Kaspar W{\"u}thrich, and Zhu Yinchu.
\newblock Exact and robust conformal inference methods for predictive machine
  learning with dependent data.
\newblock In \emph{Conference On Learning Theory}, pages 732--749. PMLR, 2018.

\bibitem[Coppersmith et~al.(2021)Coppersmith, Dempsey, Kleiman, Bentley,
  Murphy, and Nock]{coppersmith_dempsey_kleiman_bentley_murphy_nock_2021}
Daniel~D Coppersmith, Walter Dempsey, Evan Kleiman, Kate Bentley, Susan Murphy,
  and Matthew Nock.
\newblock Just-in-time adaptive interventions for suicide prevention: Promise,
  challenges, and future directions, Jan 2021.
\newblock URL \url{psyarxiv.com/eg9fx}.

\bibitem[Deshpande et~al.(2018)Deshpande, Mackey, Syrgkanis, and
  Taddy]{deshpande2018accurate}
Yash Deshpande, Lester Mackey, Vasilis Syrgkanis, and Matt Taddy.
\newblock Accurate inference for adaptive linear models.
\newblock In \emph{International Conference on Machine Learning}, pages
  1194--1203. PMLR, 2018.

\bibitem[Diaconis and Freedman(1980)]{diaconis}
P.~Diaconis and D.~Freedman.
\newblock {De Finetti's Theorem for Markov Chains}.
\newblock \emph{The Annals of Probability}, 8\penalty0 (1):\penalty0 115 --
  130, 1980.
\newblock \doi{10.1214/aop/1176994828}.
\newblock URL \url{https://doi.org/10.1214/aop/1176994828}.

\bibitem[Edgington and Onghena(2007)]{edgington2007randomization}
Eugene Edgington and Patrick Onghena.
\newblock \emph{Randomization tests}.
\newblock Chapman and Hall/CRC, 2007.

\bibitem[Fannjiang et~al.(2022)Fannjiang, Bates, Angelopoulos, Listgarten, and
  Jordan]{feedbackcovshift}
Clara Fannjiang, Stephen Bates, Anastasios~N. Angelopoulos, Jennifer
  Listgarten, and Michael~I. Jordan.
\newblock Conformal prediction for the design problem.
\newblock \emph{CoRR}, abs/2202.03613, 2022.
\newblock URL \url{https://arxiv.org/abs/2202.03613}.

\bibitem[Fisher et~al.(1937)]{fisher1937design}
Ronald~Aylmer Fisher et~al.
\newblock The design of experiments.
\newblock \emph{The design of experiments.}, \penalty0 (2nd Ed), 1937.

\bibitem[Gibbs and Candès(2021)]{gibbs2021adaptive}
Isaac Gibbs and Emmanuel Candès.
\newblock Adaptive conformal inference under distribution shift.
\newblock 2021.
\newblock Accepted.

\bibitem[Giles et~al.(2003)Giles, Kantarjian, Cortes, Garcia-Manero,
  Verstovsek, Faderl, Thomas, Ferrajoli, O’Brien, Wathen,
  et~al.]{giles2003adaptive}
Francis~J Giles, Hagop~M Kantarjian, Jorge~E Cortes, Guillermo Garcia-Manero,
  Srdan Verstovsek, Stefan Faderl, Deborah~A Thomas, Alessandra Ferrajoli,
  Susan O’Brien, Jay~K Wathen, et~al.
\newblock Adaptive randomized study of idarubicin and cytarabine versus
  troxacitabine and cytarabine versus troxacitabine and idarubicin in untreated
  patients 50 years or older with adverse karyotype acute myeloid leukemia.
\newblock \emph{Journal of clinical oncology}, 21\penalty0 (9):\penalty0
  1722--1727, 2003.

\bibitem[Godfrey et~al.(2013)Godfrey, Masquelin, and Hemmerle]{GODFREY2013795}
Alexander~G. Godfrey, Thierry Masquelin, and Horst Hemmerle.
\newblock A remote-controlled adaptive medchem lab: an innovative approach to
  enable drug discovery in the 21st century.
\newblock \emph{Drug Discovery Today}, 18\penalty0 (17):\penalty0 795--802,
  2013.
\newblock ISSN 1359-6446.
\newblock \doi{https://doi.org/10.1016/j.drudis.2013.03.001}.
\newblock URL
  \url{https://www.sciencedirect.com/science/article/pii/S135964461300069X}.

\bibitem[Hadad et~al.(2021)Hadad, Hirshberg, Zhan, Wager, and
  Athey]{hadad2021confidence}
Vitor Hadad, David~A Hirshberg, Ruohan Zhan, Stefan Wager, and Susan Athey.
\newblock Confidence intervals for policy evaluation in adaptive experiments.
\newblock \emph{Proceedings of the National Academy of Sciences}, 118\penalty0
  (15), 2021.

\bibitem[Hahn et~al.(2011)Hahn, Hirano, and Karlan]{hahn2011adaptive}
Jinyong Hahn, Keisuke Hirano, and Dean Karlan.
\newblock Adaptive experimental design using the propensity score.
\newblock \emph{Journal of Business \& Economic Statistics}, 29\penalty0
  (1):\penalty0 96--108, 2011.

\bibitem[Ham and Qiu(2022)]{ham}
Dae~Woong Ham and Jiaze Qiu.
\newblock Hypothesis testing in sequentially sampled data: Adaprt to maximize
  power beyond iid sampling, 2022.
\newblock URL \url{https://arxiv.org/abs/2205.02430}.

\bibitem[Harrison(2012)]{harrison}
Matthew~T. Harrison.
\newblock Conservative hypothesis tests and confidence intervals using
  importance sampling.
\newblock \emph{Biometrika}, 99\penalty0 (1):\penalty0 57--69, 2012.
\newblock ISSN 00063444, 14643510.
\newblock URL \url{http://www.jstor.org/stable/41720672}.

\bibitem[Hemerik and Goeman(2018)]{hemerik}
Jesse Hemerik and Jelle Goeman.
\newblock Exact testing with random permutations.
\newblock \emph{TEST}, 27\penalty0 (4):\penalty0 811--825, 2018.
\newblock \doi{10.1007/s11749-017-0571-1}.
\newblock URL \url{https://doi.org/10.1007/s11749-017-0571-1}.

\bibitem[Howard et~al.(2021)Howard, Ramdas, McAuliffe, and
  Sekhon]{howard2021time}
Steven~R Howard, Aaditya Ramdas, Jon McAuliffe, and Jasjeet Sekhon.
\newblock Time-uniform, nonparametric, nonasymptotic confidence sequences.
\newblock \emph{The Annals of Statistics}, 49\penalty0 (2):\penalty0
  1055--1080, 2021.

\bibitem[Hu and Lei(2020)]{hu2020distribution}
Xiaoyu Hu and Jing Lei.
\newblock A distribution-free test of covariate shift using conformal
  prediction.
\newblock \emph{arXiv preprint arXiv:2010.07147}, 2020.

\bibitem[Huang and Janson(2020)]{huang2020}
Dongming Huang and Lucas Janson.
\newblock Relaxing the assumptions of knockoffs by conditioning.
\newblock \emph{Ann. Statist.}, 48\penalty0 (5):\penalty0 3021--3042, 10 2020.
\newblock \doi{10.1214/19-AOS1920}.
\newblock URL \url{https://doi.org/10.1214/19-AOS1920}.

\bibitem[Huo and Fu(2017)]{huo2017risk}
Xiaoguang Huo and Feng Fu.
\newblock Risk-aware multi-armed bandit problem with application to portfolio
  selection.
\newblock \emph{Royal Society open science}, 4\penalty0 (11):\penalty0 171377,
  2017.

\bibitem[Kaufmann and Koolen(2021)]{kaufmann2021mixture}
Emilie Kaufmann and Wouter~M Koolen.
\newblock Mixture martingales revisited with applications to sequential tests
  and confidence intervals.
\newblock \emph{Journal of Machine Learning Research}, 22\penalty0
  (246):\penalty0 1--44, 2021.

\bibitem[Lai and Robbins(1985)]{LAI19854}
T.L Lai and Herbert Robbins.
\newblock Asymptotically efficient adaptive allocation rules.
\newblock \emph{Advances in Applied Mathematics}, 6\penalty0 (1):\penalty0
  4--22, 1985.
\newblock ISSN 0196-8858.
\newblock \doi{https://doi.org/10.1016/0196-8858(85)90002-8}.
\newblock URL
  \url{https://www.sciencedirect.com/science/article/pii/0196885885900028}.

\bibitem[Lai and Wei(1982)]{lai1982least}
Tze~Leung Lai and Ching~Zong Wei.
\newblock Least squares estimates in stochastic regression models with
  applications to identification and control of dynamic systems.
\newblock \emph{The Annals of Statistics}, 10\penalty0 (1):\penalty0 154--166,
  1982.

\bibitem[Lee and Lee(2018)]{lee2018markov}
Hyun-Rok Lee and Taesik Lee.
\newblock Markov decision process model for patient admission decision at an
  emergency department under a surge demand.
\newblock \emph{Flexible Services and Manufacturing Journal}, 30\penalty0
  (1):\penalty0 98--122, 2018.

\bibitem[Lehmann et~al.(2005)Lehmann, Romano, and Casella]{lehmann2005testing}
Erich~Leo Lehmann, Joseph~P Romano, and George Casella.
\newblock \emph{Testing statistical hypotheses}, volume~3.
\newblock Springer, 2005.

\bibitem[Lei and Wasserman(2014)]{lei2014distribution}
Jing Lei and Larry Wasserman.
\newblock Distribution-free prediction bands for non-parametric regression.
\newblock \emph{Journal of the Royal Statistical Society: Series B (Statistical
  Methodology)}, 76\penalty0 (1):\penalty0 71--96, 2014.

\bibitem[Li et~al.(2010)Li, Chu, Langford, and Schapire]{li2010contextual}
Lihong Li, Wei Chu, John Langford, and Robert~E Schapire.
\newblock A contextual-bandit approach to personalized news article
  recommendation.
\newblock In \emph{Proceedings of the 19th international conference on World
  wide web}, pages 661--670, 2010.

\bibitem[Li et~al.(2022)Li, Wang, Lu, Zhang, and Li]{li2022electronic}
Tianhao Li, Zhishun Wang, Wei Lu, Qian Zhang, and Dengfeng Li.
\newblock Electronic health records based reinforcement learning for treatment
  optimizing.
\newblock \emph{Information Systems}, 104:\penalty0 101878, 2022.

\bibitem[Ma et~al.(2015)Ma, Hu, and Zhang]{ma2015testing}
Wei Ma, Feifang Hu, and Lixin Zhang.
\newblock Testing hypotheses of covariate-adaptive randomized clinical trials.
\newblock \emph{Journal of the American Statistical Association}, 110\penalty0
  (510):\penalty0 669--680, 2015.

\bibitem[Mahajan and Gupta(2010)]{mahajan2010adaptive}
Rajiv Mahajan and Kapil Gupta.
\newblock Adaptive design clinical trials: Methodology, challenges and
  prospect.
\newblock \emph{Indian journal of pharmacology}, 42\penalty0 (4):\penalty0 201,
  2010.

\bibitem[Mary et~al.(2015)Mary, Gaudel, and Preux]{mary2015bandits}
J{\'e}r{\'e}mie Mary, Romaric Gaudel, and Philippe Preux.
\newblock Bandits and recommender systems.
\newblock In \emph{International Workshop on Machine Learning, Optimization and
  Big Data}, pages 325--336. Springer, 2015.

\bibitem[Mate et~al.(2020)Mate, Killian, Xu, Perrault, and
  Tambe]{mate2020collapsing}
Aditya Mate, Jackson Killian, Haifeng Xu, Andrew Perrault, and Milind Tambe.
\newblock Collapsing bandits and their application to public health
  intervention.
\newblock \emph{Advances in Neural Information Processing Systems},
  33:\penalty0 15639--15650, 2020.

\bibitem[Mate et~al.(2021)Mate, Madaan, Taneja, Madhiwalla, Verma, Singh,
  Hegde, Varakantham, and Tambe]{rmab1}
Aditya Mate, Lovish Madaan, Aparna Taneja, Neha Madhiwalla, Shresth Verma,
  Gargi Singh, Aparna Hegde, Pradeep Varakantham, and Milind Tambe.
\newblock Field study in deploying restless multi-armed bandits: Assisting
  non-profits in improving maternal and child health, 2021.
\newblock URL \url{https://arxiv.org/abs/2109.08075}.

\bibitem[Nahum-Shani et~al.(2016)Nahum-Shani, Smith, Spring, Collins,
  Witkiewitz, Tewari, and Murphy]{jitai}
Inbal Nahum-Shani, Shawna Smith, Bonnie Spring, Linda Collins, Katie
  Witkiewitz, Ambuj Tewari, and Susan Murphy.
\newblock Just-in-time adaptive interventions (jitais) in mobile health: Key
  components and design principles for ongoing health behavior support.
\newblock \emph{Annals of Behavioral Medicine}, 52, 09 2016.
\newblock \doi{10.1007/s12160-016-9830-8}.

\bibitem[Nettasinghe et~al.(2022)Nettasinghe, Chatterjee, Tipireddy, and
  Halappanavar]{hmm}
Buddhika Nettasinghe, Samrat Chatterjee, Ramakrishna Tipireddy, and Mahantesh
  Halappanavar.
\newblock Extending conformal prediction to hidden markov models with exact
  validity via de finetti's theorem for markov chains, 2022.
\newblock URL \url{https://arxiv.org/abs/2210.02271}.

\bibitem[Pallmann et~al.(2018)Pallmann, Bedding, Choodari-Oskooei, Dimairo,
  Flight, Hampson, Holmes, Mander, Odondi, Sydes, Villar, Wason, Weir, Wheeler,
  Yap, and Jaki]{adaptiveclinical}
Philip Pallmann, Alun~W. Bedding, Babak Choodari-Oskooei, Munyaradzi Dimairo,
  Laura Flight, Lisa~V. Hampson, Jane Holmes, Adrian~P. Mander, Lang'o Odondi,
  Matthew~R. Sydes, Sof{\'\i}a~S. Villar, James M.~S. Wason, Christopher~J.
  Weir, Graham~M. Wheeler, Christina Yap, and Thomas Jaki.
\newblock Adaptive designs in clinical trials: why use them, and how to run and
  report them.
\newblock \emph{BMC Medicine}, 16\penalty0 (1):\penalty0 29, 2018.
\newblock \doi{10.1186/s12916-018-1017-7}.
\newblock URL \url{https://doi.org/10.1186/s12916-018-1017-7}.

\bibitem[Papadopoulos et~al.(2007)Papadopoulos, Vovk, and Gammerman]{split}
Harris Papadopoulos, Volodya Vovk, and Alex Gammerman.
\newblock Conformal prediction with neural networks.
\newblock In \emph{19th IEEE International Conference on Tools with Artificial
  Intelligence(ICTAI 2007)}, volume~2, pages 388--395, 2007.
\newblock \doi{10.1109/ICTAI.2007.47}.

\bibitem[Pitman(1937)]{pitman}
E.~J.~G. Pitman.
\newblock Significance tests which may be applied to samples from any
  populations.
\newblock \emph{Supplement to the Journal of the Royal Statistical Society},
  4\penalty0 (1):\penalty0 119--130, 1937.
\newblock \doi{https://doi.org/10.2307/2984124}.
\newblock URL
  \url{https://rss.onlinelibrary.wiley.com/doi/abs/10.2307/2984124}.

\bibitem[Pocock and Simon(1975)]{pocock1975sequential}
Stuart~J Pocock and Richard Simon.
\newblock Sequential treatment assignment with balancing for prognostic factors
  in the controlled clinical trial.
\newblock \emph{Biometrics}, pages 103--115, 1975.

\bibitem[Popova et~al.(2018)Popova, Isayev, and Tropsha]{popova2018deep}
Mariya Popova, Olexandr Isayev, and Alexander Tropsha.
\newblock Deep reinforcement learning for de novo drug design.
\newblock \emph{Science advances}, 4\penalty0 (7):\penalty0 eaap7885, 2018.

\bibitem[Rabideau and Wang(2021)]{rabideau2021randomization}
Dustin~J Rabideau and Rui Wang.
\newblock Randomization-based confidence intervals for cluster randomized
  trials.
\newblock \emph{Biostatistics}, 22\penalty0 (4):\penalty0 913--927, 2021.

\bibitem[Rosenberger et~al.(2019)Rosenberger, Uschner, and Wang]{rosenberger}
William~F. Rosenberger, Diane Uschner, and Yanying Wang.
\newblock Randomization: The forgotten component of the randomized clinical
  trial.
\newblock \emph{Statistics in Medicine}, 38\penalty0 (1):\penalty0 1--12, 2019.
\newblock \doi{https://doi.org/10.1002/sim.7901}.
\newblock URL \url{https://onlinelibrary.wiley.com/doi/abs/10.1002/sim.7901}.

\bibitem[Schafer et~al.(1999)Schafer, Konstan, and
  Riedl]{schafer1999recommender}
J~Ben Schafer, Joseph Konstan, and John Riedl.
\newblock Recommender systems in e-commerce.
\newblock In \emph{Proceedings of the 1st ACM conference on Electronic
  commerce}, pages 158--166, 1999.

\bibitem[Shafer and Vovk(2008)]{shafer2008tutorial}
Glenn Shafer and Vladimir Vovk.
\newblock A tutorial on conformal prediction.
\newblock \emph{Journal of Machine Learning Research}, 9\penalty0 (3), 2008.

\bibitem[Simon(1979)]{simon1979restricted}
Richard Simon.
\newblock Restricted randomization designs in clinical trials.
\newblock \emph{Biometrics}, pages 503--512, 1979.

\bibitem[Su{\'a}rez-Hern{\'a}ndez et~al.(2019)Su{\'a}rez-Hern{\'a}ndez, Torras,
  and Alenya]{suarez2019practical}
Alejandro Su{\'a}rez-Hern{\'a}ndez, Carme Torras, and Guillem Alenya.
\newblock Practical resolution methods for mdps in robotics exemplified with
  disassembly planning.
\newblock \emph{IEEE Robotics and Automation Letters}, 4\penalty0 (3):\penalty0
  2282--2288, 2019.

\bibitem[Tewari and Murphy(2017)]{tewari2017ads}
Ambuj Tewari and Susan~A Murphy.
\newblock From ads to interventions: Contextual bandits in mobile health.
\newblock In \emph{Mobile Health}, pages 495--517. Springer, 2017.

\bibitem[Tibshirani et~al.(2020)Tibshirani, Barber, Candes, and
  Ramdas]{tibshirani2020conformal}
Ryan~J. Tibshirani, Rina~Foygel Barber, Emmanuel~J. Candes, and Aaditya Ramdas.
\newblock Conformal prediction under covariate shift, 2020.

\bibitem[Vovk(2013)]{vovk2013transductive}
Vladimir Vovk.
\newblock Transductive conformal predictors.
\newblock In \emph{IFIP International Conference on Artificial Intelligence
  Applications and Innovations}, pages 348--360. Springer, 2013.

\bibitem[Vovk et~al.(2005)Vovk, Gammerman, and Shafer]{alrw}
Vladimir Vovk, Alex Gammerman, and Glenn Shafer.
\newblock \emph{Algorithmic Learning in a Random World}.
\newblock Springer-Verlag, Berlin, Heidelberg, 2005.
\newblock ISBN 0387001522.

\bibitem[Vovk et~al.(2009)Vovk, Nouretdinov, and Gammerman]{vovk}
Vladimir Vovk, Ilia Nouretdinov, and Alex Gammerman.
\newblock {On-line predictive linear regression}.
\newblock \emph{The Annals of Statistics}, 37\penalty0 (3):\penalty0 1566 --
  1590, 2009.
\newblock \doi{10.1214/08-AOS622}.
\newblock URL \url{https://doi.org/10.1214/08-AOS622}.

\bibitem[Watkins(1989)]{watkins1989learning}
Christopher John Cornish~Hellaby Watkins.
\newblock Learning from delayed rewards.
\newblock 1989.

\bibitem[Xu and Xie(2021)]{xu2021conformal}
Chen Xu and Yao Xie.
\newblock Conformal prediction interval for dynamic time-series.
\newblock In \emph{International Conference on Machine Learning}, pages
  11559--11569. PMLR, 2021.

\bibitem[Zhang et~al.(2021)Zhang, Janson, and Murphy]{zhang2021}
Kelly Zhang, Lucas Janson, and Susan Murphy.
\newblock Statistical inference with m-estimators on adaptively collected data.
\newblock December 2021.

\bibitem[Zhang et~al.(2022)Zhang, Janson, and Murphy]{KZ-ea:2022}
Kelly Zhang, Lucas Janson, and Susan Murphy.
\newblock Statistical inference after adaptive sampling in non-markovian
  environments.
\newblock \emph{arXiv preprint arXiv:2202.07098}, 2022.

\bibitem[Zhang et~al.(2020)Zhang, Janson, and Murphy]{zhang2020inference}
Kelly~W. Zhang, Lucas Janson, and Susan Murphy.
\newblock Inference for batched bandits.
\newblock December 2020.

\bibitem[Zhang and Zhao(2022)]{what-is-randomization-test}
Yao Zhang and Qingyuan Zhao.
\newblock What is a randomization test?, 2022.
\newblock URL \url{https://arxiv.org/abs/2203.10980}.

\end{thebibliography}

\appendix
    \section{Proof of Theorem~\ref{weighted-mc-rand-valid}}
\label{AppendixA}
We consider the case of discrete data $\data$ and discrete resamples $\sampledData^{(1)}, \ldots, \sampledData^{(\numMCSamples)}$ and apply Bayes' theorem. Define $\mathfrak{d} = (d_0, \ldots, d_\numMCSamples)$ to be be some list of data sets, $\mathfrak{d}_{-0}$ to be the list $(d_1, \ldots, d_\numMCSamples)$, $\mathfrak{D}_{-0}$ to denote the list $(\sampledData^{(1)}, \ldots, \sampledData^{(m)})$, and $\orb(\mathfrak{d}) := \{\mathfrak{d}^\pi: \pi \in \Pi_{\zeroM}\}$, where $\mathfrak{d}^\pi := (d_{\pi(0)}, \ldots, d_{\pi(\numMCSamples)})$; we also use $(\mathfrak{d}_{-0})^\pi$ to denote the permuted list $(d_{\pi(1)}, \ldots, d_{\pi(\numMCSamples)})$ for any $\pi \in \Pi_{\zeroM}$. Then, by Bayes' theorem, we have that
\begin{align*}\mathbb{P}\left(\data = d_i|\sampledSet \in \orb(\mathfrak{d})\right) &= \frac{\mathbb{P}\left(\sampledSet \in \orb(\mathfrak{d})|\data=d_i\right)\density(d_i)}{\sum_{\text{distinct }d_j \in \mathfrak{d}}\mathbb{P}\left(\sampledSet \in \orb(\mathfrak{d})|\data=d_j\right)\density(d_j)}\\
&= \frac{\mathbb{P}\left(\sampledSet_{-0} \in \{(\mathfrak{d}_{-0})^\pi: \pi \in \Pi_{\zeroM} \text{ such that }d_{\pi(0)}= d_{i}\}|\data=d_i\right)\density(d_i)}{\sum_{\text{distinct }d_j \in \mathfrak{d}}\mathbb{P}\left(\sampledSet_{-0} \in \{(\mathfrak{d}_{-0})^\pi: \pi \in \Pi_{\zeroM} \text{ such that }d_{\pi(0)}= d_{j}\}|\data=d_j\right)\density(d_j)}\\
&= \frac{|\{(\mathfrak{d}_{-0})^\pi: \pi \in \Pi_{\zeroM} \text{ such that }d_{\pi(0)}= d_{i}\}|\left(\prod_{k \in \zeroM\backslash\{i\}} \proposalNoTilde(d_k|d_i)\right)\density(d_i)}{\sum_{\text{distinct }d_j \in \mathfrak{d}}|\{(\mathfrak{d}_{-0})^\pi: \pi \in \Pi_{\zeroM} \text{ such that }d_{\pi(0)}= d_{j}\}|\left(\prod_{k \in \zeroM\backslash\{j\}} \proposalNoTilde(d_k|d_j)\right)\density(d_j)}.
\end{align*} where we say that $d \in \mathfrak{d}$ if $d$ is an element of the list $\mathfrak{d}$ and where the last equality follows from the conditional i.i.d.-ness of the resamples given $\data$.

Letting $m_{d_i}(\mathfrak{d})$ denote the number of times $d_i$ appears in the list $\mathfrak{d}$ we have that \begin{align*}|\{(\mathfrak{d}_{-0})^\pi: \pi \in \Pi_{\zeroM} \text{ with }d_{\pi(0)}= d_{i}\}| &= \frac{m!}{(m_{d_i}(\mathfrak{d})-1)!{\displaystyle \prod_{\text{distinct } d \in \mathfrak{d} \text{ not equal to }d_i} m_d(\mathfrak{d})!}}\\
&= m_{d_i}(\mathfrak{d}) \cdot \frac{m!}{{\displaystyle \prod_{\text{distinct } d \in \mathfrak{d}} m_d(\mathfrak{d})!}}\end{align*}

and hence, \begin{align*}\mathbb{P}\left(\data = d_i|\sampledSet \in \orb(\mathfrak{d})\right) &= \frac{m_{d_i}(\mathfrak{d})f(d_i)\prod_{k \in \zeroM\backslash\{i\}} \proposalNoTilde(d_k|d_i)}{\sum_{j=0}^mf(d_j)\prod_{k \in \zeroM\backslash\{j\}} \proposalNoTilde(d_k|d_j)}\\
&= \frac{m_{d_i}(\mathfrak{d})\hat{f}(d_i)\prod_{k \in \zeroM\backslash\{i\}} \proposalNoTilde(d_k|d_i)}{\sum_{j=0}^m\hat{f}(d_j)\prod_{k \in \zeroM\backslash\{j\}} \proposalNoTilde(d_k|d_j)}\\
&= m_{d_i}(\mathfrak{d})\weight_{\mathfrak{d}}(d_i),\end{align*} 
where the second-to-last inequality is because of the proportionality hypothesis $\mathcal{H}^\propto_0$. 
Equivalently, we can think of $\data$'s conditional distribution given $\orb(\sampledSet)$ as a draw from the $m+1$ elements of $\sampledSet$, with weight on each element given by the $\weight_{\sampledSet}$ function applied to that element.
Defining the set $\mathfrak{S} := \{S(\sampledData)\rt \sampledData \in \sampledSet\}$\footnote{Just as above, we say that $\sampledData \in \sampledSet$ if $\sampledData$ is an element of the list $\sampledSet$.}, $S(\data)$ can be viewed as a draw from the weighted distribution on $\mathfrak{S}$ with the total weight of each element $S(\sampledData^{(i)})$ equal to \[\sum_{j: S(\sampledData^{(j)}) = S(\sampledData^{(i)})}\frac{\hat{\density}(\sampledData^{(j)})\prod_{k \in \zeroM\backslash\{j\}} \proposalNoTilde(\sampledData^{(k)}|\sampledData^{(j)})}{\sum_{j'=0}^{\numMCSamples}\hat{\density}(\sampledData^{(j')})\prod_{k' \in \zeroM\backslash\{j'\}} \proposalNoTilde(\sampledData^{(k')}|\sampledData^{(j')})} = \sum_{j: S(\sampledData^{(j)}) = S(\sampledData^{(i)})}\weight_{\sampledSet}(\sampledData^{(j)}).\]

Finally, note that \begin{align*}\mathbb{P}(p \leq \alpha ) &= \mathbb{P}\left(\sum_{i=0}^\numMCSamples\weight_{\sampledSet}(\sampledData^{(i)})\mathbf{1}[S(\sampledData^{(i)}) \geq S(\data)] \leq \alpha\right)\\ &= \mathbb{P}\left(S(\data) > \inf\{s \in \mathfrak{S}: \sum_{\sampledData \in \sampledSet: S(\sampledData) \leq s}\weight_{\sampledSet}(\sampledData) \geq 1-\alpha\}\right).\end{align*} Then, we have that \begin{equation}\label{validity-final}\mathbb{P}\left(S(\data) > \inf\{s \in \mathfrak{S}: \sum_{\sampledData \in \sampledSet: S(\sampledData) \leq s}\weight_{\sampledSet}(\sampledData) \geq 1-\alpha\}\;\middle|\;\mathfrak{S}\right) \leq \alpha,\end{equation} since 1.~the infimum in equation \eqref{validity-final} is the $(1-\alpha)^{\textnormal{th}}$ conditional quantile of the weighted distribution over $\mathfrak{S}$ and 2.~as noted above, $S(\data)$ is conditionally a draw from this weighted distribution. Inequality \eqref{validity-final} then also holds unconditionally, as desired.

\subsection{Connections to \cite{harrison}}\label{harrison-connection}
\subsubsection{Equivalence of the tests}
\cite{harrison} consider a problem in which the analyst has obtained a sample $\data \sim P$ from some (null) target distribution $P$ and draws samples $\sampledData^{(1)}, \ldots, \sampledData^{(\numMCSamples)} \overset{i.i.d.}{\sim} Q$ from some known proposal distribution $Q$, independently from $\data$. Then, assuming that the importance weights $v(d) := \frac{dP}{dQ}(d)$ are known, it is shown that the p-value $$\tilde{p}_*(X,\mathfrak{D}) = \sum_{i=0}^\numMCSamples\frac{ v(\sampledData^{(i)})}{\sum_{j=0}^\numMCSamples v(\sampledData^{(i)})}\mathbf{1}[S(\sampledData^{(i)}) \geq S(\sampledData^{(0)})],$$ where, as usual, $\sampledData^{(0)} := \data$ and $S$ is any test statistic is valid \cite[Theorem 2]{harrison}. It is further shown in their work that a conditional version of this test, which first conditions on a statistic $A(\data)$ of the data, replaces $P$ with its corresponding conditional distribution given $A(\data)$, and allows $Q$ to be any distribution depending on $A(\data)$ (but independent from $\data$ given $A(\data)$), is still valid (in fact, conditionally valid given $A(\data)$). 

This is precisely the form of the special case of our test presented in Remark~\ref{comp-iid} with $A(\data)$ being a statistic of the observed data for which $A(\sampledData^{(i)}) = A(\data)$ and $q(\cdot|\sampledData^{(i)}, A(\sampledData^{(i)})) = q(\cdot|A(\data)),$ for all $i \in [\numMCSamples]$. Conditional on such a choice of $A(\data)$, $\data$ follows the distribution $P_{A(\data)} := f(\cdot|A(\data))$, and $\sampledData^{(1)}, \ldots, \sampledData^{(\numMCSamples)}$ are i.i.d.~draws from $Q_{A(\data)} := q(\cdot|A(\data))$ which are (jointly) independent from $\data$. Thus, as discussed in Remark~\ref{comp-iid}, using the fact that the conditional probability $q(\sampledData^{(i)}|\sampledData^{(i)})$ may now be instead written as $q(\sampledData^{(i)}|A(\data))$, the weight $\weight_{\sampledSet}(\sampledData^{(i)})$ involved in our p-value calculation is equal (under $\mathcal{H}^\propto_0$) to $$\frac{f(\sampledData^{(i)})/q(\sampledData^{(i)}|A(\data))}{\sum_{j=0}^{\numMCSamples}f(\sampledData^{(j)})/q(\sampledData^{(i)}|A(\data))} = \frac{f(\sampledData^{(i)}|A(\data))/q(\sampledData^{(i)}|A(\data))}{\sum_{j=0}^{\numMCSamples}f(\sampledData^{(j)}|A(\data))/q(\sampledData^{(i)}|A(\data))} = \frac{v_{A(\data)}(\sampledData^{(i)})}{\sum_{j=0}^\numMCSamples v_{A(\data)}(\sampledData^{(j)})},$$ where $v_{A(\data)}(d)$ is the conditional importance weight $\frac{dP_{A(\data)}}{dQ_{A(\data)}}(d)$. Thus our p-value $p$ is precisely equal to the p-value $\tilde{p}_*(X,\mathfrak{D})$ resulting from \cite{harrison}'s conditional importance sampling procedure.  

\subsubsection{Practical considerations}
Using \cite{harrison}'s conditional importance sampling procedure, we may characterize the resampling (i.e., proposal) distributions $\proposalNoTilde$ discussed in Section~\ref{proposal-chap} by simply specifying a statistic $A(\data)$ that is invariant under all resamples and for which $\proposalNoTilde(\cdot|\sampledData^{(i)}, A(\sampledData^{(i)}))$ depends only on $A(\sampledData^{(i)}) = A(\data)$. In particular, for non-stationarity testing, each of $\text{uniform}_{\pi}$, $\text{imitation}_{\pi}$, $\text{re-imitation}_{\pi}$, and $\text{cond-imitation}_{\pi}$ depend only on the observed $\data$ through the unordered tuples $(C_i, X_i, Y_i)$ (which are the same as in the original data in all resamples), and so our method can be viewed as applying \cite{harrison}'s procedure with $A(\data)$ equal to the \emph{multiset} $\{(C_1, X_1, Y_1),\allowdisplaybreaks \ldots,\allowdisplaybreaks (C_\horizon, X_\horizon, Y_\horizon) \}$. For conditional independence testing, $\text{imitation}_{X}$ depends only on the $g$-evaluations of the actions (which too are unchanged during resampling) and so in this case $A(\data) = ((C_1, g(X_1), Y_1), \ldots, (C_\horizon, g(X_\horizon), Y_\horizon))$, whereas $\text{uniform}_{\pi}\text{+}\text{imitation}_X$, $\text{restricted-uniform}_{\pi}\text{+}\text{imitation}_X,$ and $\text{combined}_{\pi,X}$ are also further agnostic to the order of the data (where we replace the actions with their $g$-evaluations) and do not change the multiset of unordered datapoints, so that for these three resampling distributions we have that $A(\data) = \{(C_1, g(X_1), Y_1), \ldots, (C_\horizon, g(X_\horizon), Y_\horizon)\}$, where we once again use the brackets $\{\cdot\}$ to denote multiset.

\subsubsection{Comparison of proof techniques}
Finally, we comment on differences in proof techniques between the special case of our Theorem~\ref{weighted-mc-rand-valid} discussed in Remark~\ref{comp-iid} and Theorem 2 of \cite{harrison}. 

The main difference is that our proof is via the proof of Theorem~\ref{weighted-mc-rand-valid}, which is more general than Theorem 2 of \cite{harrison}. In particular, the key to our proof given above is to \emph{condition} on the multiset $\mathfrak{S} := \{S(\tilde{\data}): \tilde{\data} \in \mathfrak{D}\}$ (or equivalently, on the unordered elements of $\mathfrak{D}$ prior to their $S$-evaluations), and to show, via Bayes' theorem that, conditional on $\mathfrak{S}$, the event that $p \leq \alpha$ is equal to the event that $S(\data)$ lies above its $(1-\alpha)$ conditional quantile, which occurs with probability no greater than $\alpha$.

On the other hand, the statement of Theorem 2 given in \cite{harrison} is essentially weaker than that of Theorem~\ref{weighted-mc-rand-valid} and so, rather than by conditioning, their proof is via a change of measure argument. This proof technique, however, allows their Theorem 2 to be slightly more general than our statement in Remark~\ref{comp-iid} in that they need not restrict attention to discrete distributions to avoid measure-theoretic pathologies involving (regular) conditional probability distributions, but rather need only assume that $P \ll Q$. 

Restricting to the case in which the proposal distribution $Q$ is independent from $P$ (from which the conditional importance sampling result that is equivalent to our Theorem~\ref{weighted-mc-rand-valid} is a corollary), \cite{harrison} show that \begin{equation}\label{rn-deriv}\frac{dP_{\mathfrak{D}}}{dP_{\mathfrak{U}}}(\mathfrak{d}) = \frac{(m+1)\frac{dP}{dQ}(d_0)}{\sum_{j=0}^m\frac{dP}{dQ}(d_j)}\end{equation} where $\mathfrak{U}$ is a uniform permutation of $\mathfrak{D}$, $P_{\mathfrak{D}}$ and $P_{\mathfrak{U}}$ denote the laws of $\mathfrak{D}$ and $\mathfrak{U}$, respectively, and $\mathfrak{d} := (d_0, \ldots, d_m)$. Using equation~\eqref{rn-deriv} as well as the fact that $\mathfrak{U}$'s law is invariant under permutation, probabilities of events under $\mathfrak{D}$'s distribution (and in particular, those involving the p-value $\tilde{p}_*(X,\mathfrak{D})$, which is a deterministic function of $\mathfrak{D}$) may be transformed into events involving (permutations of) $\mathfrak{U}$ which, combined with a deterministic inequality that may be viewed as an unconditional version of inequality~\eqref{validity-final} (the argument for which is the same as it is simply a statement about weighted quantiles) then shows that the p-value is valid. We refer the reader to \cite{harrison} for details.

\section{Conditional independence testing in a stationary MDP}\label{appendixC-1}

In this section, we give a proof of Proposition~\ref{cond-indep-prop-mdp} regarding the restricted permutation set in Environment~\ref{stat-mdp-env}:

\begin{proof}
Let $\pi_i$ be any permutation in \[\Pi_4 := \{\pi \in \Pi_{[\horizon]}: Y_{\pi(t)}=C_{\pi(t+1)}, \forall t \in [\horizon-1], \textnormal{ and } C_{\pi(1)} = C_1\}.\] Then
\begin{align*}
    \hat{f}(\sampledData^{(i)}) &= \prod_{t=1}^\horizon \decisionProb(\tilde{X}^{(i)}_t|\tilde{C}^{(i)}_{t}, \tilde{H}^{(i)}_{t-1})\\
    &= \frac{\prod_{t=1}^\horizon \mathbb{P}(Y_{\pi_i(t)}|C_{\pi_i(t)}, X_{\pi_i(t)})\decisionProb(\tilde{X}^{(i)}_t|\tilde{C}^{(i)}_{t}, \tilde{H}^{(i)}_{t-1})}{\prod_{t=1}^\horizon \mathbb{P}(Y_{t}|C_{t}, X_{t})} && \text{because of equation \eqref{stat-assum}}\\
    &= \frac{\mathbb{P}(C_{\pi_i(1)})\prod_{t=1}^\horizon \mathbb{P}(C_{\pi_i(t+1)}|C_{\pi_i(t)}, X_{\pi_i(t)})\decisionProb(\tilde{X}^{(i)}_t|\tilde{C}^{(i)}_{t}, \tilde{H}^{(i)}_{t-1})}{\mathbb{P}(C_1)\prod_{t=1}^\horizon \mathbb{P}(C_{t+1}|C_{t}, X_{t})} && \text{since $\pi_i \in \Pi_4$}\\
    &= \frac{\mathbb{P}(C_{\pi_i(1)})\prod_{t=1}^\horizon \mathbb{P}(C_{\pi_i(t+1)}|C_{\pi_i(t)}, \tilde{X}^{(i)}_t)\decisionProb(\tilde{X}^{(i)}_t|\tilde{C}^{(i)}_{t}, \tilde{H}^{(i)}_{t-1})}{\mathbb{P}(C_1)\prod_{t=1}^\horizon \mathbb{P}(C_{t+1}|C_{t}, X_{t})} && \text{by $\mathcal{H}^{\CondIndep}_0$ and that $g(\tilde{X}^{(i)}_t) = g(X_{\pi_i(t)})$}\\
    &= \frac{\mathbb{P}(\tilde{C}^{(i)}_1)\prod_{t=1}^\horizon \mathbb{P}(\tilde{C}^{(i)}_t|\tilde{C}^{(i)}_t, \tilde{X}^{(i)}_t)\decisionProb(\tilde{X}^{(i)}_t|\tilde{C}^{(i)}_{t}, \tilde{H}^{(i)}_{t-1})}{\mathbb{P}(C_1)\prod_{t=1}^\horizon \mathbb{P}(C_{t+1}|C_{t}, X_{t})} && \text{by equation~\eqref{fixed}}\\
    &= \frac{1}{\mathbb{P}(C_1)\prod_{t=1}^\horizon \mathbb{P}(C_{t+1}|C_{t}, X_{t})} \cdot f(\sampledData^{(i)}), && \text{due to equation \eqref{markov-assum}}
\end{align*} and so the proportionality hypothesis is satisfied with $K = \frac{1}{\mathbb{P}(C_1)\prod_{t=1}^\horizon \mathbb{P}(C_{t+1}|C_{t}, X_{t})}$.
\end{proof}

\section{Inference in a general adaptive data collection process}\label{appendixC-2}

In this section, we relax all structural assumptions made in Section~\ref{app-chap} and assume that the data are gathered according to any adaptive data collection process (i.e., we make no assumptions on the joint distribution of $(C_t,X_t,Y_t)_{t=1}^T$ other than that the sequence $(X_t)_{t=1}^T$ is non-anticipating with respect to the filtration $\mathcal{F}_t := \sigma\left(\mathcal{H}_{t-1}\cup\{C_t\}\right)$). The hypothesis we are interested in testing is related to that discussed in Section~\ref{distributional}, except here it involves \emph{sequences} of actions. In fact, we consider $\horizon$ functions $g_1, \ldots, g_\horizon$, where $g_t$ takes as input the sequence of first $t$ actions $X_{1:t} := (X_1, \ldots, X_t)$; the null hypothesis we are interested in testing is that of simultaneous independence, conditional on these $g$-evaluations: \begin{equation*}\mathcal{H}^{\indep, g_1, \ldots, g_\horizon}_0: \left[C_t \indep X_{1:t-1}\mid (C_{1:t-1},g_{t-1}(X_{1:t-1}), Y_{1:t-1})\right] \\ \text{ and } \left[Y_t \indep X_{1:t}\mid (C_{1:t},g_t(X_{1:t}), Y_{1:t-1})\right], \forall t \in [\horizon],\end{equation*} where we define \[C_{1:0} = X_{1:0} = g_0(X_{1:0}) = Y_{1:0} = \varnothing.\] Informally, $\mathcal{H}^{\indep, g_1, \ldots, g_\horizon}_0$ states that the actions $X_t$ only influence future contexts and responses via their values filtered through the functions $g_t$. As noted in Remark~\ref{no-markov}, the setting in which all $g_t$ are constant has been covered in prior work by \cite{bojinov2019time}; however, to the best of our knowledge, the case of non-constant $g_t$ is novel. We now describe and prove the validity of the testing procedure for the null hypothesis $\mathcal{H}^{\indep, g_1, \ldots, g_\horizon}_0$.

\begin{proposition}
Suppose that the resampling distribution $\proposalNoTilde$ randomizes only the action sequence $X_{1:\horizon}$ \emph{conditional} on $(g_t(X_{1:t}))_{t=1}^{\horizon}$. That is, in each resample $\sampledData^{(i)}$, we have that \begin{equation}\label{fixed-appendix}\left(\tilde{C}^{(i)}_t,g_t(\tilde{X}^{(i)}_{1:t}),\tilde{Y}^{(i)}_t\right) = \left(C_{t},g_t(X_{1:t}),Y_{t}\right), \;\; \forall t \in [\horizon].\end{equation} Then the null hypothesis $\mathcal{H}^{\indep, g_1, \ldots, g_\horizon}_0$ implies the proportionality hypothesis $\mathcal{H}^{\propto}_0$.
\end{proposition}
\begin{proof}
We have that \begin{align*}
    &\hat{f}(\sampledData^{(i)}) \\
    &= \prod_{t=1}^\horizon \decisionProb(\tilde{X}^{(i)}_t|\tilde{C}^{(i)}_{t}, \tilde{H}^{(i)}_{t-1})\\
    &= \frac{\prod_{t=1}^\horizon \mathbb{P}(\tilde{Y}^{(i)}_t|\tilde{C}^{(i)}_{1:t}, g_t(\tilde{X}_{1:t}^{(i)}), \tilde{Y}^{(i)}_{1:t-1})\decisionProb(\tilde{X}^{(i)}_t|\tilde{C}^{(i)}_{t}, \tilde{H}^{(i)}_{t-1})\mathbb{P}(\tilde{C}^{(i)}_t|\tilde{C}^{(i)}_{1:t-1}, g_{t-1}(\tilde{X}_{1:t-1}^{(i)}), \tilde{Y}^{(i)}_{1:t-1})}{\prod_{t=1}^T \mathbb{P}(Y_t|C_{1:t}, g_t(X_{1:t}), Y_{1:t-1})\mathbb{P}(C_t|C_{1:t-1}, g_{t-1}(X_{1:t-1}), Y_{1:t-1})} && \text{equation \eqref{fixed-appendix}}\\
    &= \frac{\prod_{t=1}^\horizon \mathbb{P}(\tilde{Y}^{(i)}_t|\tilde{C}^{(i)}_{1:t}, g_t(\tilde{X}_{1:t}^{(i)}), \tilde{Y}^{(i)}_{1:t-1}, \tilde{X}_{1:t}^{(i)})\decisionProb(\tilde{X}^{(i)}_t|\tilde{C}^{(i)}_{t}, \tilde{H}^{(i)}_{t-1})\mathbb{P}(\tilde{C}^{(i)}_t|\tilde{C}^{(i)}_{1:t-1}, g_{t-1}(\tilde{X}_{1:t-1}^{(i)}), \tilde{Y}^{(i)}_{1:t-1}, \tilde{X}_{1:t-1}^{(i)})}{{\prod_{t=1}^T \mathbb{P}(Y_t|C_{1:t}, g_t(X_{1:t}), Y_{1:t-1})\mathbb{P}(C_t|C_{1:t-1}, g_{t-1}(X_{1:t-1}), Y_{1:t-1})}} && \text{due to $\mathcal{H}^{\indep, g_1, \ldots, g_\horizon}_0$}\\
    &= \frac{\prod_{t=1}^\horizon \mathbb{P}(\tilde{Y}^{(i)}_t|\tilde{C}^{(i)}_{1:t}, \tilde{X}_{1:t}^{(i)}, \tilde{Y}^{(i)}_{1:t-1})\decisionProb(\tilde{X}^{(i)}_t|\tilde{C}^{(i)}_{t}, \tilde{H}^{(i)}_{t-1})\mathbb{P}(\tilde{C}^{(i)}_t|\tilde{C}^{(i)}_{1:t-1}, \tilde{X}_{1:t-1}^{(i)}, Y^{(i)}_{1:t-1})}{{\prod_{t=1}^T \mathbb{P}(Y_t|C_{1:t}, g_t(X_{1:t}), Y_{1:t-1})\mathbb{P}(C_t|C_{1:t-1}, g_{t-1}(X_{1:t-1}), Y_{1:t-1})}}\\
    &= \frac{1}{{\prod_{t=1}^T \mathbb{P}(Y_t|C_{1:t}, g_t(X_{1:t}), Y_{1:t-1})\mathbb{P}(C_t|C_{1:t-1}, g_{t-1}(X_{1:t-1}), Y_{1:t-1})}} \cdot f(\sampledData^{(i)}),
\end{align*} which satisfies the proportionality hypothesis with $K = \frac{1}{{\prod_{t=1}^T \mathbb{P}(Y_t|C_{1:t}, g_t(X_{1:t}), Y_{1:t-1})\mathbb{P}(C_t|C_{1:t-1}, g_{t-1}(X_{1:t-1}), Y_{1:t-1})}}$.
\end{proof}

One concrete setting in which the above randomization procedure could be used to test against the hypothesis $\mathcal{H}^{\indep, g_1, \ldots, g_\horizon}_0$ is in a completely general mobile health setting in which no environmental assumptions are made and there are only two treatments: a control and an experimental treatment. A hypothesis which may be of interest to test is that, conditional on the prior sequence of contexts and responses, each of the next response and context is independent of the action sequence given the \emph{number} of experimental treatments taken up until that time. In such a setting, one could use our framework under the above randomization scheme to test the null hypothesis $\mathcal{H}^{\indep, g_1, \ldots, g_\horizon}_0$ where $g_t(X_{1:t})$ is an indicator that the experimental treatment appears at least half of the time in the treatment sequence $X_{1:t}$; i.e., $g_t(X_{1:t}) = \mathbf{1}[\sum_{s=1}^tX_s \geq t/2]$ where $X = 1$ denotes the experimental treatment and $X=0$ denotes the control.

\section{Inference in scale families}\label{simultaneous-inference-appendix}
In this section, we describe how the test discussed in Section~\ref{confidence-inversion} can also be used for inference in semiparametric scale families.

Instead of assuming that each reward distribution is a member of the same location family, it may be more natural in some situations to assume a semiparametric scale family. As such, letting $\theta_x$ now denote action $x$'s scale parameter (i.e., we assume that $Y_t\mid X_t=x\sim h_{0}(y/\theta_x)$\footnote{Recall that we only consider discrete random variables, per Remark~\ref{discrete-remark}, and hence no Jacobian is required.}), we may instead test against the null hypothesis $\mathcal{H}^{\text{Scale},\delta,x,x'}$ that $\frac{\theta_{x'}}{\theta_{x}} = \delta$. To do so, we simply revise the reward-modification trick discussed in Section~\ref{confidence-inversion} for the location family to instead replace $Y_t$ with $Y_t \cdot \delta \cdot \mathbf{1}[X_t=x]$ and again perform the usual test for conditional independence from Section~\ref{distributional} using $g(X_t) = \begin{cases}
\{x,x'\} \text{ if } X_t \in \{x, x'\}\\
X_t \text{ otherwise}
\end{cases}$.

Similar to as discussed in Section~\ref{confidence-inversion}, the above test can be inverted to construct (simultaneous) confidence intervals.
    \section{Non-conditionally-i.i.d.~resampling and sharing samples}
\label{AppendixD}

In this section, we discuss how to handle non-conditionally-i.i.d.~resampling procedures and how this more general procedure can be used to share samples between different values $y \in \mathcal{Y}$ in the construction of conformal prediction regions.

\subsection{Non-conditionally-i.i.d.~resampling}\label{non-iid-appendix}
In this section, we describe a more general version of the weighted MC randomization test of Algorithm~\ref{weighted-mc-rand} that can be used in the setting of non-conditionally-i.i.d.~resampling, as discussed in Remark~\ref{non-iid}. In particular, the Remark states that, letting $\proposal$ denote the joint conditional distribution of the resamples $(\sampledData^{(1)}, \ldots, \sampledData^{(m)})$, one may redefine the weights to be \begin{equation}\label{secondweight}\weightNoTilde_{\sampledSet}(\sampledData^{(i)}) = \frac{\hat{f}(\sampledData^{(i)})\sum_{\pi \in \Sigma: \pi(0) = i}\proposal((\mathfrak{D}_{-0})^\pi|\sampledData^{(i)})}{\sum_{j \in \zeroM}\hat{f}(\sampledData^{(j)})\sum_{\pi' \in \Sigma:\pi'(0)=j}\proposal((\mathfrak{D}_{-0})^{\pi'}|\sampledData^{(j)})},\end{equation} and the p-value \[p := \sum_{i=0}^\numMCSamples\weightNoTilde_\sampledSet(\sampledData^{(i)})\mathbf{1}[S(\sampledData^{(i)}) \geq S(\data)]\] will be valid.

Recall that the key idea behind Algorithm~\ref{weighted-mc-rand}---and the proof of Theorem~\ref{weighted-mc-rand-valid}---was to condition on the event $(\sampledData^{(0)}, \ldots, \sampledData^{(\numMCSamples)}) \in\{(d_{\pi(0)}, \ldots, d_{\pi(\numMCSamples)}): \pi \in \Pi_{\zeroM}\}$ for some list $\mathfrak{d} = (d_0, \ldots, d_{\numMCSamples})$ and to apply Bayes' theorem as well as the proportionality hypothesis to derive the weight formula (equation \eqref{weight-eqn}). The entire permutation set $\Pi_{\zeroM}$ need not, however, be considered. Our generalization involves using any arbitrary subset of permutations $\Sigma$; when $\Sigma$ is small, our method is computationally tractable for non-conditionally-i.i.d.~resampling schemes.

More specifically, define \[\pseudoorb_{\Sigma}(\mathfrak{d}) := \{(d_{\pi(0)}, \ldots, d_{\pi(\numMCSamples)}): \pi \in \Sigma\}\] to be the \emph{$\Sigma$-pseudo-orbit}\footnote{We use the prefix ``pseudo'' since $\Sigma$ need not be a group.} of the list $\mathfrak{d}$. Then, by conditioning on $\pseudoorb_{\Sigma}(\sampledSet)$, we will be able to show that the weighting given by equation~\eqref{secondweight} will be valid. We first prove a result similar to Theorem~\ref{weighted-mc-rand-valid} in the case of distinct data $\data$ and resamples $\sampledData^{(1)}, \ldots, \sampledData^{(m)}$ using exactly the same proof strategy:

\begin{lemma}\label{distinct-lemma}
    Let $\Sigma$ be any subset of the full set of permutations on $\zeroM$, $\Pi_{\zeroM}$. Define $\Sigma_i$ to be $\{\pi \in \Sigma: \pi(0) = i\}$ and assume that $\proposal$ is such that $\sampledData^{(0)}, \ldots, \sampledData^{(m)}$ are all distinct. Then, with weights 
    \[w^{\proposal, \Sigma}_{\sampledSet,i}(\sampledData^{(i)}) := \frac{\hat{f}(\sampledData^{(i)})\sum_{\pi \in \Sigma_i}\proposal((\mathfrak{D}_{-0})^\pi|\sampledData^{(i)})}{\sum_{j=0}^\numMCSamples\hat{f}(\sampledData^{(j)})\sum_{\pi' \in \Sigma_j}\proposal((\mathfrak{D}_{-0})^{\pi'}|\sampledData^{(j)})},\]
    we have that \[p := \sum_{i=0}^\numMCSamples w^{\proposal, \Sigma}_{\sampledSet,i}(\sampledData^{(i)})\mathbf{1}[S(\sampledData^{(i)}) \geq S(\data)]\] stochasically dominates the uniform distribution under $\mathcal{H}^{\propto}_0$.
\end{lemma}
\begin{proof}[Proof sketch]
We give a sketch of the proof as much of it is the same as that of Theorem~\ref{weighted-mc-rand-valid}. Using Bayes' theorem, we have that \begin{align*}
    &\mathbb{P}(\data = d_i|(\sampledData^{(0)}, \ldots, \sampledData^{(\numMCSamples)}) \in \pseudoorb_\Sigma(\mathfrak{d})) \\
    &= \frac{\mathbb{P}((\sampledData^{(0)}, \ldots, \sampledData^{(\numMCSamples)}) \in \pseudoorb_\Sigma(\mathfrak{d})|\data = d_i)f(d_i)}{\sum_{j=0}^\numMCSamples \mathbb{P}((\sampledData^{(0)}, \ldots, \sampledData^{(\numMCSamples)}) \in \pseudoorb_\Sigma(\mathfrak{d})|\data = d_j)f(d_j)}\\
    &= \frac{f(d_i)\sum_{\pi \in \Sigma_i}\mathbb{P}((\mathfrak{D}_{-0}) = (d_{\pi(1)}, \ldots,  d_{\pi(\numMCSamples)})|\data = d_i)}{\sum_{j=0}^\numMCSamples f(d_j)\sum_{\pi' \in \Sigma_j}\mathbb{P}((\mathfrak{D}_{-0}) = (d_{\pi'(1)},\ldots, d_{\pi'(\numMCSamples)})|\data = d_j)}.
\end{align*}
Thus, conditional on $\pseudoorb_\Sigma(\sampledSet)$, $S(\data)$'s distribution is indeed the weighted distribution on the multiset $\{S(\sampledData): \sampledData \in \sampledSet\}$ with weight of the $i^{\textnormal{th}}$ element equal to $w^{\proposal, \Sigma}_{\sampledSet, i}(\sampledData^{(i)})$ under $\mathcal{H}^\propto_0$. The rest of the proof, in this setting of distinct resamples, follows in precisely the same manner as that of Theorem~\ref{weighted-mc-rand-valid}.
\end{proof}

We now extend to the case of repeated samples via a coupling argument. Define $\mathcal{D}' := \mathcal{D}\times \zeroM$ so that, for each point $d$ in $\mathcal{D}$ it has $m+1$ ``clones'' in $\mathcal{D'}$ given by $(d,0), \ldots, (d,m)$. We also define a density $\density'$ on $\mathcal{D}'$ given by $\density'((d,i)) := (\numMCSamples+1)^{-1}\density(d)$ for all $d \in \mathcal{D}, i \in \zeroM$. Lastly, we extend the test statistic $S$ to a test statistic $S'$ on $\mathcal{D}'$ which also takes each clone to the value of its original: $S'((d,i)) = S(d)$ for all $d \in \mathcal{D}, i \in \zeroM$. 

We will couple $p$ to a random variable $p'{}$ obtained by a procedure on the space $\mathcal{D}'$ which draws distinct elements $\tilde{D}'{}^{(i)}$, to be defined below. We call the original process (that may draw repeated elements and acts only on $\mathcal{D}$) $\mathcal{P}$ and the coupled process $\mathcal{P}'$. The observed dataset that the coupled process $\mathcal{P}'$ will see and use is a cloned version $(D,j)$ of the true observed data $\data$, where $j$ is sampled uniformly at random from the set $\zeroM$; importantly $\mathcal{P}'$ treats this cloned dataset $D' := (D,j)$ as though it were the observed dataset. Finally, we use $\proposal'$ to denote the conditional resampling distribution of $\mathcal{P}'$, induced by $\proposal$. That is, for a list $\sampledSet_{-0}' := (\tilde{D}'{}^{(1)}, \ldots, \tilde{D}'{}^{(\numMCSamples)})$ of resamples in $\mathcal{D}'$, $\proposal'(\sampledSet_{-0}'|\data')$ denotes the joint conditional probability of $\mathcal{P}'$ having obtained the list of resamples $\sampledSet_{-0}'$ given that it observed the dataset $\data'$. Algorithm~\ref{alg:cap} describes how $\mathcal{P}'$ is defined with respect to $\mathcal{P}$.

\begin{algorithm}[!ht]
\KwInput{$\mathcal{P}'$'s observed dataset $D' := (D,j)$ where $j \sim \text{Unif}(\zeroM)$}
$\tilde{D}'{}^{(0)} \gets D'$\\
$\sampledSet'{}^{(0)} \gets (\tilde{D}'^{(0)})$\\
\For{$i=1, \ldots, \numMCSamples$}{
$\tilde{D}^{(i)} \gets i^{\text{th}}$ element of $\sampledSet$ in the process $\mathcal{P}$\\
$j \gets \text{Unif}(\{j' \in \zeroM: (\tilde{D}^{(i)},j') \not\in \sampledSet'{}^{(i-1)}\})$\\
$\tilde{D}'{}^{(i)} \gets (\tilde{D}^{(i)},j)$\\
$\sampledSet'{}^{(i)} \gets \textnormal{concatenation of } \sampledSet'{}^{(i-1)} \textnormal{ with } \tilde{D}'{}^{(i)}$\\
}
$\sampledSet' \gets \sampledSet'{}^{(\numMCSamples)}$\\
\KwOutput{\[p'{} := \frac{\sum_{i=0}^\numMCSamples\mathbf{1}[S'(\tilde{D}'{}^{(i)}) \geq S'(\data')]f'(\tilde{D}'{}^{(i)})\sum_{\pi \in \Sigma_i}\proposal'((\mathfrak{D}'_{-0})^{\pi}|\tilde{D}'{}^{(i)})}{\sum_{j=0}^\numMCSamples f'(\tilde{D}'{}^{(j)})\sum_{\pi' \in \Sigma_j}\proposal'((\mathfrak{D}'_{-0})^{\pi'}|\tilde{D}'{}^{(i)})},\] }
\caption{Coupling of $\mathcal{P}'$ to $\mathcal{P}$}\label{alg:cap}
\end{algorithm}

At a high level, the coupled process $\mathcal{P}'$ draws distinct elements by sampling an element's clone number $j$ uniformly at random from all remaining yet-unselected values; using these distinct resamples, it then calculates a p-value in precisely the same way as Algorithm~\ref{weighted-mc-rand} with weights $\weightNoTilde_{\sampledSet'}(\sampledData'{}^{(i)})$.

Since the draws of $\mathcal{P}'$ are guaranteed to be distinct, Lemma~\ref{distinct-lemma} will guarantee $p'{}$'s stochastic domination of the uniform distribution as long as $f'$ is the true density of $D'$, the observed data for $\mathcal{P}'$ (since $\proposal'$ is the conditional resampling distribution and because $f'$ is trivially proportional to itself, thereby satisfying the proportionality hypothesis in this coupled model). For $d' \in \mathcal{D}'$, let $C(d')$ denote the second component of $d'$  (i.e., the \emph{clone number} of $d'$) and $d$ be the first component (i.e., the uncloned version of $d'$ in $\mathcal{D}$); then we see that $f'$ is $\data'$'s true density:
\begin{align*}
    \mathbb{P}(D' = d')
    &= \mathbb{P}(C(D') = C(d')|D = d)\mathbb{P}(D = d)\\
    &= \mathbb{P}(C(D') = C(d'))\mathbb{P}(D = d) && \text{since the clone number $j$ is sampled independently of $\data$}\\
    &= (m+1)^{-1}f(d) && \text{because $j \sim \text{Unif}(\zeroM)$ to clone $\data$}\\
    &= f'(d')
\end{align*} as desired.

We complete the coupling by showing that \begin{equation}\label{d-prop}\proposal'((\mathfrak{d}'_{-0})^\pi|d'_{i}) = \kappa\proposal((\mathfrak{d}_{-0})^\pi|d_{i}), \forall i \in \zeroM, \pi \in \Sigma_i \textnormal{ for some }\kappa >0 \textnormal{ depending on neither }i \textnormal{ nor }\pi,\end{equation}
where $\mathfrak{d}$ is some list $(d_0, \ldots, d_\numMCSamples)$ of elements in $\mathcal{D}$ and $\mathfrak{d}' := (d'_0, \ldots, d'_\numMCSamples)$ is a list of elements of $\mathcal{D}'$ comprising distinct clones of the $d_i$ (i.e., the first component of $d'_i$ is equal to $d_i$, and each element of $\mathfrak{d}'$ is distinct), $(\mathfrak{d}'_{-0})^{\pi} := (d'_{\pi(1)}, \ldots, d'_{\pi(m)})$, and $(\mathfrak{d}_{-0})^{\pi} := (d_{\pi(1)}, \ldots, d_{\pi(m)})$. To see why this is true, note that if $\mathcal{P}'$ observes $d'_i$ as the original dataset, then $\mathcal{P}$ must have seen its uncloned version: $d_i$. Thus, applying the cloning number function $C$ defined above to $(\mathfrak{d}'_{-0})^{\pi}$ elementwise, we have that 
\begin{align*}
    \proposal'((\mathfrak{d}'_{-0})^\pi|d'_{i}) &= \mathbb{P}(\sampledSet'_{-0} = (\mathfrak{d}'_{-0})^{\pi}|\data' = d'_i)\\
    &= \mathbb{P}(C(\sampledSet_{-0}') = C((\mathfrak{d}'_{-0})^{\pi}), \sampledSet_{-0} = (\mathfrak{d}_{-0})^{\pi}|C(\data') = C(d'_i), \data = d_i)\\
    &= \mathbb{P}(C(\sampledSet_{-0}') = C((\mathfrak{d}_{-0}')^{\pi})|\sampledSet_{-0} = (\mathfrak{d}_{-0})^{\pi}, \data' = d'_i)\mathbb{P}(\sampledSet_{-0} = (\mathfrak{d}_{-0})^{\pi}|\data = d_i) && \text{as $\sampledSet_{-0} \indep C(D') \mid D$ }.
\end{align*}
To see that this is proportional to $\proposal((\mathfrak{d}_{-0})^{\pi}|d_{i})$ (which is precisely the second term in the last line above), 
let $m_d((\mathfrak{d}_{-0})^{\pi})$ denote the number of elements in $(\mathfrak{d}_{-0})^{\pi}$ with first component equal to $d$, and observe that we may write the first term in the last line above as 
\begin{align*}
    &\mathbb{P}(C(\sampledSet_{-0}') = C((\mathfrak{d}_{-0}')^{\pi})|\sampledSet_{-0} = (\mathfrak{d}_{-0})^{\pi}, \data' = d'_i) \\
    &= \left(\prod_{\text{distinct }d \in \mathfrak{d} \text{ not equal to }d_i}\frac{1}{\prod_{k=1}^{m_d((\mathfrak{d}'_{-0})^{\pi})}(m+1-k+1)}\right)\cdot \frac{1}{\prod_{k=1}^{m_{d_i}((\mathfrak{d}'_{-0})^{\pi})}(m+1-k)}\\
    &= \left(\prod_{\text{distinct }d \in \mathfrak{d} \text{ not equal to }d_i}\frac{(m+1-m_d((\mathfrak{d}'_{-0})^{\pi}))!}{(m+1)!}\right)\cdot \frac{(m-m_{d_i}((\mathfrak{d}'_{-0})^{\pi}))!}{m!}.
\end{align*}
Now note that $m_d((\mathfrak{d}'_{-0})^{\pi})$ does not depend on the choice of $\pi \in \Sigma_i$ and so we need only consider the permutation $\pi^*_i \in \Sigma_i$ given by \[\pi^*_i: k \mapsto \begin{cases}
i \textnormal{ if } k = 0\\
k-1 \textnormal{ if } 0 < k \leq i\\
k \textnormal{ if } i < k \leq m
\end{cases}\] so that, for any $\pi \in \Sigma_i$, $m_d((\mathfrak{d}'_{-0})^{\pi}) = m_d((\mathfrak{d}'_{-0})^{\pi^*_i}) =  m_d(\mathfrak{d}'_{-i})$, where $\mathfrak{d}'_{-i} := (d'_0, \ldots, d'_{i-1}, d'_{i+1}, \ldots, d'_m)$. As such, combining this fact with what has been shown above, gives that \[\mathbb{P}(C(\sampledSet_{-0}') = C((\mathfrak{d}'_{-0})^{\pi^*_i})|\sampledSet_{-0} = (\mathfrak{d}_{-0})^{\pi^*_i}, \data' = d'_i) = \left(\prod_{\text{distinct }d \in \mathfrak{d} \text{ not equal to }d_i}\frac{(m+1-m_d(\mathfrak{d}'_{-i}))!}{(m+1)!}\right)\cdot \frac{(m-m_{d_i}(\mathfrak{d}'_{-i}))!}{m!}.\]
Note that the right-hand side above depends on $d_i$ only through its value, not its subscript, and hence the equation above takes the same value for any $j\neq i$ for which $d_j=d_i$. 
%Now, it is clear that the above is the same for distinct $i,j$ with $d_i = d_j$. 
In the case of distinct $i,j$ with $d_i \neq d_j$, the above gives that
\begin{align*}&\mathbb{P}(C(\sampledSet_{-0}') = C((\mathfrak{d}_{-0}')^{\pi^*_i})|\sampledSet_{-0} = (\mathfrak{d}_{-0})^{\pi^*_i}, \data' = d'_i) \\
&= \left(\prod_{\text{distinct }d \in \mathfrak{d} \text{ equal to neither }d_i \text{ nor } d_j}\frac{(m+1-m_d(\mathfrak{d}'_{-i}))!}{(m+1)!}\right)\cdot \frac{(m+1-m_{d_j}(\mathfrak{d}'_{-i}))!}{(m+1)!}\cdot \frac{(m-m_{d_i}(\mathfrak{d}'_{-i}))!}{m!}\\
&= \left(\prod_{\text{distinct }d \in \mathfrak{d} \text{ equal to neither }d_i \text{ nor } d_j}\frac{(m+1-m_d(\mathfrak{d}'_{-j}))!}{(m+1)!}\right)\cdot \frac{(m+1-m_{d_j}(\mathfrak{d}'_{-i}))!}{(m+1)!}\cdot \frac{(m-m_{d_i}(\mathfrak{d}'_{-i}))!}{m!}\\
&= \left(\prod_{\text{distinct }d \in \mathfrak{d} \text{ equal to neither }d_i \text{ nor } d_j}\frac{(m+1-m_d(\mathfrak{d}'_{-j}))!}{(m+1)!}\right)\cdot \frac{(m-m_{d_j}(\mathfrak{d}'_{-j}))!}{(m+1)!}\cdot \frac{(m+1-m_{d_i}(\mathfrak{d}'_{-j}))!}{m!}\\
&= \left(\prod_{\text{distinct }d \in \mathfrak{d} \text{ not equal to }d_j}\frac{(m+1-m_d(\mathfrak{d}'_{-j}))!}{(m+1)!}\right)\cdot \frac{(m-m_{d_j}(\mathfrak{d}'_{-j}))!}{m!}\\
&= \mathbb{P}(C(\sampledSet_{-0}') = C(\mathfrak{d}'_{-j})|\sampledSet_{-0} = \mathfrak{d}_{-j}, \data' = d'_j)\\
&= \mathbb{P}(C(\sampledSet_{-0}') = C((\mathfrak{d}_{-0}')^{\pi^*_j})|\sampledSet_{-0} = (\mathfrak{d}_{-0})^{\pi^*_j}, \data' = d'_j),\end{align*}
as desired, where the third equality follows from the fact that $m_{d_j}(\mathfrak{d}'_{-j}) + 1 = m_{d_j}(\mathfrak{d}'_{-i})$ for any pair $i,j$ with $d_i \neq d_j$. By the same argument made for $\pi^*_i$, the last line of the above remains true if we replace $\pi^*_j$ with any $\pi \in \Sigma_j$ and thus, we have shown that not only does $\mathbb{P}(C(\sampledSet_{-0}') = C((\mathfrak{d}'_{-0})^{\pi})|\sampledSet_{-0} = (\mathfrak{d}_{-0})^\pi, \data' = d'_i)$ not depend on the choice of $\pi \in \Sigma_i$, for any given $i$, but it also does not depend on $i \in \zeroM$, and hence we have the desired proportionality of equation \eqref{d-prop}.

Finally, since $f'(\sampledData'{}^{(i)}) = (m+1)^{-1}f(\tilde{D}^{(i)})$ by definition, and $\hat{f}(\sampledData^{(i)}) = Kf(\sampledData^{(i)})\; \forall i \in \zeroM$ for some $K$ not depending on $i$ under the proportionality hypothesis, we have that $\hat{f}(\sampledData^{(i)}) = K(m+1)f'(\tilde{D}'{}^{(i)}), \forall i \in \zeroM$ and hence the two are proportional. As such, we have that \[\frac{\sum_{i=0}^\numMCSamples\mathbf{1}[S'(\tilde{D}'{}^{(i)}) \geq S'(\data')]f'(\tilde{D}'{}^{(i)})\sum_{\pi \in \Sigma_i}\proposal'((\mathfrak{D}'_{-0})^{\pi}|\tilde{D}'{}^{(i)})}{\sum_{j=0}^\numMCSamples f'(\tilde{D}'{}^{(j)})\sum_{\pi' \in \Sigma_j}\proposal'((\mathfrak{D}'_{-0})^{\pi'}|\tilde{D}'{}^{(j)})}\] \[= \frac{\sum_{i=0}^\numMCSamples\mathbf{1}[S(\tilde{D}{}^{(i)}) \geq S(\data)]\hat{f}(\tilde{D}{}^{(i)})\sum_{\pi \in \Sigma_i}\proposal((\mathfrak{D}_{-0})^{\pi}|\tilde{D}{}^{(i)})}{\sum_{j=0}^\numMCSamples \hat{f}(\tilde{D}{}^{(j)})\sum_{\pi' \in \Sigma_j}\proposal((\mathfrak{D}_{-0})^{\pi'}|\tilde{D}{}^{(j)})}\] and so $p'{} = p$. This combined with the fact shown above that $p'$ is a valid p-value implies the desired validity of $p$ under $\mathcal{H}^{\propto}_0$.

\subsection{Sharing samples between $y \in \mathcal{Y}$}\label{sharing-appendix}
We now show how the above framework can be used to share samples between different grid values $y \in \mathcal{Y}$ as discussed in Remark~\ref{sharing}. We describe this sample sharing in the absence of the rounding procedure of Remark~\ref{sharing} which considers only a small grid $\mathcal{Y}' \subseteq \mathcal{Y}$ (i.e., we will obtain samples for \emph{each} $y$ in the full support $\mathcal{Y}$ and share these samples amongst all other elements of $\mathcal{Y}$), but both of these methods can be combined to construct conformal prediction regions even more efficiently. 

In more detail, recall that the naive interval construction described in Section~\ref{invert-chap} runs $|\mathcal{Y}|$ \emph{independent} non-stationarity tests, using our weighted MC randomization testing framework for conditionally i.i.d.~resamples using $\proposalNoTilde$, at each $y \in \mathcal{Y}$. To formalize this notion, set $(\data_{[1:\horizon-1]}, C_\horizon, X_\horizon)$ to denote the entire dataset except the last response and let $U_y$ denote the exogenous randomness used by the weighted MC randomization test when determining if $y$ is in the acceptance region. Then we can define an acceptance function $\varphi(U_y, (\data_{[1:\horizon-1]}, C_\horizon, X_\horizon), y)$ which is $1$ (i.e., accepts) if the test, using $U_y$ as exogenous randomness, states that $y$ is in the acceptance region upon seeing $(\data_{[1:\horizon-1]}, C_\horizon, X_\horizon)$, and is otherwise $0$ (i.e., rejects). In the naive gridding procedure, the $U_y$ are all jointly independent and are distributed identically to the exogeneous randomness $U$ that is used in the usual non-stationarity test. As such, it is clear that $\mathbb{E}[\varphi(U_{Y_\horizon}, (\data_{[1:\horizon-1]}, C_\horizon, X_\horizon), Y_\horizon)] \geq 1-\alpha$ due to the validity of the non-stationarity test proved in Section~\ref{non-stationarity} and so the corresponding prediction region attains coverage at least that of the nominal rate. The $U_y$, however, need not be independent. Instead, for example, if we have $U_y = U$ for all $y \in \mathcal{Y}$, where, again, $U$ is the independent exogenous randomness used by the non-stationarity test, then our prediction region would once again control miscoverage at the nominal rate. More generally, as long as each $U_y$ (and $U$) is generated independently from $\data$ and we have the following equality in distribution: \[(U_{Y_\horizon}, Y_\horizon) \overset{d}{=} (U, Y_\horizon),\] the resulting prediction regions will be valid. The process of sharing samples between $y \in \mathcal{Y}$ is based upon this idea.

While described as resampling datasets $\sampledData^{(i)}$ conditional on $\data$ in Section~\ref{non-stationarity}, the non-stationarity tests we describe really sample permutations $\pi_i$, conditionally on $\data$, and then set $\sampledData^{(i)}$ equal to $\data^{\pi_i} := ((C_{\pi_i(1)},\allowbreak X_{\pi_i(1)},\allowbreak Y_{\pi_i(1)}),\allowbreak \ldots,\allowbreak (C_{\pi_i(\horizon)},\allowbreak X_{\pi_i(\horizon)},\allowbreak Y_{\pi_i(\horizon)}))$; we use $\proposalNoTilde_{\Pi}$ to denote the corresponding (to $\proposalNoTilde$) resampling distribution over permutations. In the naive interval construction procedure which uses independent exogenous randomness, let $\sampledSet_{\Pi}^y$ denote the (multi)set of permutations sampled for determining if $y$ is in the acceptance region. We claim that all these samples can be shared so that each $y \in \mathcal{Y}$ instead uses the entire set of shared permutations $\sampledSet_{\Pi} := \bigcup_{y \in \mathcal{Y}}\sampledSet_{\Pi}^y$ as opposed to just $\sampledSet_{\Pi}^y$:

\begin{proposition}
Let $\proposalNoTilde$ be any conditionally i.i.d.~resampling distribution and let $\proposalNoTilde_\Pi$ denote the corresponding resampling distribution over permutations. Furthermore define, for each $y \in \mathcal{Y}$, a conditional distribution $\proposalNoTilde'_{\Pi,y}$ over permutations which draws its samples conditionally independently from $Y_\horizon$ given $(\data_{[1:\horizon-1]}, C_\horizon, X_\horizon)$ and is defined by \[\proposalNoTilde'_{\Pi,y}(\cdot | ((C_1, X_1, Y_1), \ldots, (C_\horizon, X_\horizon, Y_\horizon))) = \proposalNoTilde_{\Pi}\left(\cdot|((C_1, X_1, Y_1), \ldots, (C_\horizon, X_\horizon, y))\right).\]

To construct a prediction region, suppose that, for each $y \in \mathcal{Y}$, a sampled (multi)set $\sampledSet_{\Pi}^y = \{\pi_{1,y}, \ldots, \pi_{\numMCSamples,y}\}$ of permutations is generated by sampling each permutation conditionally i.i.d.~from $\proposalNoTilde'_{\Pi,y}(\cdot | \data)$\footnote{Since $\proposalNoTilde'_{\Pi,y}$ does not depend on $Y_\horizon$, we can indeed sample from this distribution during prediction region construction wherein $Y_\horizon$ is unobserved}, but the full (multi)set $\sampledSet_{\Pi}$ of $\numMCSamples|\mathcal{Y}|$ permutations is used to determine $y$'s membership in the acceptance region. Specifically, defining $\data_y :=((C_1, X_1, Y_1), \ldots, (C_\horizon, X_\horizon, y))$, writing $\mathcal{Y} = \{y_1, \ldots, y_{|\mathcal{Y}|}\}$, and defining $\Lambda := \{(0,0)\} \cup \left([\numMCSamples]\times [|\mathcal{Y}|]\right)$, we construct an acceptance region $\mathfrak{A}$ from $\sampledSet_{\Pi}$ via the rule: \begin{equation}\label{rule}y \in \mathfrak{A} \iff \sum_{(i,j) \in \Lambda} w^{\proposalNoTilde,y}_{\sampledSet_\Pi,i,y_j}((\data_y)^{\pi_{i,y_j}})\mathbf{1}[S((\data_y)^{\pi_{i,y_j}}) \geq S(\data_y)] > \alpha,\end{equation} where

\[w^{\proposalNoTilde,y}_{\sampledSet_\Pi,i,y_j}((\data_y)^{\pi_{i,y_j}}) := \frac{\displaystyle \hat{f}((\data_y)^{\pi_{i,y_j}})\proposalNoTilde'_{\Pi,y_j}(\pi_{i,y_j}^{-1}|(\data_y)^{\pi_{i,y_j}})\prod_{(\tilde{i},\tilde{j}) \in \Lambda\backslash\{(i,j)\}}\proposalNoTilde'_{\Pi,y_{\tilde{j}}}(\pi_{\tilde{i},y_{\tilde{j}}} \circ \pi_{i,y_j}^{-1}|(\data_y)^{\pi_{i,y_j}})}{\displaystyle \sum_{(i',j') \in \Lambda}\hat{f}((\data_y)^{\pi_{i',y_{j'}}})\proposalNoTilde'_{\Pi,y_{j'}}(\pi_{i',y_{j'}}^{-1}|(\data_y)^{\pi_{i',y_{j'}}})\prod_{(\tilde{i}',\tilde{j}') \in \Lambda\backslash\{(i',j')\}}\proposalNoTilde'_{\Pi,y_{\tilde{j}'}}(\pi_{\tilde{i}', y_{\tilde{j}'}} \circ \pi_{i',y_{j'}}^{-1}|(\data_y)^{\pi_{i',y_{j'}}})},\] and we take $\pi_{0,y_0}$ to simply be the identity permutation.

Then the resultant prediction region $\mathfrak{A}$ controls miscoverage at the nominal rate under the proportionality hypothesis $\mathcal{H}^\propto_0$.
\end{proposition}

Before giving a proof we make a brief computational/procedural note that, similar to equation~\eqref{q-prop} of Remark~\ref{comp-iid}, when $\proposalNoTilde(\cdot|(\data_y)^{\pi_i,y_j})$ does not depend on $(i,j) \in \Lambda$ for any $y \in \mathcal{Y}$, we have that \begin{align*}
   w^{\proposalNoTilde,y}_{\sampledSet_\Pi,i,y_j}((\data_y)^{\pi_{i,y_j}})
    &= \frac{\displaystyle \hat{f}((\data_y)^{\pi_{i,y_j}})\proposalNoTilde'_{\Pi,y_j}(\pi_{i,y_j}^{-1}|(\data_y)^{\pi_{i,y_j}})\prod_{(\tilde{i},\tilde{j}) \in \Lambda\backslash\{(i,j)\}}\proposalNoTilde'_{\Pi,y_{\tilde{j}}}(\pi_{\tilde{i},y_{\tilde{j}}} \circ \pi_{i,y_j}^{-1}|(\data_y)^{\pi_{i,y_j}})}{\displaystyle \sum_{(i',j') \in \Lambda}\hat{f}((\data_y)^{\pi_{i',y_{j'}}})\proposalNoTilde'_{\Pi,y_{j'}}(\pi_{i',y_{j'}}^{-1}|(\data_y)^{\pi_{i',y_{j'}}})\prod_{(\tilde{i}',\tilde{j}') \in \Lambda\backslash\{(i',j')\}}\proposalNoTilde'_{\Pi,y_{\tilde{j}'}}(\pi_{\tilde{i}', y_{\tilde{j}'}} \circ \pi_{i',y_{j'}}^{-1}|(\data_y)^{\pi_{i',y_{j'}}})}\\
    &= \frac{\displaystyle \hat{f}((\data_y)^{\pi_{i,y_j}})\proposalNoTilde'_{\Pi,y_j}(\pi_{i,y_j}^{-1}|(\data_y)^{\pi_{i,y_j}})/\proposalNoTilde'_{\Pi,y_{j}}(\pi_{i,y_{j}} \circ \pi_{i,y_j}^{-1}|(\data_y)^{\pi_{i,y_j}})}{\displaystyle \sum_{(i',j') \in \Lambda}\hat{f}((\data_y)^{\pi_{i',y_{j'}}})\proposalNoTilde'_{\Pi,y_{j'}}(\pi_{i',y_{j'}}^{-1}|(\data_y)^{\pi_{i',y_{j'}}})\proposalNoTilde'_{\Pi,y_{j'}}(\pi_{i', y_{j'}} \circ \pi_{i',y_{j'}}^{-1}|(\data_y)^{\pi_{i',y_{j'}}})}\\
    &= \frac{\displaystyle \hat{f}((\data_y)^{\pi_{i,y_j}})\proposalNoTilde'_{\Pi,y_j}(\pi_{i,y_j}^{-1}|(\data_y)^{\pi_{i,y_j}})/\proposalNoTilde'_{\Pi,y_{j}}(\textnormal{id}|(\data_y)^{\pi_{i,y_j}})}{\displaystyle \sum_{(i',j') \in \Lambda}\hat{f}((\data_y)^{\pi_{i',y_{j'}}})\proposalNoTilde'_{\Pi,y_{j'}}(\pi_{i',y_{j'}}^{-1}|(\data_y)^{\pi_{i',y_{j'}}})\proposalNoTilde'_{\Pi,y_{j'}}(\textnormal{id}|(\data_y)^{\pi_{i',y_{j'}}})},
\end{align*} where $\textnormal{id}$ denotes the identity permutation. Hence, computation of $w^{\proposalNoTilde,y}_{\sampledSet_\Pi,i,y_j}((\data_y)^{\pi_{i,y_j}})$ across all $(i,j) \in \Lambda$ becomes tractable in only $O(\numMCSamples|\mathcal{Y}|)$ operations. 

\begin{proof}
Consider a non-stationarity test in which we draw $\numMCSamples|\mathcal{Y}|$ samples and the resampling distribution $\breve{\proposalNoTilde}$ we consider ignores $Y_\horizon$, and instead draws these $\numMCSamples|\mathcal{Y}|$ samples in $|\mathcal{Y}|$ groups, the $j^{\text{th}}$ of which consists of $\numMCSamples$ resamples $\data^{\pi_{1,y_j}}, \ldots, \data^{\pi_{\numMCSamples,y_j}}$ where the $\pi_{i,y_j}$ are drawn conditionally i.i.d.~from $\proposalNoTilde'_{\Pi,y_j}\left(\cdot|\data\right)$ for each $j \in [|\mathcal{Y}|]$, and where $\mathcal{Y} = \{y_1, \ldots, y_{|\mathcal{Y}|}\}$. It is important to note that $\breve{\proposalNoTilde}$ is \emph{not} a conditionally i.i.d~resampling scheme and thus, as discussed in Section~\ref{conformal-sims}, the workaround discussed in Section~\ref{non-iid-appendix} must be employed. Defining $\Sigma = \{(0 \; \; \ell): \ell \in [0:\numMCSamples|\mathcal{Y}|]\}$ and letting $i \in \zeroM$ index with any given group and $j \in \begin{cases}[|\mathcal{Y}|] \text{ if } i > 0\\
\{0\} \text{ if } i = 0
\end{cases}$ index over groups, notice that \[w^{\proposalNoTilde,Y_\horizon}_{\sampledSet_\Pi,i,y_j}(\data^{\pi_{i,y_j}}) = w^{\breve{\proposalNoTilde}, \Sigma}_{\sampledSet,(i,j)}(\sampledData^{(i,j)}),\] where $\sampledSet$ denotes the full set of $\numMCSamples|\mathcal{Y}|$ resamples and we double index resamples as $\data^{(i,j)} = \data^{\pi_{i,y_j}}$ for all $(i,j)$ in $\Lambda$. This is because, letting $\proposalNoTilde'_{y_j}$ denote the induced (by $\proposalNoTilde'_{\Pi,y_j}$) distribution on each resample given by $\proposalNoTilde'_{y_j}(\data^\pi|\data) = \proposalNoTilde'_{\Pi,y_j}(\pi|\data)$ and looking at the rightmost terms in the numerator of $w^{\proposalNoTilde,y}_{\sampledSet_\Pi,i,y_j}((\data_y)^{\pi_{i,y_j}})$'s definition, we have that
\begin{align*}
     &\proposalNoTilde'_{\Pi,y_j}(\pi_{i,y_j}^{-1}|\data^{\pi_{i,y_j}})\prod_{(\tilde{i},\tilde{j}) \in \Lambda\backslash\{(i,j)\}}\proposalNoTilde'_{\Pi,y_{\tilde{j}}}(\pi_{\tilde{i},y_{\tilde{j}}} \circ \pi_{i,y_j}^{-1}|\data^{\pi_{i,y_j}})\\
    &= \proposalNoTilde'_{y_j}(\data|\data^{\pi_{i,y_j}})\prod_{(\tilde{i},\tilde{j}) \in \Lambda\backslash\{(i,j)\}}\proposalNoTilde'_{y_{\tilde{j}}}(\data^{\pi_{\tilde{i},y_{\tilde{j}}}}|\data^{\pi_{i,y_j}})\\
    &= \proposalNoTilde'_{y_j}(\data|\sampledData^{(i,j)})\prod_{(\tilde{i},\tilde{j}) \in \Lambda\backslash\{(i,j)\}}\proposalNoTilde'_{y_{\tilde{j}}}(\sampledData^{(\tilde{i},\tilde{j})}|\sampledData^{(i,j)})\\
    &= \sum_{\pi \in \Sigma_{(i,j)}} \breve{\proposalNoTilde}(((\sampledData^{(0,0)},\sampledData^{(1,1)}, \ldots, \sampledData^{(\numMCSamples,1)}, \ldots, \sampledData^{(1,|\mathcal{Y}|)}, \ldots, \sampledData^{(\numMCSamples,|\mathcal{Y}|)})^\pi)_{-0} | \sampledData^{(i,j)}),
\end{align*} where $\Sigma_{(i,j)}$ is the single element subset of $\Sigma$ containing solely the permutation on $[0:\numMCSamples|\mathcal{Y}|]$ that swaps $0$ and $m(j-1)+i$. Thus, the numerators of $w^{\proposalNoTilde,Y_\horizon}_{\sampledSet_\Pi,i,y_j}(\data^{\pi_{i,y_j}})$ and  $w^{\breve{\proposalNoTilde}, \Sigma}_{\sampledSet,(i,j)}(\sampledData^{(i,j)})$ are equal, and precisely the same argument shows that the denominators are also equal and thus $w^{\proposalNoTilde,Y_\horizon}_{\sampledSet_\Pi,i,y_j}(\data^{\pi_{i,y_j}}) = w^{\breve{\proposalNoTilde}, \Sigma}_{\sampledSet,(i,j)}(\sampledData^{(i,j)})$.

Hence, the p-value \[\sum_{(i,j) \in \Lambda} w^{\proposalNoTilde,y}_{\sampledSet_\Pi,i,y_j}(\data^{\pi_{i,y_j}})\mathbf{1}[S(\data^{\pi_{i,y_j}}) \geq S(\data)],\] corresponding to line~\eqref{rule} is a valid p-value by the main result of Section~\ref{non-iid-appendix}. Letting $U$ denote the exogenous randomness used by this test, notice that the exogenous random variables $U_y$ used by each $y \in \mathcal{Y}$ in the prediction region construction of line~\eqref{rule} are all (deterministically) equal---since the only randomness in determining if $y \in \mathfrak{A}$ is in the random sampling of permutations $\sampledSet_\Pi$, which does not depend on the specific choice of $y \in \mathcal{Y}$---and marginally identically distributed to $U$. Since each of $U$ and $U_y$ is generated independently from $Y_\horizon$, we have that $(U_{Y_\horizon}, Y_\horizon) \overset{d}{=} (U, Y_\horizon)$ and hence the corresponding prediction region $\mathfrak{A}$ is valid, as desired.
\end{proof}
    \section{Pseudocode for resampling distributions}\label{pseudocode}
\label{AppendixE}

In this appendix section, we provide pseudocode for all resampling procedures described in Section~\ref{proposal-chap}. Throughout this section, we use the notation $\text{Cat}(p)$, for any $p \in [0,1]^d$ with $\sum_{i=1}^dp_i = 1$, to denote the categorical distribution on $[d]$ with probability of sampling $i$ equal to $p_i$.

\subsection{Non-stationarity testing in a $C$-stationary strongly non-reactive environment}
We begin with the resampling procedures for non-stationarity testing in a $C$-stationary strongly non-reactive environment (Environment~\ref{exch-non-react-env}).
\begin{algorithm}[!ht]
  
  \KwInput{Data sequence $\data$}
  Set $\sampledData$ to the empty list and set $R \gets [\horizon]$\\
  \For{$t=1, \ldots, T$}{
    Sample \[i \sim \text{Cat}\left(\left(\frac{\decisionProb(X_j|\sampledData, C_j)\mathbf{1}[j \in R]}{\sum_{j'=1}^T \decisionProb(X_{j'}|\sampledData, C_{j'})\mathbf{1}[j' \in R]}\right)_{j=1}^\horizon\right),\] if $\decisionProb(\cdot|\sampledData)$'s support intersects $R$; otherwise, terminate the sampling procedure\\
    Append $Z_i$ to $\sampledData$ and set $R \gets R \backslash \{i\}$
  }
  \KwOutput{$\sampledData$}

\caption{$\text{imitation}_{\pi}$}
\label{sim1-descr}
\end{algorithm}

\begin{algorithm}[!ht]
  
  \KwInput{Data sequence $\data$; probability distribution $\mathbb{P}_{U_t}$ denoting the distribution of the $t^{\text{th}}$ exogenous random variable $U_t$ generated by $\decisionMaker$}
  Set $\sampledData$ to the empty list and set $R \gets [\horizon]$\\
  \For{$t=1, \ldots, T$}{
    Sample \[\tilde{U}_t \sim \mathbb{P}_{U_t}(\cdot | \exists s \in R: X_s = \delta_t(C_s, \tilde{H}_{t-1}, \tilde{U}_1, \ldots, \tilde{U}_{t}))\] if the conditioning event is non-empty; otherwise, terminate the sampling procedure\\
    Sample $i \sim \text{Unif}\left(\{s \in R: X_s = \delta_t(C_s, \tilde{H}_{t-1}, \tilde{U}_1, \ldots, \tilde{U}_{t})\}\right)$\\
    Append $Z_i$ to $\sampledData$ and set $R \gets R \backslash \{i\}$
  }
  \KwOutput{$\sampledData$}

\caption{$\text{re-imitation}_{\pi}$}
\label{sim2-descr}
\end{algorithm}

\begin{algorithm}[!ht]
  
  \KwInput{Data sequence $\data$ as well as the exogenous randomness $U_1, \ldots, U_\horizon$ used to generate it}
  Set $\sampledData$ to the empty list and set $R \gets [\horizon]$\\
  \For{$t=1, \ldots, T$}{
        Sample $i \sim \text{Unif}\left(\{s \in R: X_s = \delta_t(C_s, \tilde{H}_{t-1}, U_1, \ldots, U_t)\}\right)$ if the set is non-empty; otherwise, terminate the sampling procedure\\
    Append $Z_i$ to $\sampledData$ and set $R \gets R \backslash \{i\}$
  }
  \KwOutput{$\sampledData$}

\caption{$\text{cond-imitation}_{\pi}$}
\label{sim3-descr}
\end{algorithm}

\subsection{Non-stationarity testing in an MDP}
As discussed in Section~\ref{non-stat-mdp-proposal}, the datasets used in non-stationarity testing in an MDP are augmented with an additional action $X_{\horizon+1}$, and thus we may view these datasets as a list of $\horizon+1$ state action pairs: $\data = ((C_1, X_1), \ldots, (C_{\horizon+1}, X_{\horizon+1}))$. Under this framework, all resampling procedures in this section utilize the following function $\phi$, which takes the state-action pair $(c,x)$ to the set of indices which follow it: \[\phi(c,x) = \{t \in [2:\horizon+1]: (C_{t-1}, X_{t-1}) = (c,x)\}.\] Additionally, using this view of the dataset $\data$, we use $Z_t$ to denote the $t^{\text{th}}$ state-action pair $(C_t, X_t)$; $\tilde{Z}_t$ denotes the $t^{\text{th}}$ state-action pair of the resampled dataset $\sampledData$. Using the function $\phi$, we now present the corresponding $\textnormal{uniform}_{\pi}$, $\textnormal{imitation}_{\pi}$, $\textnormal{re-imitation}_{\pi}$, and $\textnormal{cond-imitation}_{\pi}$ for an MDP under this setup of the dataset $\data$.

\begin{algorithm}[!ht]
  
  \KwInput{Data sequence $\data$}
  Set $\sampledData \gets ((C_1, X_1)) $ and set $R \gets [1:\horizon+1]$\\
  \For{$t=2, \ldots, \horizon+1$}{
    Sample \[i \sim \text{Unif}\left(R \cap \phi(\tilde{Z}_{t-1})\right)\] if the set is non-empty; otherwise, terminate the sampling procedure\\
    Append $Z_i$ to $\sampledData$ and set $R \gets R \backslash \{i\}$
  }
  \KwOutput{$\sampledData$}

\caption{$\text{uniform}_{\pi}$ in an MDP}
\label{u-descr-mdp}
\end{algorithm}

\begin{algorithm}[!ht]
  
  \KwInput{Data sequence $\data$}
  Set $\sampledData \gets ((C_1, X_1)) $ and set $R \gets [1:\horizon+1]$\\
  \For{$t=2, \ldots, \horizon+1$}{
    Sample \[i \sim \text{Cat}\left(\left(\frac{\decisionProb(X_j|\sampledData, C_j)\mathbf{1}[j \in R \cap \phi(\tilde{Z}_{t-1})]}{\sum_{j'=2}^{T+1} \decisionProb(X_{j'}|\sampledData, C_{j'})\mathbf{1}[j' \in R \cap \phi(\tilde{Z}_{t-1})]}\right)_{j=2}^{\horizon+1}\right),\] if $\exists j \in R \cap \phi(\tilde{Z}_{t-1})$ such that $\decisionProb(X_j|\sampledData, C_j) > 0$; otherwise, terminate the sampling procedure\\
    Append $Z_i$ to $\sampledData$ and set $R \gets R \backslash \{i\}$
  }
  \KwOutput{$\sampledData$}

\caption{$\text{imitation}_{\pi}$ in an MDP}
\label{sim1-descr-mdp}
\end{algorithm}

\begin{algorithm}[!ht]
  
  \KwInput{Data sequence $\data$; probability distribution $\mathbb{P}_{U_t}$ denoting the distribution of the $t^{\text{th}}$ exogenous random variable $U_t$ generated by $\decisionMaker$}
  Set $\sampledData \gets ((C_1, X_1)) $ and set $R \gets [1:\horizon+1]$\\
  \For{$t=2, \ldots, T+1$}{
    Sample \[\tilde{U}_t \sim \mathbb{P}_{U_t}(\cdot | \exists s \in R \cap \phi(\tilde{Z}_{t-1}): X_s = \delta_t(C_s, \tilde{H}_{t-1}, \tilde{U}_1, \ldots, \tilde{U}_{t}))\] if the conditioning event is non-empty; otherwise, terminate the sampling procedure\\
    Sample $i \sim \text{Unif}\left(\{s \in R\cap \phi(\tilde{Z}_{t-1}): X_s = \delta_t(C_s, \tilde{H}_{t-1}, \tilde{U}_1, \ldots, \tilde{U}_{t})\}\right)$\\
    Append $Z_i$ to $\sampledData$ and set $R \gets R \backslash \{i\}$
  }
  \KwOutput{$\sampledData$}

\caption{$\text{re-imitation}_{\pi}$ in an MDP}
\label{sim2-descr-mdp}
\end{algorithm}

\begin{algorithm}[!ht]
  
  \KwInput{Data sequence $\data$ as well as the exogenous randomness $U_1, \ldots, U_{\horizon+1}$ used to generate it}
  Set $\sampledData \gets ((C_1, X_1)) $ and set $R \gets [1:\horizon+1]$\\
  \For{$t=2, \ldots, T+1$}{
        Sample $i \sim \text{Unif}\left(\{s \in R \cap \phi(\tilde{Z}_{t-1}) : X_s = \delta_t(C_s, \tilde{H}_{t-1}, U_1, \ldots, U_t)\}\right)$ if the set is non-empty; otherwise, terminate the sampling procedure\\
    Append $Z_i$ to $\sampledData$ and set $R \gets R \backslash \{i\}$
  }
  \KwOutput{$\sampledData$}

\caption{$\text{cond-imitation}_{\pi}$ in an MDP}
\label{sim3-descr-mdp}
\end{algorithm}

\subsection{Conditional independence testing}
We now give pseudocode for our resampling procedures for conditional independence testing discussed in Section~\ref{distributional-proposal}. We first show pseudocode for $\text{restricted-uniform}_{\pi}$ resampling, which, although not used on its own for conditional independence testing, makes up the first stage of the $\text{restricted-uniform}_{\pi}\text{+}\text{imitation}_X$ resampling scheme. We then go on to present the $\text{imitation}_X$ resampling procedure, which is also a key ingredient that is used in the $\text{uniform}_{\pi}\text{+}\text{imitation}_X$ and $\text{restricted-uniform}_{\pi}\text{+}\text{imitation}_X$ resampling procedures, which are presented subsequently. Finally, we show pseudocode for $\text{combined}_{\pi,X}$ sampling, which combines the permutation and randomization of $X_t$'s into a single stage.

\begin{algorithm}[!ht]
  
  \KwInput{Data sequence $\data$}
  Set $\sampledData$ to the empty list\\
  Set $\Gamma \gets \{\pi' \in \Pi_{[\horizon]}: g(X_{\pi'(t)}) = g(X_{t}), \forall t \in [T]\}$\\
  Sample $\pi \sim \textnormal{Unif}(\Gamma)$\\
  $\sampledData \gets (Z_{\pi(t)})_{t=1}^T$\\
  \KwOutput{$\sampledData$}

\caption{$\text{restricted-uniform}_{\pi}$}
\label{ru-descr}
\end{algorithm}

\begin{algorithm}[!ht]
  \KwInput{Data sequence $\data$}
 Set $\sampledData$ to the empty list\\
  \For{$t=1, \ldots, T$}{
    Sample \[\tilde{X}_{t} \sim \decisionProb(\cdot|\sampledData, C_t, g(X_t)),\] if there exists $x \in \mathcal{X}$ with $\decisionProb(x|\sampledData, C_t, g(X_t)) > 0$; otherwise, terminate the sampling procedure\\
    Append $\tilde{Z}_t := (C_t, \tilde{X}_{t},Y_t)$ to $\sampledData$
  }
  \KwOutput{$\sampledData$}
\caption{$\text{imitation}_X$}
\label{s-descr}
\end{algorithm}

\begin{algorithm}[!ht]
\KwInput{Data sequence $\data$}
  Sample $\data'$ according to the $\text{uniform}_{\pi}$ distribution applied to $\data$\\
  Sample $\sampledData$ according to the $\text{imitation}_X$ distribution (Algorithm~\ref{s-descr}) applied to $\data'$\\
  \KwOutput{$\sampledData$}
\caption{$\text{uniform}_{\pi}\text{+}\text{imitation}_X$}
\label{us-descr}
\end{algorithm}

\begin{algorithm}[!ht]
\KwInput{Data sequence $\data$}
  Sample $\data'$ according to the $\text{restricted-uniform}_{\pi}$ distribution (Algorithm~\ref{ru-descr}) applied to $\data$\\
  Sample $\sampledData$ according to the $\text{imitation}_X$ distribution (Algorithm~\ref{s-descr}) applied to $\data'$\\
  \KwOutput{$\sampledData$}
\caption{$\text{restricted-uniform}_{\pi}\text{+}\text{imitation}_X$}
\label{rus-descr}
\end{algorithm}

\begin{algorithm}[!ht]
  \KwInput{Data sequence $\data$}
 Set $\sampledData$ to the empty list and set $R \gets [\horizon]$\\
  \For{$t=1, \ldots, T$}{
  Sample \[i \sim \text{Cat}\left(\left(\frac{\sum_{x \in \mathcal{X}: g(x) = g(X_j)}\decisionProb(x|\sampledData, C_j)\mathbf{1}[j \in R]}{\sum_{j'=1}^T \sum_{x' \in \mathcal{X}: g(x') = g(X_{j'})}\decisionProb(x'|\sampledData, C_{j'})\mathbf{1}[j' \in R]}\right)_{j=1}^\horizon\right),\] if there exists $j \in R$ such that $\sum_{x \in \mathcal{X}: g(x) = g(X_j)}\decisionProb(x|\sampledData, C_j) > 0$; otherwise, terminate the sampling procedure\\
  Set $R \gets R\backslash\{i\}$\\
    Sample \[\tilde{X}_{t} \sim \decisionProb(\cdot|\sampledData, C_i, g(X_i)),\] if there exists $x \in \mathcal{X}$ with $\decisionProb(x|\sampledData, C_i, g(X_i)) > 0$; otherwise, terminate the sampling procedure\\
    Append $\tilde{Z}_i := (C_i, \tilde{X}_{t},Y_i)$ to $\sampledData$
  }
  \KwOutput{$\sampledData$}
\caption{$\text{combined}_{\pi,X}$}
\label{c-descr}
\end{algorithm}

    \section{Supplementary simulation results}
\label{AppendixF}

\subsection{MCMC plots}\label{mcmc-appendix}
\begin{figure}[!htb]
     \centering
     \includegraphics[width=6.5in, height=2.8in]{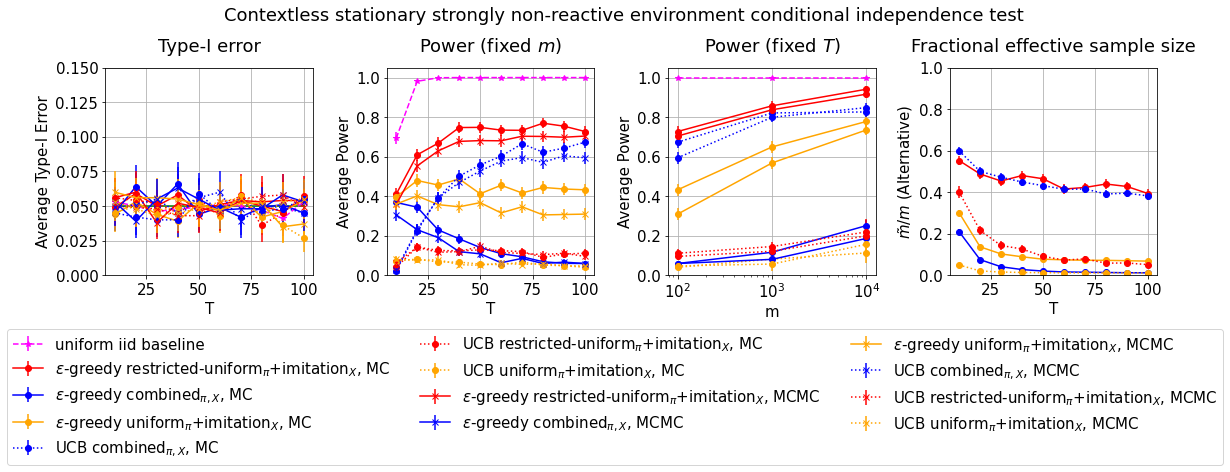}
     \caption{Type-I error rate (leftmost) and power (second from left) of both weighted MC and unweighted MCMC randomization tests at fixed $\numMCSamples=100$ and varying $\horizon$ as well as power at fixed $\horizon=100$ and varying $\numMCSamples$ (third from left) and fractional effective sample size plots at fixed $\numMCSamples=100$ and varying $\horizon$ (rightmost) in a contextless stationary strongly non-reactive environment on data gathered via $\epsilon$-greedy, UCB, and the uniform i.i.d.~baseline.}
     %$\mathcal{X}= \{-1,1\}$ and $Y_i |X_i \sim \mathcal{N}(X_i,1), i \in [m-1]$. Data is collected using $\pi^{\egreedy}$ and $\pi^{\UCB}$, with $Y_m |X_m \sim \mathcal{N}(X_m,1)$ under the null $\mathcal{H}_0^{\Stat}$ and $Y_m |X_m \sim \mathcal{N}(4X_m,1)$ under the alternative $\mathcal{H}_1^{\Stat}$.}
     \label{fig:dist_test_full}
 \end{figure}

\begin{figure}[!htb]
     \centering
     \includegraphics[width=6.5in, height=2.8in]{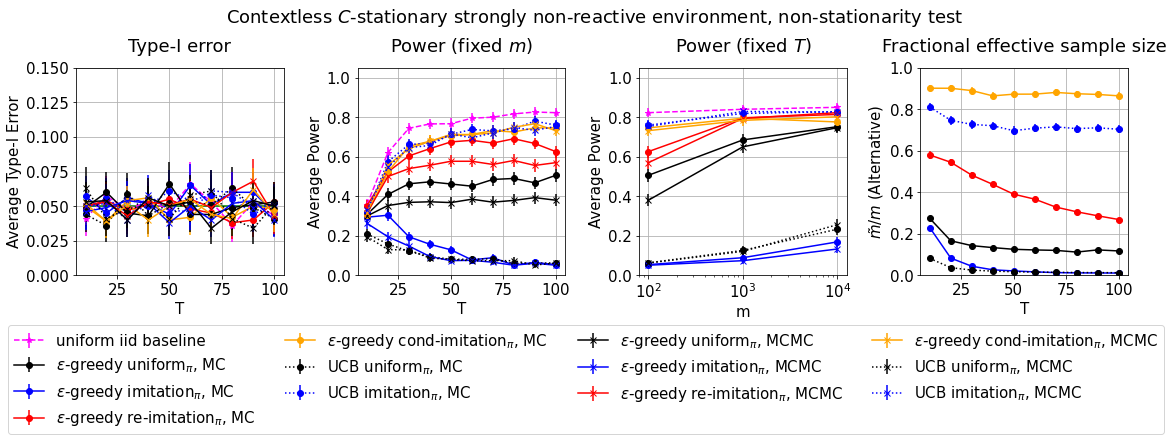}
     \caption{Type-I error rate (leftmost) and power (second from right) of both weighted MC and unweighted MCMC randomization tests at fixed $\numMCSamples=100$ and varying $\horizon$ as well as power for fixed $\horizon=100$ and varying $\numMCSamples$ (third from right) and fractional effective sample size at fixed $\numMCSamples=100$ and varying $\horizon$ (rightmost) in a contextless $C$-stationary strongly non-reactive environment with data gathered via $\epsilon$-greedy, UCB, and the uniform i.i.d.~baseline.}
     %$\mathcal{X}= \{-1,1\}$ and $Y_i |X_i \sim \mathcal{N}(X_i,1), i \in [m-1]$. Data is collected using $\pi^{\egreedy}$ and $\pi^{\UCB}$, with $Y_m |X_m \sim \mathcal{N}(X_m,1)$ under the null $\mathcal{H}_0^{\Stat}$ and $Y_m |X_m \sim \mathcal{N}(4X_m,1)$ under the alternative $\mathcal{H}_1^{\Stat}$.}
     \label{fig:nonstat_full}
 \end{figure}

\begin{figure}[!htb]
     \centering
     \includegraphics[width=6.5in, height=2.8in]{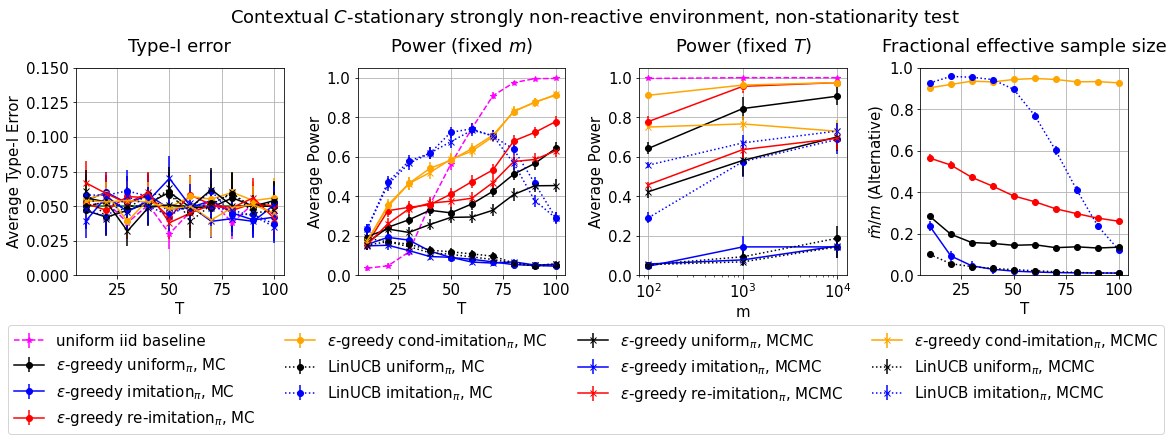}
     \caption{Type-I error rate (leftmost) and power (second from right) of both weighted MC and unweighted MCMC randomization tests at fixed $\numMCSamples=100$ and varying $\horizon$ as well as power for fixed $\horizon=100$ and varying $\numMCSamples$ (third from right) and fractional effective sample size at fixed $\numMCSamples=100$ and varying $\horizon$ (rightmost) in a contextless $C$-stationary strongly non-reactive environment with data gathered via $\epsilon$-greedy, LinUCB, and the uniform i.i.d.~baseline.}
     %$\mathcal{X}= \{-1,1\}$ and $Y_i |X_i \sim \mathcal{N}(X_i,1), i \in [m-1]$. Data is collected using $\pi^{\egreedy}$ and $\pi^{\UCB}$, with $Y_m |X_m \sim \mathcal{N}(X_m,1)$ under the null $\mathcal{H}_0^{\Stat}$ and $Y_m |X_m \sim \mathcal{N}(4X_m,1)$ under the alternative $\mathcal{H}_1^{\Stat}$.}
     \label{fig:nonstat2_full}
 \end{figure}

\begin{figure}[!htb]
     \centering
     \includegraphics[width=6.7in, height=3.in]{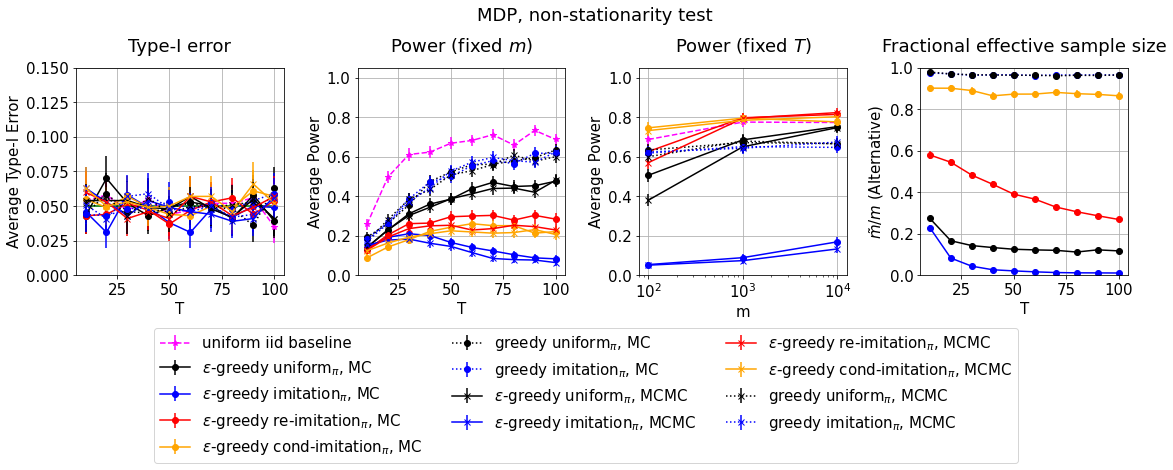}
     \caption{Type-I error rate (leftmost) and power (second from right) of both weighted MC and unweighted MCMC randomization tests at fixed $\numMCSamples=100$ and varying $\horizon$ as well as power for fixed $\horizon=100$ and varying $\numMCSamples$ (third from right) and fractional effective sample size at fixed $\numMCSamples=100$ and varying $\horizon$ (rightmost) in a contextless $C$-stationary strongly non-reactive environment with data gathered via $\epsilon$-greedy and greedy $Q$-learning.}
     %$\mathcal{X}= \{-1,1\}$ and $Y_i |X_i \sim \mathcal{N}(X_i,1), i \in [m-1]$. Data is collected using $\pi^{\egreedy}$ and $\pi^{\UCB}$, with $Y_m |X_m \sim \mathcal{N}(X_m,1)$ under the null $\mathcal{H}_0^{\Stat}$ and $Y_m |X_m \sim \mathcal{N}(4X_m,1)$ under the alternative $\mathcal{H}_1^{\Stat}$.}
     \label{fig:nonstat3_full}
 \end{figure}

\begin{figure}[!htb]
     \centering
     \includegraphics[width=6.4in, height=3.2in]{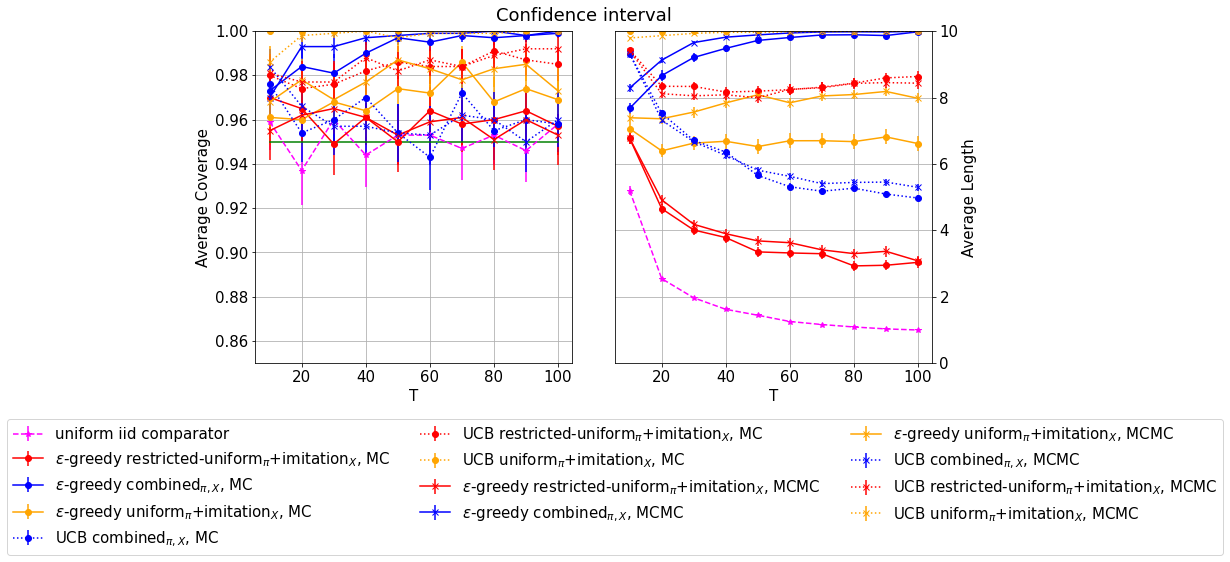}
     \caption{Coverage and average length of confidence intervals for $b_0$ using both weighted MC and unweighted MCMC randomization tests with data gathered via $\epsilon$-greedy, UCB, and the uniform i.i.d.~baseline.}
     %$\mathcal{X}= \{-1,1\}$ and $Y_i |X_i \sim \mathcal{N}(X_i,1), i \in [m-1]$. Data is collected using $\pi^{\egreedy}$ and $\pi^{\UCB}$, with $Y_m |X_m \sim \mathcal{N}(X_m,1)$ under the null $\mathcal{H}_0^{\Stat}$ and $Y_m |X_m \sim \mathcal{N}(4X_m,1)$ under the alternative $\mathcal{H}_1^{\Stat}$.}
     \label{fig:conf_full}
 \end{figure}

\begin{figure}[!htb]
     \centering
     \includegraphics[width=6.4in, height=3.2in]{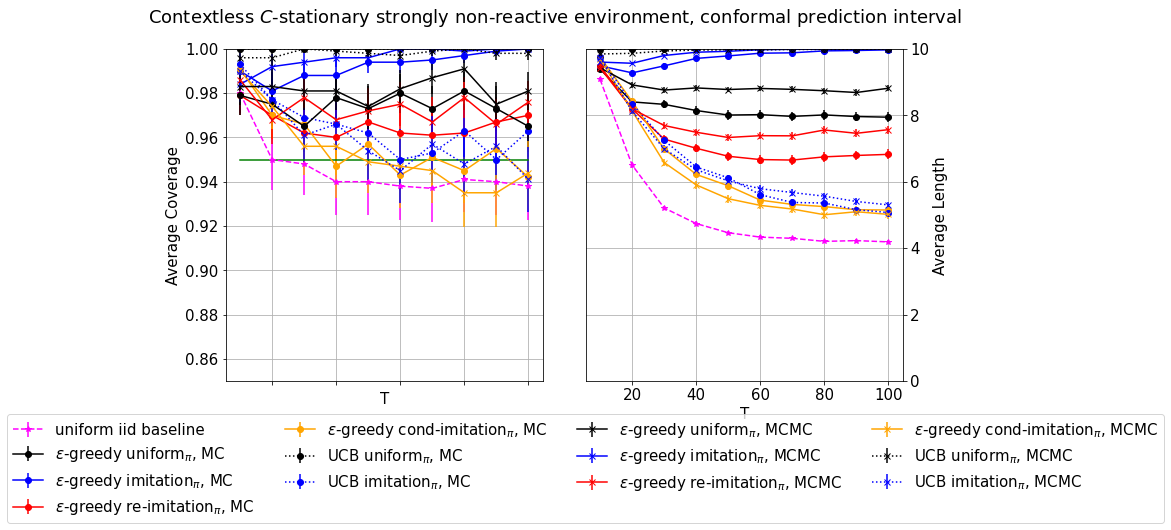}
     \caption{Coverage and average length of conformal prediction intervals for $Y_\horizon$ using both weighted MC and unweighted MCMC randomization tests with data gathered via $\epsilon$-greedy, UCB, and the uniform i.i.d.~baseline.}
     %$\mathcal{X}= \{-1,1\}$ and $Y_i |X_i \sim \mathcal{N}(X_i,1), i \in [m-1]$. Data is collected using $\pi^{\egreedy}$ and $\pi^{\UCB}$, with $Y_m |X_m \sim \mathcal{N}(X_m,1)$ under the null $\mathcal{H}_0^{\Stat}$ and $Y_m |X_m \sim \mathcal{N}(4X_m,1)$ under the alternative $\mathcal{H}_1^{\Stat}$.}
     \label{fig:conformal1_full}
 \end{figure}

In this section, we present the power, Type-I error, coverage, and length plots for the unweighted MCMC randomization test and its inversion, for the inferential tasks discussed in Section~\ref{sim-chap}, in Figures~\ref{fig:dist_test_full}--\ref{fig:conformal1_full}.

\subsection{Computation times}\label{comp-time-appendix}
\begin{figure}[!htb]
     \centering
     \includegraphics[width=6.2in, height=3in]{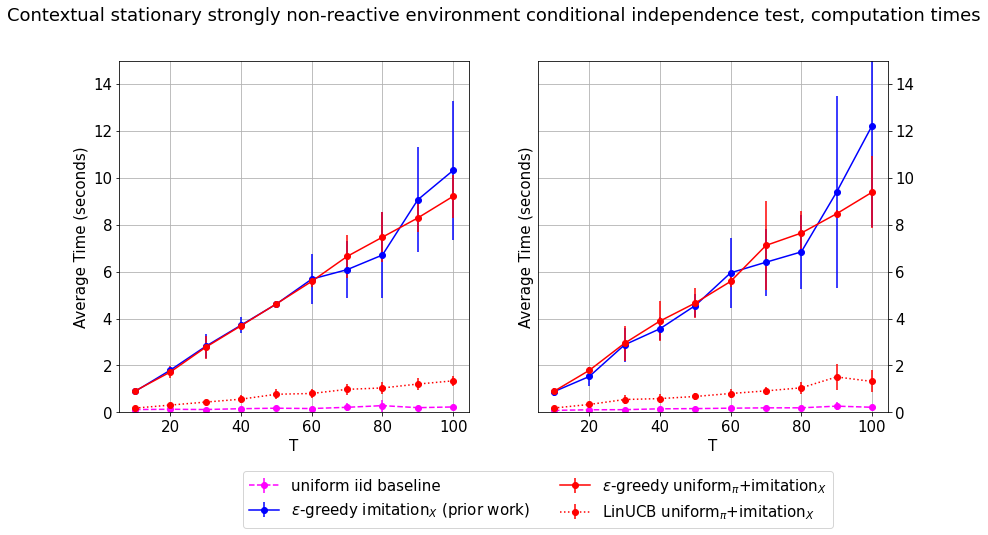}
     \caption{Computation times under the null (left) and alternative (right) distributions of randomization tests at fixed $\numMCSamples = 100$ and varying $\horizon$ in a contextual stationary strongly non-reactive environment on data gathered via $\epsilon$-greedy and LinUCB. Note that the computation times for LinUCB $\text{imitation}_X$ are omitted as both Type-I error and power curves are simply generated i.i.d.~$\text{Bern}(0.05)$.}
     %$\mathcal{X}= \{-1,1\}$ and $Y_i |X_i \sim \mathcal{N}(X_i,1), i \in [m-1]$. Data is collected using $\pi^{\egreedy}$ and $\pi^{\UCB}$, with $Y_m |X_m \sim \mathcal{N}(X_m,1)$ under the null $\mathcal{H}_0^{\Stat}$ and $Y_m |X_m \sim \mathcal{N}(4X_m,1)$ under the alternative $\mathcal{H}_1^{\Stat}$.}
     \label{fig:cont_dist_test_time}
 \end{figure}

\begin{figure}[!htb]
     \centering
     \includegraphics[width=6.3in, height=3.3in]{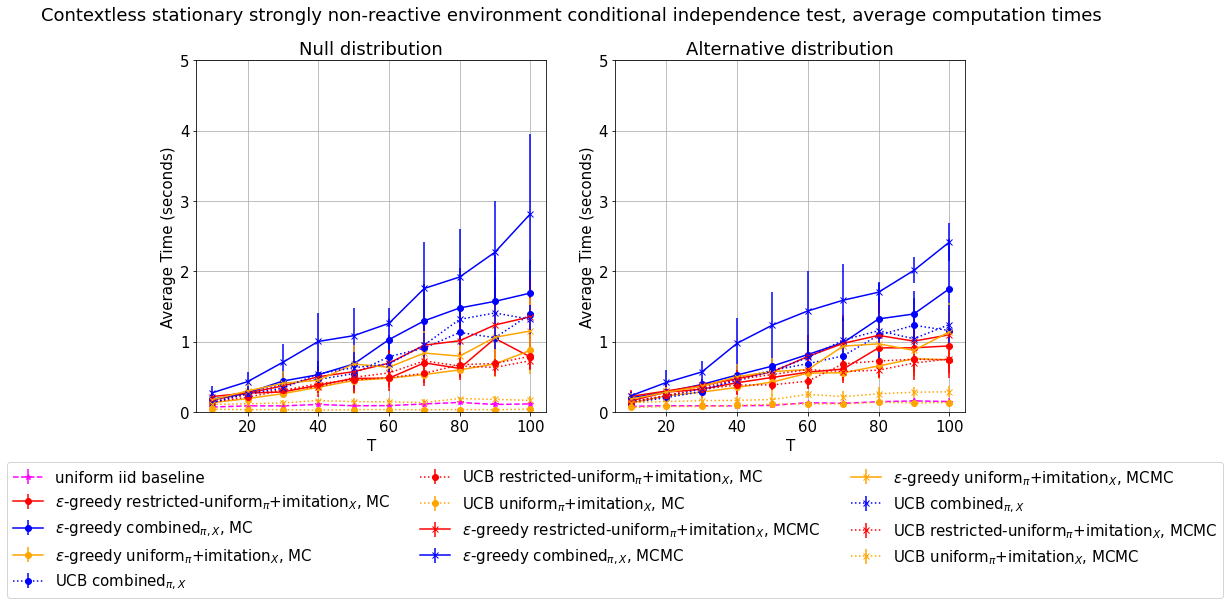}
     \caption{Computation times under the null (left) and alternative (right) distributions of both weighted MC and unweighted MCMC randomization tests at fixed $\numMCSamples=100$ and varying $\horizon$ in a contextless stationary strongly non-reactive environment on data gathered via $\epsilon$-greedy, UCB, and the uniform i.i.d.~baseline.}
     %$\mathcal{X}= \{-1,1\}$ and $Y_i |X_i \sim \mathcal{N}(X_i,1), i \in [m-1]$. Data is collected using $\pi^{\egreedy}$ and $\pi^{\UCB}$, with $Y_m |X_m \sim \mathcal{N}(X_m,1)$ under the null $\mathcal{H}_0^{\Stat}$ and $Y_m |X_m \sim \mathcal{N}(4X_m,1)$ under the alternative $\mathcal{H}_1^{\Stat}$.}
     \label{fig:dist_test_time}
 \end{figure}

\begin{figure}[!htb]
     \centering
     \includegraphics[width=6.3in, height=3.3in]{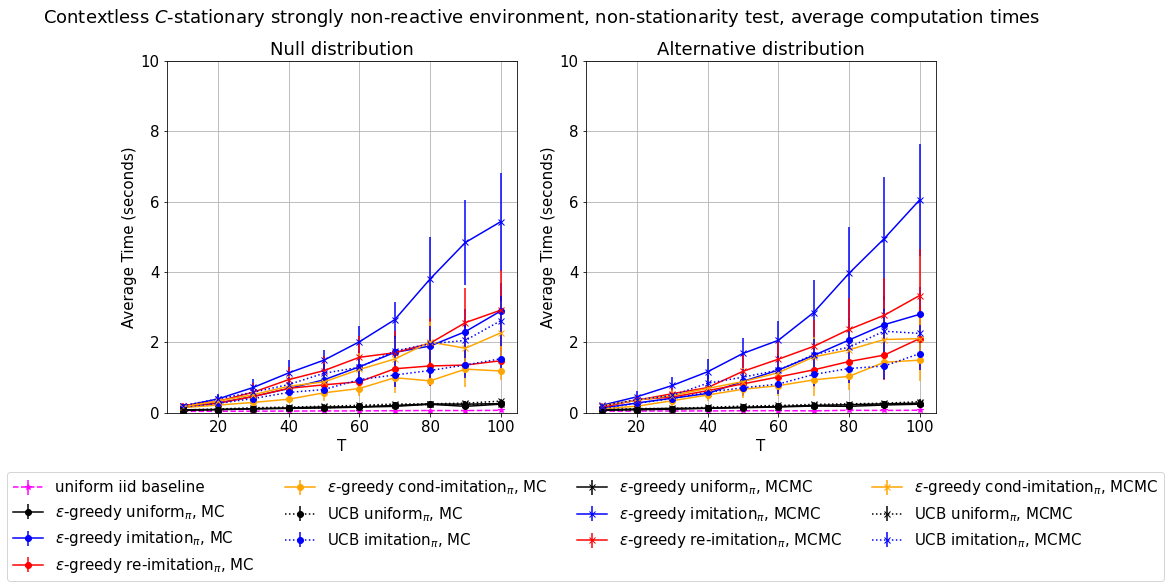}
     \caption{Computation times under the null (left) and alternative (right) distributions of both weighted MC and unweighted MCMC randomization tests at fixed $\numMCSamples=100$ and varying $\horizon$ in a contextless $C$-stationary strongly non-reactive environment with data gathered via $\epsilon$-greedy, UCB, and the uniform i.i.d.~baseline.}
     %$\mathcal{X}= \{-1,1\}$ and $Y_i |X_i \sim \mathcal{N}(X_i,1), i \in [m-1]$. Data is collected using $\pi^{\egreedy}$ and $\pi^{\UCB}$, with $Y_m |X_m \sim \mathcal{N}(X_m,1)$ under the null $\mathcal{H}_0^{\Stat}$ and $Y_m |X_m \sim \mathcal{N}(4X_m,1)$ under the alternative $\mathcal{H}_1^{\Stat}$.}
     \label{fig:nonstat_time}
 \end{figure}

\begin{figure}[!htb]
     \centering
     \includegraphics[width=6.3in, height=3.3in]{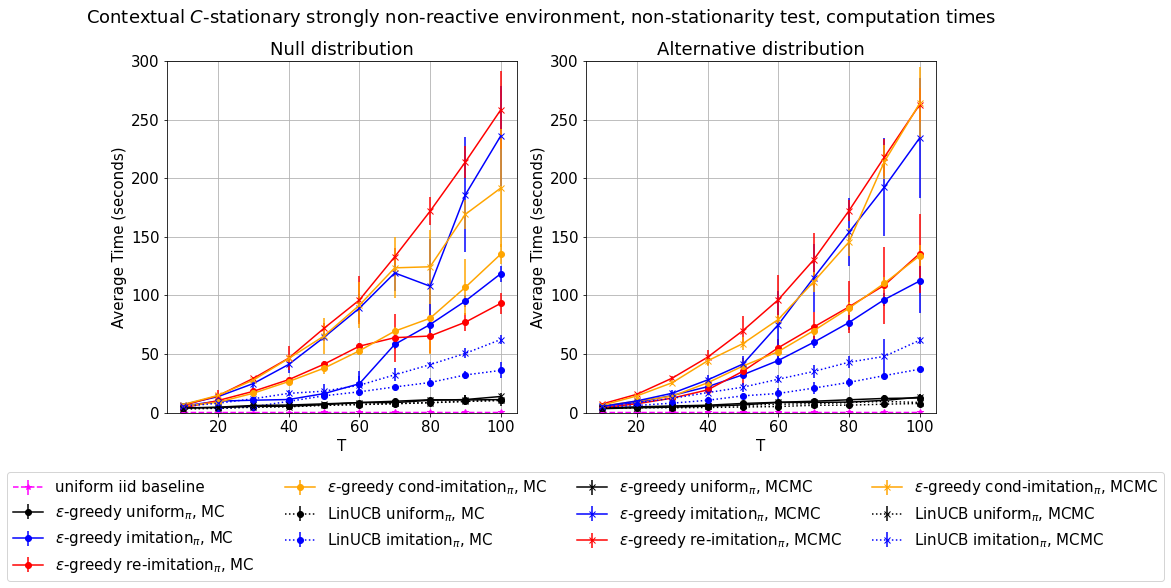}
     \caption{Computation times under the null (left) and alternative (right) distributions of both weighted MC and unweighted MCMC randomization tests at fixed $\numMCSamples=100$ and varying $\horizon$ in a contextless $C$-stationary strongly non-reactive environment with data gathered via $\epsilon$-greedy, LinUCB, and the uniform i.i.d.~baseline.}
     %$\mathcal{X}= \{-1,1\}$ and $Y_i |X_i \sim \mathcal{N}(X_i,1), i \in [m-1]$. Data is collected using $\pi^{\egreedy}$ and $\pi^{\UCB}$, with $Y_m |X_m \sim \mathcal{N}(X_m,1)$ under the null $\mathcal{H}_0^{\Stat}$ and $Y_m |X_m \sim \mathcal{N}(4X_m,1)$ under the alternative $\mathcal{H}_1^{\Stat}$.}
     \label{fig:nonstat2_time}
 \end{figure}

\begin{figure}[!htb]
     \centering
     \includegraphics[width=5.3in, height=3.15in]{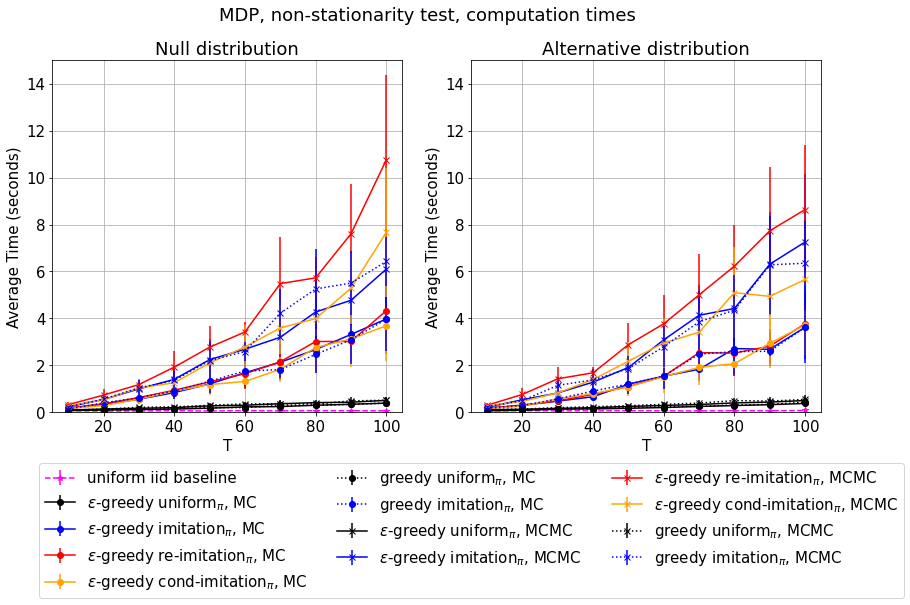}
     \caption{Computation times under the null (left) and alternative (right) distributions of both weighted MC and unweighted MCMC randomization tests at fixed $\numMCSamples=100$ and varying $\horizon$ in a contextless $C$-stationary strongly non-reactive environment with data gathered via $\epsilon$-greedy and greedy $Q$-learning.}
     %$\mathcal{X}= \{-1,1\}$ and $Y_i |X_i \sim \mathcal{N}(X_i,1), i \in [m-1]$. Data is collected using $\pi^{\egreedy}$ and $\pi^{\UCB}$, with $Y_m |X_m \sim \mathcal{N}(X_m,1)$ under the null $\mathcal{H}_0^{\Stat}$ and $Y_m |X_m \sim \mathcal{N}(4X_m,1)$ under the alternative $\mathcal{H}_1^{\Stat}$.}
     \label{fig:nonstat3_time}
 \end{figure}

\begin{figure}[!htb]
     \centering
     \includegraphics[width=6.5in, height=3.5in]{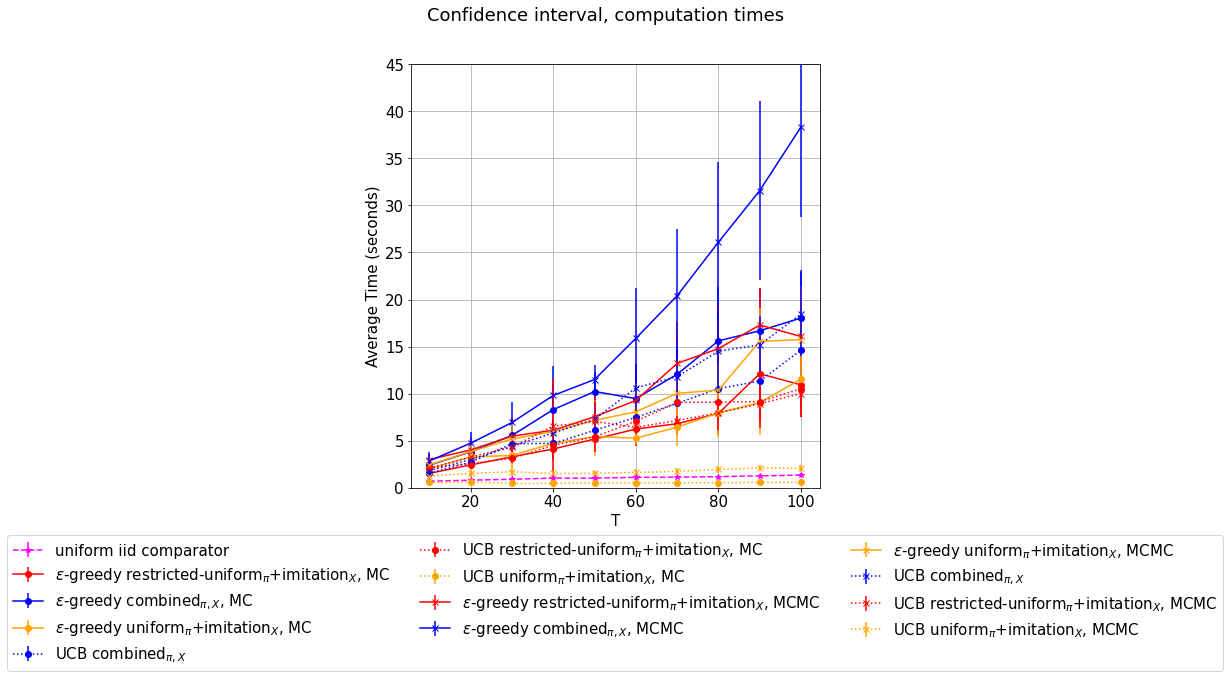}
     \caption{Computation time of construction of confidence interval for $b_0$ using both weighted MC and unweighted MCMC randomization tests with data gathered via $\epsilon$-greedy, UCB, and the uniform i.i.d.~baseline.}
     %$\mathcal{X}= \{-1,1\}$ and $Y_i |X_i \sim \mathcal{N}(X_i,1), i \in [m-1]$. Data is collected using $\pi^{\egreedy}$ and $\pi^{\UCB}$, with $Y_m |X_m \sim \mathcal{N}(X_m,1)$ under the null $\mathcal{H}_0^{\Stat}$ and $Y_m |X_m \sim \mathcal{N}(4X_m,1)$ under the alternative $\mathcal{H}_1^{\Stat}$.}
     \label{fig:conf_time}
 \end{figure}

\begin{figure}[!htb]
     \centering
     \includegraphics[width=6.5in, height=3.5in]{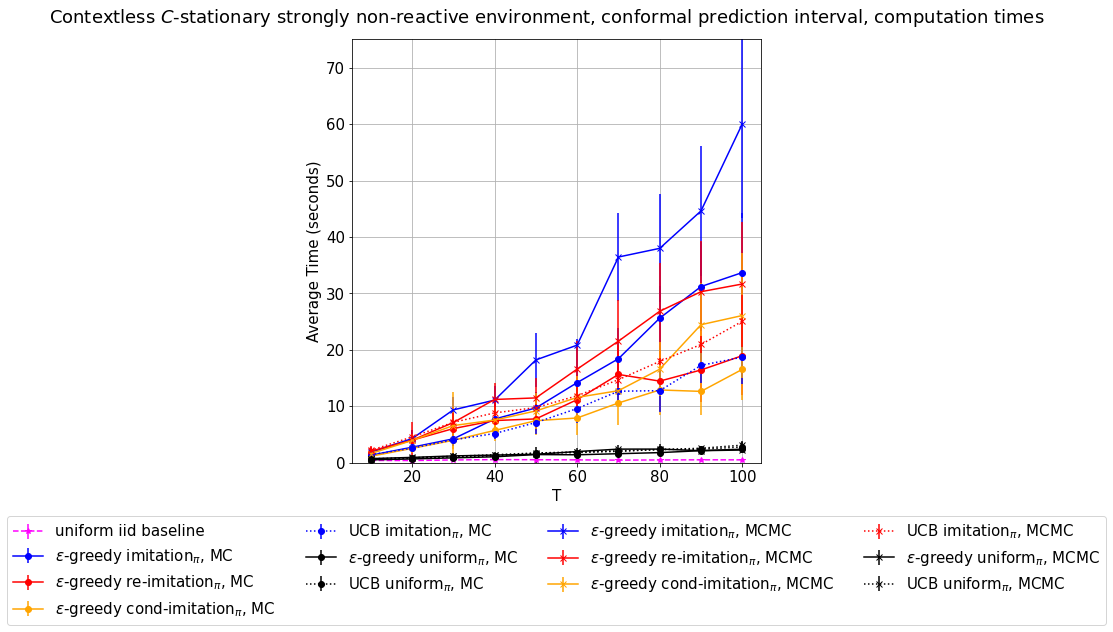}
     \caption{Computation time of construction of conformal prediction interval for $Y_\horizon$ using both weighted MC and unweighted MCMC randomization tests with data gathered via $\epsilon$-greedy, UCB, and the uniform i.i.d.~baseline.}
     %$\mathcal{X}= \{-1,1\}$ and $Y_i |X_i \sim \mathcal{N}(X_i,1), i \in [m-1]$. Data is collected using $\pi^{\egreedy}$ and $\pi^{\UCB}$, with $Y_m |X_m \sim \mathcal{N}(X_m,1)$ under the null $\mathcal{H}_0^{\Stat}$ and $Y_m |X_m \sim \mathcal{N}(4X_m,1)$ under the alternative $\mathcal{H}_1^{\Stat}$.}
     \label{fig:conformal1_time}
 \end{figure}

\begin{figure}[!htb]
     \centering
     \includegraphics[width=6.3in, height=3.3in]{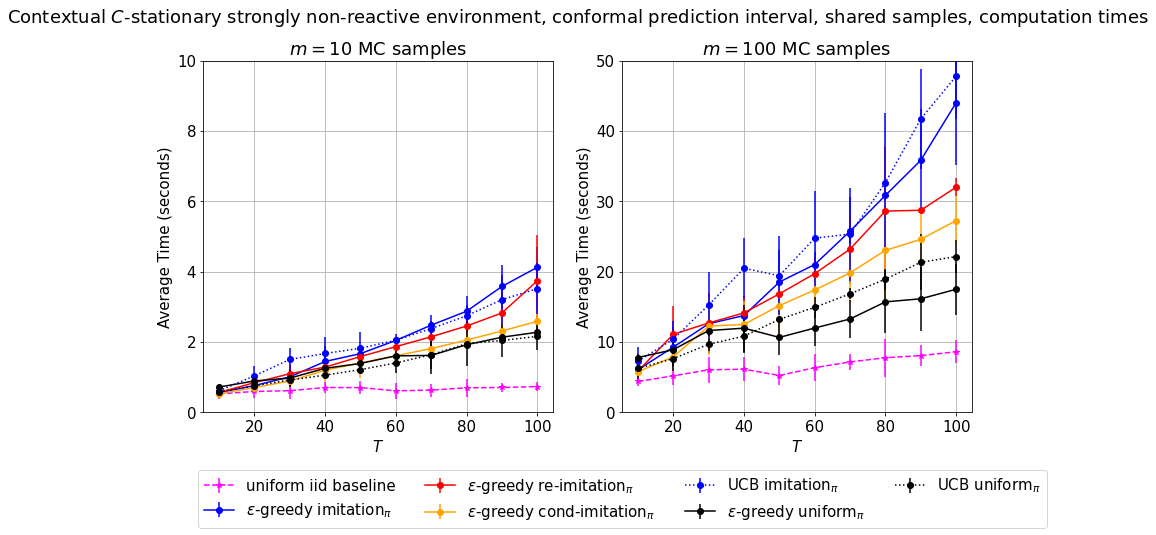}
     \caption{Computation time of construction of conformal prediction interval for $Y_\horizon$ using the MC randomization test with data gathered via $\epsilon$-greedy, UCB, and the uniform i.i.d.~baseline.}
     %$\mathcal{X}= \{-1,1\}$ and $Y_i |X_i \sim \mathcal{N}(X_i,1), i \in [m-1]$. Data is collected using $\pi^{\egreedy}$ and $\pi^{\UCB}$, with $Y_m |X_m \sim \mathcal{N}(X_m,1)$ under the null $\mathcal{H}_0^{\Stat}$ and $Y_m |X_m \sim \mathcal{N}(4X_m,1)$ under the alternative $\mathcal{H}_1^{\Stat}$.}
     \label{fig:conformal2_time}
 \end{figure}

 In this section, we plot the computation time curves (to compute $p$) for all resampling algorithms, environments, adaptive assignment algorithms, and types of randomization test (i.e., weighted MC or unweighted MCMC) discussed in Section~\ref{sim-chap} in Figures~\ref{fig:cont_dist_test_time}--\ref{fig:conformal2_time}.

 \subsection{Auxiliary simulation results}\label{auxiliary-sims}
\begin{figure}
     \centering
     \includegraphics[width=4.8in, height=3.3in]{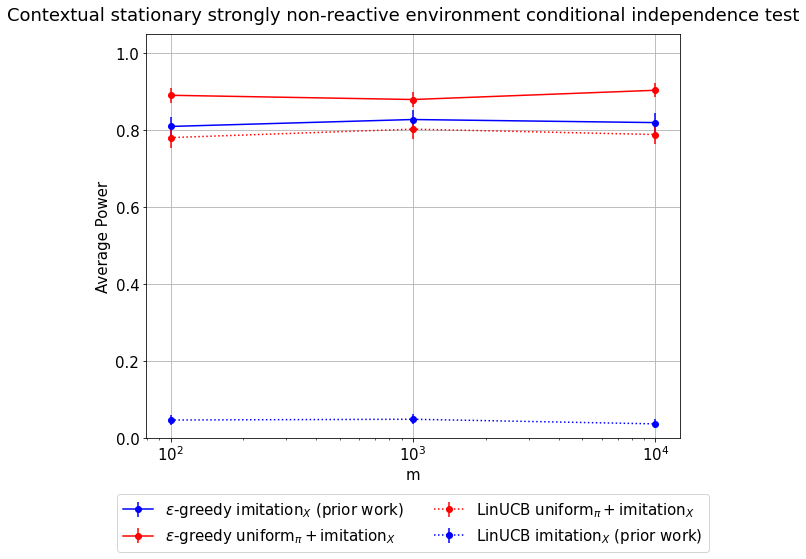}
     \caption{Power of randomization tests at fixed $\horizon = 100$ and varying $\numMCSamples$ in a contextual stationary strongly non-reactive environment on data gathered via $\epsilon$-greedy and LinUCB.}
     %$\mathcal{X}= \{-1,1\}$ and $Y_i |X_i \sim \mathcal{N}(X_i,1), i \in [m-1]$. Data is collected using $\pi^{\egreedy}$ and $\pi^{\UCB}$, with $Y_m |X_m \sim \mathcal{N}(X_m,1)$ under the null $\mathcal{H}_0^{\Stat}$ and $Y_m |X_m \sim \mathcal{N}(4X_m,1)$ under the alternative $\mathcal{H}_1^{\Stat}$.}
     \label{fig:cont_dist_test_100T}
 \end{figure}

\begin{figure}
     \centering
     \includegraphics[width=6.7in, height=3.3in]{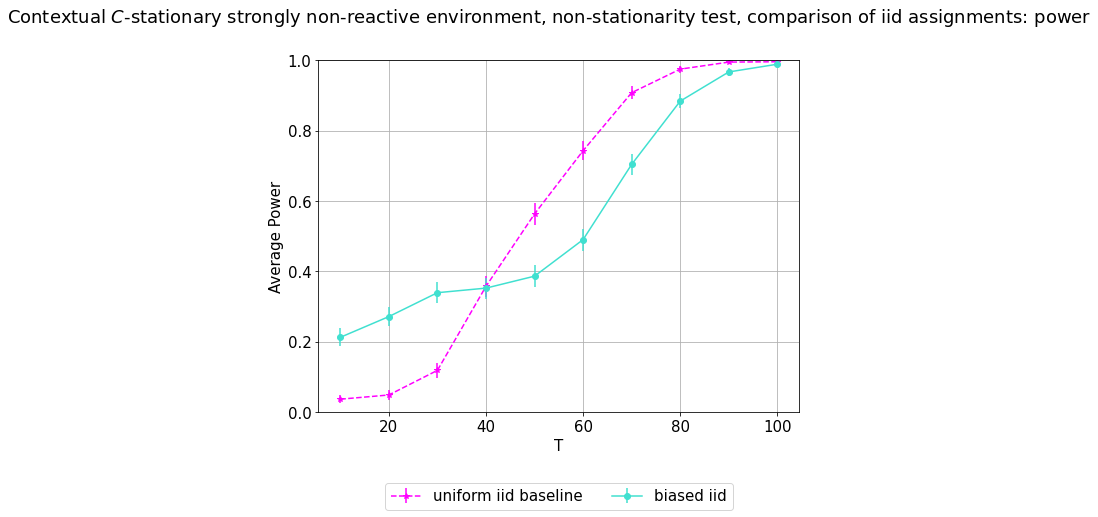}
     \caption{Power comparison of uniform i.i.d.~baseline versus biased i.i.d.~baseline that selects actions independently from $2\textnormal{Bern}(0.1)-1$}
     %$\mathcal{X}= \{-1,1\}$ and $Y_i |X_i \sim \mathcal{N}(X_i,1), i \in [m-1]$. Data is collected using $\pi^{\egreedy}$ and $\pi^{\UCB}$, with $Y_m |X_m \sim \mathcal{N}(X_m,1)$ under the null $\mathcal{H}_0^{\Stat}$ and $Y_m |X_m \sim \mathcal{N}(4X_m,1)$ under the alternative $\mathcal{H}_1^{\Stat}$.}
     \label{fig:comparison}
 \end{figure}

 In this section we present all remaining auxiliary simulation results. Figure~\ref{fig:cont_dist_test_100T} displays the power plot at fixed $\horizon = 100$ and varying $\numMCSamples$ for the conditional independence test in the contextual stationary strongly non-reactive environment on data gathered via $\epsilon$-greedy and LinUCB discussed in Section~\ref{prior-work-comparison}. On the other hand, Figure~\ref{fig:comparison} illustrates that the phenomenon of shift in relative performance of the uniform i.i.d.~baseline in comparison to both $\epsilon$-greedy and LinUCB, described in Section~\ref{context-stat-bandit} also occurs when the baseline is compared to a biased i.i.d.~adaptive assignment algorithm which selects actions at each timestep independently from $2\text{Bern}(0.1)-1$.

%\small
\end{document}